\documentclass[12pt,preprint]{aastex}
\usepackage{apjfonts,graphicx,graphics}
\usepackage{amsmath}
\usepackage{natbib}
\newcommand{\simless}
     {\ensuremath{\lower
3pt\hbox{$\rlap{\raise5pt\hbox{$\char'074$}}\mathchar"7218$}}}
\newcommand{\simgreat}
     {\ensuremath{\lower
3pt\hbox{$\rlap{\raise5pt\hbox{$\char'076$}}\mathchar"7218$}}}
\newcommand{\simgt}{\lower.5ex\hbox{$\; \buildrel > \over \sim \;$}}
\newcommand{\simlt}{\lower.5ex\hbox{$\; \buildrel < \over \sim \;$}}
\shorttitle{Magnetic Field Interpretation}
\shortauthors{Koch, Tang, \& Ho}
\begin{document}
\title{Interpreting the Role of the Magnetic Field from Dust Polarization Maps}
\author{
Patrick M. Koch\altaffilmark{1},
Ya-Wen Tang\altaffilmark{2,3,1}, and
Paul T. P. Ho\altaffilmark{1,4}
}
\altaffiltext{1}{Academia Sinica, Institute of Astronomy and
 Astrophysics, Taipei, Taiwan}
\altaffiltext{2}{Universit\'e de Bordeaux, Observatoire Aquitain des Sciences de l'Univers,
2 rue de l'Observatoire, BP 89, F-33271 Floirac Cedex, France}
\altaffiltext{3}{CNRS, UMR 5804, Laboratoire d'Astrophysique de Bordeaux,
2 rue de l'Observatoire, BP 89, F-33271 Floirac Cedex, France}
\altaffiltext{4}{Harvard-Smithsonian Center for Astrophysics, 60
 Garden Street, Cambridge, MA 02138, USA}

\email{pmkoch@asiaa.sinica.edu.tw}
%
%
\begin{abstract}

Dust polarization observations from the Submillimeter Array (SMA) and the Caltech
Submillimeter Observatory (CSO) are analyzed with the goal of providing a general tool
to interpret the role of the magnetic field in molecular clouds. Magnetic field and dust
emission gradient orientations are observed to show distinct patterns and features. The
angle $\delta$ between these two orientations can be interpreted as a magnetic field
alignment deviation, assuming the emission gradient orientation to coincide with the 
density gradient orientation in the magnetohydrodynamics (MHD) force equation. In 
SMA high-resolution (collapsing) cores, additional symmetry properties in $\delta$ can
reveal accretion and outflow zones. All these observational findings suggest the angle 
$\delta$ to be a relevant quantity that can assess the role of the magnetic field. Indeed,
when comparing this angle with the (projection-free) magnetic field significance $\Sigma_B$
\citep{koch12a}, it is demonstrated that $|\delta|$ yields 
an approximation
to the change in $\Sigma_B$.
Thus, changes in the magnetic field alignment deviation $\delta$ trace changes 
in the role of the magnetic field. The angle $\delta$ is observationally straightforward
to determine, providing a tool to distinguish between zones of minor or significant 
magnetic field impact. This is 
exemplified by
the CSO M$+0.25+0.01$, Mon R2, CO$+0.02-0.02$, 
M$-0.02-0.07$ sources and by the SMA high-resolution data from W51 e2, W51 North, Orion BN/KL
and g5.89. 
Additional CSO sources are analyzed, providing further support of this result.
Finally, based on the different features found in our sample 
of 31 sources in total,
covering sizes
from large-scale complexes to collapsing cores, a schematic evolutionary scenario is proposed.
Here, the significance of the magnetic field evolves both with position and scale, 
and can be assessed with the angle $\delta$.

\end{abstract}
%
%
\keywords{ISM: clouds --- ISM: magnetic fields, polarization
--- ISM: individual
          (W51 e2, W51 North, Orion BN/KL, G5.89-0.39, M$+0.25+0.01$, Mon R2, CO$+0.02-0.02$, 
M$-0.02-0.07$) --- Methods: polarization}
%
%

\section{Introduction}     \label{intro}

Magnetic field observations toward star forming regions are becoming increasingly 
more important as the relevance of the magnetic field in the star formation process is being
recognized in the literature \citep[e.g.,][]{crutcher12}. On the observational side, recent instrumentation
progress with improved sensitivities and polarization capabilities is advancing this field. 
Common observing techniques leading to some magnetic field information include Zeeman splitting, 
dust polarization, molecular line polarization and synchrotron emission. Any of these techniques
has shortcomings and advantages, usually revealing only partial information of a magnetic field
structure with its morphology, direction and strength. Zeeman splitting in spectral lines
\citep[e.g.,][]{crutcher09} typically leads to a line-of-sight field strength. In most cases, 
isolated positions or patches in a molecular cloud are mapped with this technique. Capturing the 
field morphology over a more extended area has been challenging. Recent observations of 
Zeeman splitting in masers are now able to start to trace the field morphology with a largely 
increased number of detections \citep{surcis11, vlemmings11}. Furthermore, results by 
\citet{etoka12} and \citet{surcis12} are consistently complementing dust polarization observations.
In this latter technique, dust grains are expected to be aligned with their shorter axis parallel
to the magnetic field lines most likely due to radiative torques \citep{draine96,draine97, lazarian07}, 
therefore revealing a plane-of-sky projected field orientation and morphology \citep[e.g.,][]{hildebrand88}.
A growing number of observations by both single dish telescopes (CSO, e.g., \citet{kirby09,dotson10,shinnaga12}, 
the James Clerk Maxwell Telescope (JCMT), e.g., \citet{matthews09}) and interferometers 
(the Berkeley-Illinois-Maryland Association (BIMA) array, e.g., \citet{lai02}, 
the SMA, e.g., \citet{girart06, tang09b, rao09} and the Combined Array for Research in Millimeter-wave
Astronomy (CARMA), \citet{hull13, stephens13}), in the (sub-)millimeter
regime with resolutions from $\sim 10\arcsec$ to sub-arcsecond, are detecting dust polarized emission
showing systematic magnetic field structures. Additionally, a comparison of dust polarization observations
at different frequencies can possibly also shed light on the nature of different grain populations
\citep{vaillancourt08, vaillancourt12}. Complementary to dust polarization, linearly polarized spectral
lines \citep{gk81, gk82} can probe the field morphology throughout the circumstellar envelope.
Recent detections in various molecular lines are giving hints of a complex field structure
in the envelope \citep{vlemmings12,girart12}. Unfortunately, both polarized line and dust
emission do not provide information on the magnetic field strength. Additional modeling is needed.
Recent techniques revealing additional magnetic field properties include e.g., polarization angle 
dispersion functions providing an estimate of the turbulent to mean magnetic field strength ratio  
\citep{hildebrand09, houde09, houde11}, comparison between co-existing ion and molecular line
spectra constraining the ambipolar diffusion scale \citep{hezareh11, li08} and the 
polarization-intensity gradient method leading to a position-dependent estimate of the 
magnetic field strength \citep{koch12a, koch12b}. Finally, observing polarized synchrotron 
emission from relativistic electrons has the potential to map morphology and field strength
in outflows and jets \citep{carrasco10}. This technique can, thus, ideally complement other magnetic field 
observations that are limited to core regions.

Dust polarization observations are probably providing the largest data sets so far to study
the role of the magnetic field in star formation regions. In particular, these observations
typically show a fairly large coverage, with polarized emission found throughout a significant
area overlapping with Stokes $I$ emission. Consequently, dust polarization maps are ideal to 
investigate systematic magnetic field features. Nevertheless, assessing the role of the magnetic 
field from such observations remains challenging. When interpreting the data, one is often left
with either relying on theoretical concepts or comparing the data with numerical simulations
\citep[e.g.,][]{padovani12}. In this work we aim at filling this gap.
We propose a phenomenological approach. Based on systematically observed features
we aim at providing guidelines for an interpretation of the role of the magnetic field. 
A particular emphasis is given to the relative importance of magnetic field versus gravity.
To that purpose, the relation between the magnetic field and emission gradient 
orientation (the angle $\delta$) -- originally pointed out in \citet{koch12a} --
is studied here on a broader sample with a further expanded complementary analysis.
We note that during the revision of this work, first numerical simulations analyzing
the relative orientations of magnetic fields with respect to density structures were
presented in \citet{soler13}. Their results indicate that both density and the level of 
magnetization leave their imprints on the distributions of relative orientations.

This present work has benefited from a growing number
of polarization observations over the past years. In particular, the high-resolution
(sub-)arcsecond observations by the SMA \citep{tang12,tang10,tang09b,tang09a,chen12,girart13,girart09,girart06, alves10, rao09}
were pioneering insights, methods and interpretation of the magnetic field that we are
now also able to apply to earlier lower-resolution observations by other instruments.
The proposed interpretation here is, thus, generally applicable to dust polarization
observations.
This study is part of the program on the SMA\footnote{
The Submillimeter Array is a joint project between the Smithsonian
Astrophysical Observatory and the Academia Sinica Institute of Astronomy
and Astrophysics, and is funded by the Smithsonian Institution and the
Academia Sinica.
}
\citep{ho04} to investigate the structure of the magnetic field from large
to small scales.

The paper is organized as follows. In Section \ref{interpretation} we present observational
results of the angle $\delta$ from 2 different instruments, from $\sim 20\arcsec$ to 
sub-arcsecond resolutions. We, additionally, provide an interpretation of the angle $\delta$
by manipulating the MHD force equation. 
Supplementary material putting this in a context with our previous work is given in 
Appendix A. 
A brief summary and comparison of the relevant 
magnetic field quantities from our previous and current study follows in Section \ref{summary_parameters}. 
Section \ref{section_tracer} establishes the angle $\delta$ as an approximation and reliable 
tracer for the changing role of the magnetic field over an observed map. 
Appendix B contains maps of a large sample in support of this finding.
As a consequence, in 
Section \ref{section_scenarios} we propose a schematic scenario where the magnetic field 
significance evolves and is revealed through distinct changes and features in the angle $\delta$.
Conclusion and summary are given in Section \ref{section_conclusion}.

\section{Intensity Gradient and Magnetic Field Orientations: the Angle $\delta$}  \label{interpretation}

Projected magnetic field orientations (rotated by 90$^{\circ}$ with respect to detected
dust polarization orientations) form an angle $\delta$ with their Stokes $I$
dust intensity gradient orientations (Figure \ref{schematic_basics}). 
Observationally, this angle $\delta$ is 
straight forward to determine. 
Based on our recent results that point toward a connection in these orientations
\citep{koch12a, koch12b}, we further investigate this finding here on a larger and 
diverse data set.
The data analyzed in Section \ref{delta_observations} 
and Appendix B
with their systematic patterns in the angle $\delta$ are suggestive
for a direct physical meaning of this angle. A first intuitive interpretation of 
$\delta$ was attempted in the context of the polarization - intensity gradient 
method \citep{koch12a}. In this new method it was assumed that the emission intensity
gradient is a measure for the result of all the forces acting in a molecular cloud.
In Section \ref{delta_mhd} we will proceed to explain the angle $\delta$ in the 
framework of ideal magneto-hydrodynamics (MHD).
Appendix A discusses in detail how this interpretation is linked to the approach 
in \citet{koch12a}.

\subsection{The Angle $\delta$ from Observations} \label{delta_observations}

We present results for the angle $\delta$, derived from both single dish and 
interferometric observations. In a first section we investigate features in $\delta$ in 4
sources where the magnetic field is probed at different physical scales. As we will argue 
later in the Sections \ref{section_tracer} and \ref{section_scenarios}, we consider these  
sources as representative for different evolutionary stages in the star formation process, 
where the role of the magnetic field can be assessed by analyzing the angle $\delta$.
In support of this argument, an additional set of 24 sources is analyzed in Appendix B.
In the second section here, we then focus on higher resolution observations which are probing
systems with collapsing cores, where the angle $\delta$ yet shows additional characteristics.

\subsubsection{$|\delta|$ over Different Scales}

Both Hertz on the CSO and SCUPOL on the JCMT have published catalogs of polarization observations toward 
56 \citep{dotson10} and 83 \citep{matthews09} regions, respectively, containing 
mostly Galactic sources and some galaxies\footnote{
The publicly available reduced data can be found at {\rm http://cdsarc.u-strasbg.fr/viz-bin/qcat?J/A+A/}
for the JCMT, and at {\rm http://iopscience.iop.org/0067-0049/186/2/406/fulltext/}
for the CSO.
}
. In the following we showcase an analysis of a few
of these sources, based on published data, to illustrate characteristic absolute value $|\delta|$-patterns.
An identical analysis is carried out for the additional 24 CSO sources in Appendix B.

On (large) scales, where a clear gravity center still remains to be formed, a $|\delta|$-map
seems irregular with no clear signatures (Figure \ref{figure_delta_scales}). 
In some areas, similar $|\delta|$-values are 
grouped together, other areas show a continuous change in $|\delta|$. Typically, specific 
$|\delta|$-values can not yet be associated with specific locations in the dust Stokes $I$ map.
This is because structures like clouds, envelopes and cores have not yet been formed.   
The complex CO$+0.02-0.02$ together with M$-0.02-0.07$, observed with a physical resolution
$\ell\sim 770$ mpc 
and covering a large area of about 12~pc $\times$ 12~pc 
(CSO, \citet{dotson10}), is such an example. 
One might speculate that such a system is 
still affected and shaped by its surroundings where larger scale dynamics can lead to 
irregular patterns
(Figure \ref{figure_delta_scales}). The correlation between magnetic field and intensity gradient
position angles ($P.A.$s) is not yet clearly established with more cloud-like groups 
(correlation coefficient $\mathcal{C}=0.71$). The 
histogram of $|\delta|$ shows a rather uniform distribution with $\langle|\delta|\rangle\approx 48^{\circ}$
(Middle bottom panel in Figure \ref{figure_delta_scales}).

On a next (smaller) scale we have selected 2 sources that show elongated structures:
Mon R2 ($\ell\sim 92$ mpc) and M$+0.25+0.01$ ($\ell\sim 770$ mpc
over an area of about 7.7~pc $\times$ 4~pc;
CSO, \citet{dotson10}). 
The top and middle rows in Figure 
\ref{figure_delta_scales_cont} reveal clear and distinct features. In the case of Mon R2 (top row), the 
magnetic field appears roughly perpendicular to the major axis of the Stokes $I$ contours. 
Here, material is likely following the field lines (driven by gravity) leading to an elongated
and possibly flattened structure with gravity also starting to pull in the field lines at the 
2 ends of the major axis. As a result, the $|\delta|$-map shows mostly small values, except 
toward the 2 ends of the major axis. Here, maximum deviations of the field from the intensity
gradient occur because the field is not yet clearly dragged in. The correlation, 
$\mathcal{C}=0.78$, for such a source is tighter than for the above system CO$+0.02-0.02$ 
with M$-0.02-0.07$. The $|\delta|$-histogram shows a clear peak at small values 
with $\langle|\delta|\rangle\approx 23^{\circ}$.
M$+0.25+0.01$ displays inverse features (middle row): the magnetic field is 
roughly along the major axis of the source. Consequently, in a $|\delta|$-map, the smallest
values are found at the 2 ends of the major axis, with the largest deviations in between 
them along the axis. The correlation is systematically off ($\mathcal{C}=0.61$), 
which then also leads to the largest average deviation, $\langle|\delta|\rangle\approx 57^{\circ}$, 
of the 3 sources described here. Table \ref{table_quantities} summarizes these results.

The collapsing core W51 e2 ($\ell\sim 24$ mpc, \citet{tang09b}) yet shows different signatures on these smallest scales
(bottom row in Figure \ref{figure_delta_scales_cont}). 
The source mostly appears symmetrical. $|\delta|$-values are generally
small with $\langle|\delta|\rangle\approx 18^{\circ}$. These 2 findings might indicate that the source is
already in a later phase of its collapse where gravity has pulled in and aligned the field
lines with the intensity gradient. There is a trend of increasing $|\delta|$-values beneath
and above a plane in northeast-southwest direction. Polarized emission is absent along this
direction. This area likely coincides with an accretion plane 
\citep{ho96, zhang97, young98, zhang98, solins04,keto08}.
We additionally explore this particular configuration in the next Section \ref{section_delta_high_resolution}.
Larger $|\delta|$-values as found in Mon R2 are missing here. This, however, might be due
to the current limitations in resolution and/or sensitivity of the observation. There are 
hints of increased $|\delta|$-values around the emission peak. Compared to Mon R2, the 
zones of larger $|\delta|$-values appear to have shrunken and moved closer to the center.
W51 e2 shows the tightest correlation between field and intensity gradient orientations 
with $\mathcal{C}=0.95$ (Table \ref{table_quantities}). 

It is striking that the $|\delta|$-maps of the above sources Mon R2 and M$+0.25+0.01$ reveal
clear opposite trends. 
Inspecting the additional 24 sources from the CSO catalog shows that these trends are also
found in other sources (Appendix B). The non-uniform and incomplete polarization coverage
in some sources obviously tends to erase clear features. 
Projection effects are unlikely mimicking these trends. 
It seems more plausible that certain selection effects
and initial conditions from larger scales lead to one or the other case, with projection 
effects possibly distorting some features. Large-scale flows might initially compress the 
field lines, leading to structures like seen in M$+0.25+0.01$. Subsequently, gravity might 
take over and start to pull in the field lines. This will then eventually lead to a more
symmetrical structure like in W51 e2. The field morphology in Mon R2 matches with an
hourglass-like scenario where initially straight field lines are later pulled in by 
gravity leading again to an e2-type configuration. Mon R2 might be showing the beginning
of such a phase. 

In summary, the sources presented here seem suggestive for an evolutionary sequence where
the role of the magnetic field leaves some signatures in the $|\delta|$-maps. 
In Section \ref{section_tracer} we will add further evidence to this idea, 
which will allow us to conclude with a schematic scenario in Section \ref{section_scenarios}.

\subsubsection{$\delta$ in High-Resolution Cores}  \label{section_delta_high_resolution}

We highlight an additional feature of $\delta$-maps, focusing on high-resolution (up to $\sim0\farcs7$)
cores observed with the SMA. Figure \ref{figure_delta_cores} shows $\delta$ for W51 e2 \citep{tang09b}, 
W51 North \citep{tang12}, G5.89-0.39\footnote{
Here and in the following sections G5.89-0.39 is abridged as g5.89.
} \citep{tang09a} and Orion BN/KL \citep{tang10}.
We note that $\delta$ is displayed in the range of $-90^{\circ}$ to $+90^{\circ}$, with the sign 
depending on whether the magnetic field is rotated counter-clockwise or clockwise with respect
to the intensity gradient. For an enhanced visual impression the data are over-gridded. The color 
coding makes clear that there are systematic changes in the sign of $\delta$ between confined areas.
This is particularly obvious for W51 e2 and the main cores in W51 North and Orion BN/KL. Nevertheless, 
even in g5.89 -- where only isolated patches of polarized emission are detected, possibly due to 
expanding HII regions \citep{tang09a} -- large areas with identical signs are found. We further 
analyze these maps by investigating the azimuthal $(az)$ dependence in $\delta$. The top panel in 
Figure \ref{figure_delta_azimuth} illustrates $\delta=\delta(az)$ for W51 e2 and the 
main cores in W51 North and Orion BN/KL.
For all sources azimuth is measured counter-clockwise from West, with its center assumed to be
at the emission peak of each source. There are seemingly ranges in 
azimuth where $\delta$ is preferentially negative or positive, and there are zones where $\delta$
changes sign. In the middle panel we have by eye re-defined the origin of the azimuth coordinate in 
order to align the features from the top panel. Starting from positive values around $40^{\circ}$
to $80^{\circ}$, all 3 sources then show smaller and smaller values, reaching similar negative
values after crossing zero. After this first zero-crossing there is a clear gap with no data
for Orion BN/KL and W51 North. Then, values jump to the positive followed by a second zero-crossing
over a smaller azimuth range. A clear gap is found here for W51 e2 between $az\approx 350^{\circ}$
and $az\approx 50^{\circ}$. We remark that between $az\approx 360^{\circ}$ and $az\approx 0^{\circ}$
all sources flip sign in $\delta$ again for a second time. Thus, aligning the features
from the top panel in Figure  \ref{figure_delta_azimuth} clearly reveals systematic changes
in $\delta$ with 2 fairly smooth zero-crossings and 2 zones where $\delta$ flips sign. 
For comparison, the bottom panel in  Figure  \ref{figure_delta_azimuth} shows $\delta$ as a 
function of azimuth for g5.89, where no clear features can be identified. 

Figure \ref{figure_schematic_delta_azimuth} proposes an explanation for the above findings
in the context of a collapsing system with accretion and outflow zones. Initially straight
field lines are dragged in and bent by gravity. With respect to the intensity gradient 
orientations, distinct zones with negative or positive $\delta$-values emerge.
The largest field curvature is expected in the accretion zones with $\delta$ flipping its
sign between maximum negative and maximum positive values 
around
the accretion 
plane. With material mostly following the field lines and the lines being less bent toward
the poles, $|\delta|$ is decreasing toward the outflow regions. This then leads to a continuous
change in $\delta$ across zero. We remark that outflows are likely to perturb the field 
morphologies around the poles. Nevertheless, due to their small mass and possibly smaller 
dust-to-gas ratio, the currently available observations with their sensitivities and 
resolutions are predominantly tracing the core's field morphology. We, thus, neglect a 
possible influence of the outflow in the proposed scenario here.  

Remarkably, despite being illustrated for a face-on pinched field structure in 
Figure \ref{figure_schematic_delta_azimuth}, all 3 sources in the middle panel in Figure 
\ref{figure_delta_azimuth} clearly reveal features matching the schematic. Nevertheless, 
we have to note that changes from positive to negative zones do not exactly occur  within 
a $180^{\circ}$ range in azimuth.  Equally, maximum and minimum values in $\delta$ vary
with source. One might speculate that irregularities and differences are due to asymmetries
and inclination/projection effects that stretch or compress certain features. In spite
of these shortcomings, we can still conclude that the magnetic field azimuthal patterns
in $\delta$ provide some clues on the state of a system by identifying characteristic
accretion and outflow areas. Furthermore, these features appear to be robust even in 
the presence of unknown inclination angles. Finally, the proposed explanation in 
Figure \ref{figure_schematic_delta_azimuth} naturally leads to a bimodal distribution
in $\delta$. This has indeed been observed for W51 e2 in Figure 2 in \citet{koch12a}.
Distributions for all high-resolution cores discussed here are displayed in 
Figure \ref{figure_histogram}.

\subsection{The Angle $\delta$ in the MHD Force Equation} \label{delta_mhd}

\begin{figure}
\begin{center}
\includegraphics[scale=1]{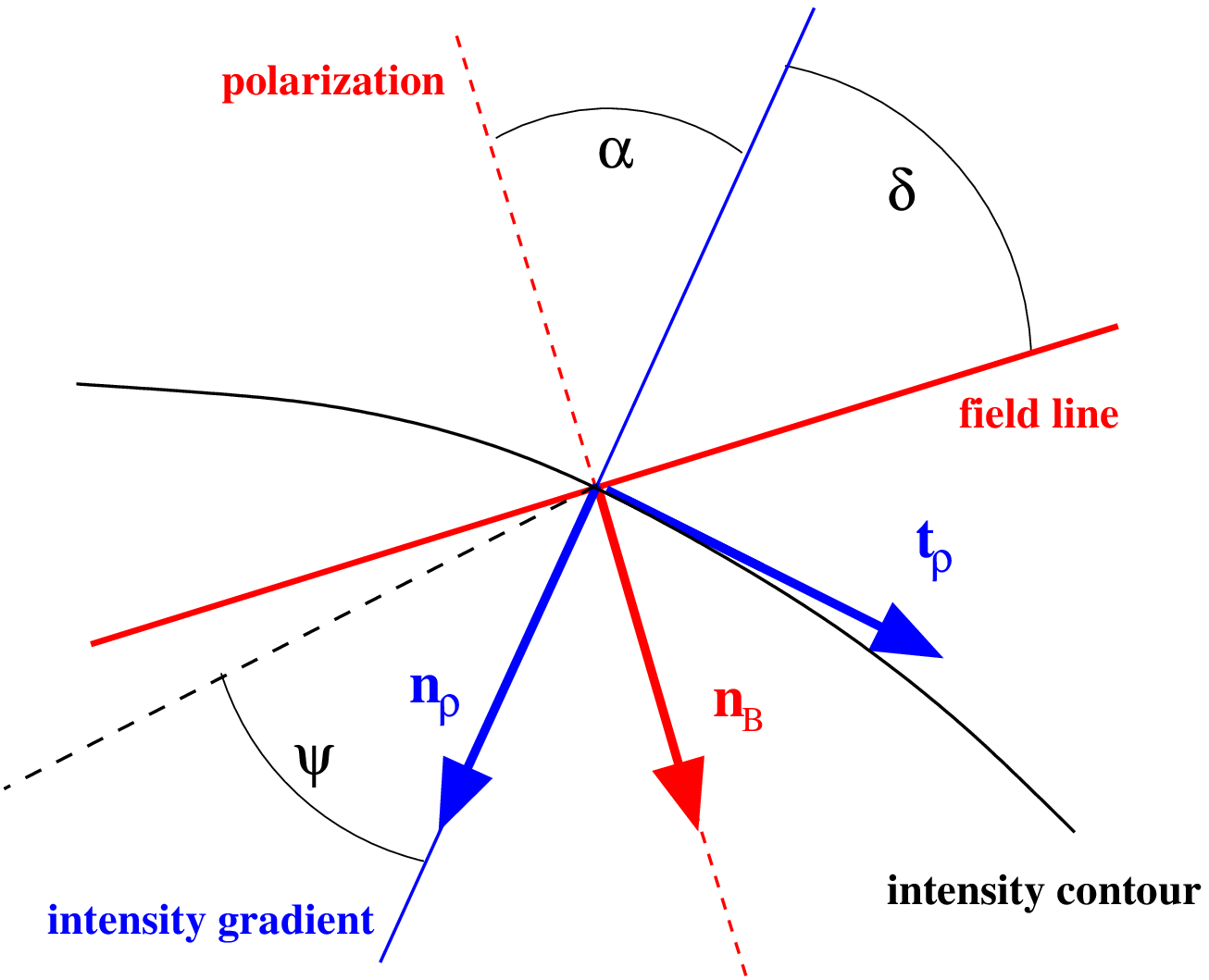}
 \caption{\label{schematic_basics} 
The angle $\delta$ between magnetic field orientation (red solid line) and  
intensity gradient orientation (blue solid line). 
The magnetic field tension force is directed normal to the field line along 
the unity vector $\mathbf{n}_B$ which is collinear to the originally detected 
dust polarization orientation (red dashed line).  The unity vector 
$\mathbf{n}_{\rho}\equiv \frac{\nabla \rho}{|\nabla \rho|}$ is normal to the 
emission intensity contour (black solid line) which leads to
$\mathbf{n}_B \cdot \mathbf{n}_{\rho} = \cos\alpha=\sin\delta$ with $\delta+\alpha=\pi/2$.
The unity vector $\mathbf{t}_{\rho}$ is tangential to the emission contour, forming
an orthonormal system together with $\mathbf{n}_{\rho}$.
The deviation between intensity gradient and gravity and/or pressure gradient orientations
(black dashed line) is indicated with the angle $\psi$.
}
\end{center}
\end{figure}

\begin{figure}
\begin{center}
\includegraphics[scale=0.45]{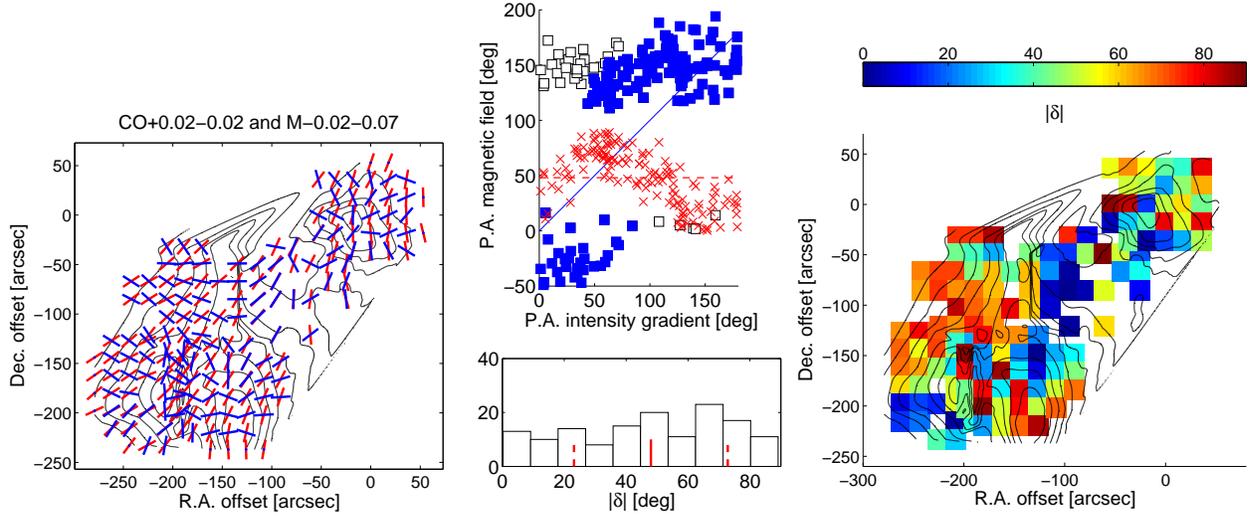} 
 \caption{\label{figure_delta_scales}
The complex CO$+0.02-0.02$ together with M$-0.02-0.07$ (CSO, \citet{dotson10}).
The left panel shows the dust Stokes $I$ continuum contours, overlaid with 
the magnetic field segments (red) and the intensity gradient segments (blue). 
The right panel displays the absolute difference, $|\delta|$, between the 2
orientations in the range from 0$^{\circ}$ to 90$^{\circ}$. The correlation 
between the magnetic field and intensity gradient orientations is illustrated 
with the blue filled squares in the middle top panel.
The black empty squares belong to pairs with $P.A.$s close to $P.A.=0$,
one $P.A.$ being on the left and the other being on the right 
hand side of the vertical. In order to properly display their
correlations, the magnetic field $P.A.$ is re-defined beyond the 
0 to 180$^{\circ}$ range for these cases (blue filled squares above 
180$^{\circ}$ and below 0$^{\circ}$).
For visual guidance added is the straight blue line, 
representing a perfect correlation.
Also shown are the absolute differences ($\le 90^{\circ}$)  
between the $P.A.$s for each pair (red crosses). Both  $P.A.$s are defined
counter-clockwise starting from north. The red dashed line marks the 
average absolute difference $\langle|\delta|\rangle$. 
Errors in the magnetic field $P.A.$s are typically a few degrees, with maximum
uncertainties up to about 10$^{\circ}$ (not shown). Uncertainties in the intensity gradient
$P.A.$s are limited to a few degrees after averaging and interpolating Stokes $I$
values. Resulting errors in $|\delta|$ then range from a few degrees up to a 
maximum of about 10$^{\circ}$.
The histogram in 
the middle bottom panel represents the distribution of $|\delta|$ with 
its mean and $\pm$-standard deviation marked with the red solid and 
dashed lines.
}
\end{center}
\end{figure}

We adopt an ideal MHD force equation as a starting point for the following derivation. 
As in \citet{koch12a}, we 
impose the slight restriction that the orthogonal field component is a slowly varying
function, i.e. $\frac{\Delta B_{\bot}}{B}\ll 1$. This will hold for any smooth large-scale 
field functions\footnote{
On very small scales this assumption might eventually fail for tangled magnetic 
fields. There is, however, no indications of such tangled fields to date from observed field morphologies.
We further remark that this formalism does not rule out the existence of turbulence.
In fact, the total level of turbulence $\Delta B$ in W51 e2 was independently estimated 
based on a structure function \citep{koch12a}. Uncertainties in observed polarization
position angles due to turbulence are found to be typically 5 to 10$^{\circ}$, 
which translates into $\Delta B\sim \pm 1$ mG for W51 e2. This is small compared to the total 
field strength estimates in most positions except for the weakest ones. Thus, current 
observational results justify the above assumption.
}.
In return, this then allows us to simplify and combine the magnetic field hydrostatic
pressure and the field tension terms. With this, the force equation becomes \citep{koch12a}:
\begin{equation}
\rho\left(\frac{\partial}{\partial t}+\mathbf{v}\cdot\mathbf{\nabla}\right)\mathbf{v} =
-\mathbf{\nabla}P -\rho\nabla\phi+\frac{1}{4\pi}\frac{1}{R}B^2 \,\,\mathbf{n}_B,
                                                                 \label{mhd_momentum}
\end{equation}
where $\rho$ and $\mathbf{v}$ are the dust density and velocity, respectively.
$B$ is the magnetic field strength. $P$ is the hydrostatic dust pressure. 
$\phi$ is the gravitational potential resulting from the 
total mass contained inside the star forming region. $\mathbf{\nabla}$ denotes
the gradient. The field tension force (last term on the right hand side) 
with the field curvature $1/R$ is directed normal to the field line
along the unity vector $\mathbf{n}_B$ (Figure \ref{schematic_basics}).
As usual, the inertial term on the left hand side in Equation (\ref{mhd_momentum}) 
describes the resulting action based on the force terms on the right hand side.

In an observed map, the detected intensity emission is typically a function of temperature
and density. We now assume that the direction of the local temperature gradient is identical 
or at least not too different from the local density gradient direction\footnote{
It has to be stressed that the temperature does {\it not} have to be constant. Neither requires
the above argument that the temperature changes more slowly than the density. The temperature 
can even change faster as long as its direction aligns roughly with the density gradient 
direction. {\it Locally} identifying $\nabla\rho$ with the intensity emission gradient
only needs the {\it local directions}
of temperature and density gradients to be aligned. Thus, regions of different temperatures
do not pose a problem in our approach, as long as temperature changes are accompanied by 
density changes toward roughly the same direction.
}. 
In this case, the observed intensity gradient direction on a map can be identified with 
the direction of the density gradient along the unity vector 
$\frac{\nabla \rho}{|\nabla \rho|}\equiv \mathbf{n}_{\rho}$. This connects
$\mathbf{n}_{\rho}$ and $\mathbf{n}_B$ as $\mathbf{n}_{\rho}\cdot\mathbf{n}_B=\cos\alpha$, 
where $\alpha$ is the angle between the intensity gradient and the originally detected
polarization orientations (Figure \ref{schematic_basics}, see also Figure 3 in \citet{koch12a}). 
$|\delta|=\pi/2-|\alpha|$ is the angle between the magnetic field and the intensity gradient 
orientations analyzed in the previous sections\footnote{
The angle $\delta$ is always defined as the smaller of the 2 complementary angles to $\pi$
at the interception of field and intensity gradient orientations; i.e. $\delta$ is always in 
the range between 0 and $+\pi/2$. Absolute values are used in the equations here because 
$\delta$ can also be given a sense of orientation (Section \ref{section_delta_high_resolution}).
}
. We are now aiming at manipulating Equation (\ref{mhd_momentum}) in a way that 
changes/variations in the morphology in a map can be attributed to the various forces.
In particular, we want to link the field tension along $\mathbf{n}_B$ to morphological 
features. To that purpose, we are projecting $\mathbf{n}_B$
onto the orthonormal system $(\mathbf{n}_{\rho},\mathbf{t}_{\rho})$, 
$\mathbf{n}_B=\sin\delta \,\,\mathbf{n}_{\rho} + \sin\alpha \,\, \mathbf{t}_{\rho}$ 
(Figure \ref{schematic_basics}), where $\mathbf{t}_{\rho}$ is a unity vector tangential
to the intensity contour. Equation (\ref{mhd_momentum}) can, thus, be written as:

\begin{equation}
\rho\left(\frac{\partial}{\partial t}+\mathbf{v}\cdot\mathbf{\nabla}\right)\mathbf{v} =
-\mathbf{\nabla}P -\rho\nabla\phi+\sin|\delta|\,\,\frac{1}{4\pi}\frac{1}{R}B^2 \,\,\mathbf{n}_{\rho}
                                 +\sin|\alpha|\,\,\frac{1}{4\pi}\frac{1}{R}B^2 \,\,\mathbf{t}_{\rho}.
                     \label{mhd_delta_vect}
\end{equation} 
Equation (\ref{mhd_delta_vect}) provides a 
direct interpretation for the angle $|\delta|$, with 
some interesting insight which we are addressing in the following. 
Without the field tension 
force, the above Equation (\ref{mhd_delta_vect}) would describe the 
dynamics governed by only the gravitational force (and a possible pressure gradient). 
The angle $|\delta|$, with 
$\sin|\delta| \in [0,1]$, quantifies the fraction of the available field tension along the 
density gradient direction. Therefore, $\sin|\delta|$ is a measure for how efficiently the magnetic
field contributes to and is responsible for an observed density (emission) morphology
along the gradient directions. 

In the context of a collapsing system, $\sin|\delta|$ measures how efficiently the magnetic 
field inhibits a collapse. In the extreme case where $\sin|\delta|\approx 0$ 
($|\delta|\approx 0^{\circ}$), an observed density (emission) gradient is aligned with the 
field. Consequently, for any field strength, the magnetic field does not influence the 
local structure along the density gradient direction.
In a close-to-circular e2-type system, the local dynamics along the gradient directions, i.e. 
in radial direction, proceed as if the field does not exist (Equation (\ref{mhd_delta_vect})). 
The dynamics in radial direction $\mathbf{n}_{\rho}$  are entirely and only driven by 
the gravitational force $\rho\nabla\phi$ (with a likely negligible pressure gradient $\nabla P$).
In other words, in this situation the field plays a very minor role compared to gravity. 
This conclusion is consistent with our result in \citet{koch12b} where 
the field-to-gravity force ratio $\Sigma_B$ is calculated to be $\sim 20\%$ in the e2 core.
Nevertheless, Equation (\ref{mhd_delta_vect}) shows that a tangential force component
depending on field strength and curvature is also present. For pulled-in relatively 
straight field lines, a small curvature (large $R$) will suppress this component. Closer to 
the center, an increasingly smaller curvature radius might lead to additional dynamics
in tangential direction. 
In the other extreme case when $\sin|\delta|\approx 1$ 
($|\delta|\approx 90^{\circ}$), the magnetic field maximally contributes to an observed
morphology along the density gradient direction. 
Besides the orientation of the field with the density gradient, the 
absolute field strength is also relevant here. With $\sin|\delta|>0$, the field will 
always work against gravity, slowing down gravity\footnote{
This is certainly the case where gravity is bending the field lines and dragging them
in, as e.g., in a collapsing system as discussed here. We, however, have to acknowledge 
that, generally speaking, $\mathbf{n}_{B}$ in Figure \ref{schematic_basics} could also
point in opposite direction if the field line is bent the other way. In such a case, 
the projection onto the same orthonormal system $(\mathbf{n}_{\rho},\mathbf{t}_{\rho})$
will lead to opposite signs in the Equation (\ref{mhd_delta_vect}). 
This ambiguity can be overcome by introducing a sign for the field curvature, $\pm1/R$, 
for a convex or concave field shape.
}
at a maximum when $|\delta|=90^{\circ}$.
At such locations we would generally expect a density gradient that is less steep with 
emission contours that are spaced further apart. 
The tangential force component in Equation (\ref{mhd_delta_vect}) is negligible here.

In summary, different values in the angle $|\delta|$ reflect regions with different dynamics.
The larger $|\delta|$, the more important the field is dynamically in working against 
gravity. This is also consistent with a decreasing force ratio $\Sigma_B$ with smaller 
radius in a collapsing core as derived in \citet{koch12b}.

\section{Magnetic Field Quantities}         \label{summary_parameters}

Here, we provide a brief summary and comparison of the magnetic 
field parameters resulting from our work so far (Section \ref{comparison}).
Besides the angle $\delta$ introduced in the previous section, we have already derived 
a technique to estimate the local magnetic field significance $\Sigma_B$ and the local
field strength $B$ in \citet{koch12a}. In Section \ref{consistent_picture} we demonstrate
how these 2 different studies are connected. A detailed discussion on the relevant 
assumptions is given in Appendix A.

\subsection{$\Delta_B$, $\Sigma_B$ and $B$ -- A Summary} \label{comparison}

The combination of the three parameters -- $\sin|\delta|$, $\Sigma_B$ and $B$ -- characterizes and 
quantifies the local role of the magnetic field: 
\begin{eqnarray}
field\,\,\, alignment\,\,\, deviation:    &\Delta_B&=\sin|\delta|, 
                                                   \label{eq_delta_align}                                      \\
relative\,\,\, field\,\,\, significance:  &\Sigma_B&=\frac{\sin\psi}{\sin\alpha}=\frac{\Delta_G}{\sqrt{1-\Delta_B^2}},
                                                         \label{eq_sigma_B}                                    \\
absolute\,\,\, field\,\,\, strength:      &B&=\sqrt{\frac{\sin\psi}{\sin\alpha}\left(\rho\nabla\phi+\nabla P\right)\cdot 4\pi R}\nonumber \\
                                           &&  =\sqrt{\Sigma_B\left(\rho\nabla\phi+\nabla P\right)\cdot 4\pi R}, \label{eq_B}
\end{eqnarray}
where we have introduced the new symbols $\Delta_B\equiv \sin|\delta| = \cos \alpha$ 
for the field deviation and $\Delta_G\equiv \sin \psi$ for the gravity alignment deviation, 
$\psi$ being the angle between the intensity gradient and the gravity and/or pressure force orientation
(Figure \ref{schematic_basics} and Figure 3 in \citet{koch12a}).
The field strength $B$ and its relative significance $\Sigma_B$ compared
to the gravity and/or pressure forces are analyzed in detail in \citet{koch12a,koch12b}. 

We stress that both $\Sigma_B$ 
and $\Delta_B$ rely only on measurable angles.
Thus, both parameters are independent of any
mass and field strength assumptions and any further modeling of a molecular cloud.
The two quantities are solely based on angles which reflect the field and 
dust morphologies that are the overall result of all the forces acting in the 
system. The relative field significance $\Sigma_B$ leads to the total field strength $B$ when 
adding the mass and density information. This breaks the degeneracy in the dimensionless 
'scale-free' quantities based on angles only, and connects them to each individual 
physical system. The order of magnitude of the field strength is, thus, largely
determined by the mass, with $\Sigma_B$ quantifying the fraction of this mass
that can be balanced by the field.  
Whereas $\Sigma_B$ compares locally the field force to the other forces, $\Delta_B$
measures how much of the maximally available field tension force is taking part 
in the dynamics along the gradient direction. 
Thus, $\Delta_B$ is constrained to an upper limit of 1, whereas
$\Sigma_B$ has no upper limit and can be larger than one if the field is 
dominating. Table \ref{table_field_quantities} compares the properties of these 
three field parameters.


\subsection{Linking $\Delta_B$ and $\Sigma_B$} \label{consistent_picture}

Figure \ref{sigma_b_inter} illustrates the connection between field and gravity deviations, 
$\Delta_B$ and $\Delta_G$, and the field significance $\Sigma_B$ based on Equation (\ref{eq_sigma_B}). 
Small values in the deviation $\Delta_B$ -- i.e., close alignment between magnetic field
and intensity gradient orientations -- minimize the contribution of the field tension 
term along the density gradient direction in Equation (\ref{mhd_delta_vect}).
This is already suggestive
for a rather minor role of the magnetic field in this situation. Nevertheless, this 
criterion alone does not yet provide any information on the field compared to the other
forces in Equation (\ref{mhd_delta_vect}). This caveat is overcome with the field significance
$\Sigma_B$ (as a result from the polarization - intensity gradient method). And indeed, 
this reveals that the field significance scales roughly as $\sim \Delta_G\equiv \sin\psi\simlt 1$
for small values in $\Delta_B$. Thus, for most of the parameter space, the field is indeed
of minor importance compared to other forces (Figure \ref{sigma_b_inter}). 
Only for large values in $\Delta_G \sim 1$, 
$\Sigma_B$ can equal unity or become slightly larger. Intuitively this is expected, 
because growing values in $\Delta_G$ indicate larger deviations in the intensity gradient
direction from the gravity center which have to be caused by a growing presence of the 
magnetic field. 

As $\Delta_B$ grows, the field becomes more relevant 
for the dynamics along the density gradient direction in Equation (\ref{mhd_delta_vect}).
Therefore, we expect the field to take on a more important role in shaping and 
influencing an observed dust continuum morphology. Additionally bringing in $\Sigma_B$, 
we can indeed confirm that the relative field significance $\Sigma_B$
is growing as $1/\sqrt{1-\Delta_B^2}$ for constant values in $\Delta_G$ (dotted 
magenta lines in Figure \ref{sigma_b_inter}). The larger $\Delta_G$ already is, the faster $\Sigma_B$
passes unity. Or, in other words, for any existing deviation in the intensity gradient
from the gravity direction, an additionally growing deviation with the field orientation
reflects a growing field importance. This growth in field significance, nevertheless, 
does not yet automatically mean that the field is dominating. The gray dotted line
in Figure \ref{sigma_b_inter} separates the $(\Delta_G, \Delta_B)$- parameter space into $\Sigma_B <1$
and $\Sigma_B > 1$. 
\footnote{Note that the boundary 
zone with $\Delta_B\equiv 1$ (except $\Delta_B= 1$ with $\Delta_G= 0$)
is, 
strictly speaking, excluded because the method is failing here.}

In summary, whereas $\sin|\delta|\equiv \Delta_B$ in Equation (\ref{mhd_delta_vect}) is already 
indicative for the role of the magnetic field, $\Sigma_B$ further sharpens and quantifies
this by additionally taking into account the information of the other forces in the system
( i.e., $\Delta_G$). It has to be stressed that introducing $\sin|\delta|$ in 
Equation (\ref{mhd_delta_vect}) and deriving $\Sigma_B$ with the polarization - intensity 
gradient method are two independent approaches. As a sole common ground, 
they start with an ideal MHD force equation, but then 
proceed with two different assumptions. A comparison and validity of these two 
assumptions is given in Appendix A.


\section{$|\delta|$ - Map: A Tracer for the Change in the Role of the Magnetic Field}
                                         \label{section_tracer}

As explained in Section \ref{interpretation}, the angle $\delta$ can be directly
linked to the (projected) MHD force Equation (\ref{mhd_delta_vect}). 
Nevertheless, the role 
and the significance of the magnetic field -- the field tension in the 
rightmost term in Equation (\ref{mhd_momentum}) -- are not yet explicitly visible due to
the (unknown) gravitational force term. Additionally, the angle $\delta$ can be 
affected by projection effects. The magnetic field-to-gravitational force ratio
$\Sigma_B$ (Section \ref{summary_parameters} and \citet{koch12a}) can overcome these
shortcomings and links the magnetic field tension to the gravity in a system via 
the polarization - intensity gradient method. Moreover, projection effects are minimized in 
$\Sigma_B$ or even cancel out. 

In the following sections, we further investigate the connection between $\delta$ and $\Sigma_B$, 
with the goal of establishing $\delta$ as a tracer for the change in the role of the 
magnetic field over an observed map. This is motivated by the aim of providing a 
simple tool ($|\delta|$-map, Section \ref{delta_observations}) to interpret polarization data.
Whereas calculating $\Sigma_B$ requires a more elaborate modeling, the angle $\delta$ 
can be readily read off from a map once intensity gradient orientations are derived.
Often, a simple inspection by eye will already allow for differentiating between clearly
small or large $\delta$-values. This, in contrast, is non-trivial for $\Sigma_B$. There
is, thus, a substantial benefit from gaining further insight in $\delta$ and its deeper
connection to $\Sigma_B$.

\subsection{Approximating $\Sigma_B$}

We start by addressing the concern of projection effects in $\delta$. For any observable, 
projection effects cancel out when looking at the relative change (logarithmic 
derivative) of this observable. Assuming the 3-dimensional (lower index '3') 
deprojected angle $\delta_3$ linked with an inclination angle $\iota_B$ to the
projected 2-dimensional (lower index '2') angle $\delta_2$, 
$\Delta_{B_3}=\sin\delta_3=\sin\delta_2\cdot \cos\iota_B$ (see Figure 13 in \citet{koch12a}),
 we can write the logarithmic derivative:
\begin{equation}
\frac{d(\Delta_{B_3})}{\Delta_{B_3}}=\frac{d(\sin\delta_2\cdot \cos\iota_B)}{\sin\delta_2\cdot \cos\iota_B}
                               =\frac{d(\sin\delta_2)}{\sin\delta_2}=\frac{d(\Delta_{B_2})}{\Delta_{B_2}}.
\end{equation}
$d$ denotes the total differential, and we have assumed $d(\cos\iota_B)\equiv 0$.
In the previous Section \ref{interpretation}, lower indices '2' were omitted for 
map-projected quantities. In practice, we are interested in changes in $\delta$ from one
pixel to its neighboring pixels in a map, in order to assess the change in the role
of the magnetic field. It is, thus, a very mild assumption to impose no change in the 
inclination angle, i.e. $d(\cos\iota_B)=0$, between adjacent pixels. If necessary, this 
condition can be relaxed to $d(\cos\iota_B)\approx 0$ or very small, which will still
leave the above equation correct on a pixel-to-pixel basis, but would allow for a 
change in inclination from one end to another end in a map, if this should seem 
plausible. The logarithmic differential for the field significance $\Sigma_B$, 
Equation (\ref{eq_sigma_B}), can then be written as:
\begin{equation}
\frac{d\Sigma_B}{\Sigma_B}=\frac{1}{\Delta_G}\cdot d(\Delta_G)+\frac{\Delta_B}{1-\Delta_B^2}\cdot d(\Delta_B),
                                                            \label{eq_diff_sigma}
\end{equation}
where we have again omitted the lower indices '2' for the map-projected quantities
$\Delta_G$ and $\Delta_B$. 
In Equation (\ref{eq_diff_sigma}) the magnetic field alignment deviation $\Delta_B$ and the 
gravity alignment deviation $\Delta_G$ are decoupled into additive terms quantifying their
relative contributions to $d\Sigma_B/\Sigma_B$. We re-define their individual 
contributions as:
\begin{equation}
\Xi_B=w_B\cdot d(\Delta_B)\,\,\,\,\,\,\,\,\,  {\rm and}\,\,\,\,\,\,\,\,\,  \Xi_G=w_G\cdot d(\Delta_G),
\end{equation}
where we have introduced the weight functions $w_B=\frac{\Delta_B}{1-\Delta_B^2}$
and $w_G=\frac{1}{\Delta_G}$ for the magnetic field and gravity, respectively.
A change in the field significance, $d\Sigma_B / \Sigma_B$, generally depends
on the 4 parameters $\Delta_B$, $\Delta_G$, $d(\Delta_B)$ and $d(\Delta_G)$. 
For simplicity, we assume here first $d(\Delta_B)\approx d(\Delta_G)\approx const$, i.e. we
assume roughly similar (small) changes in the field and gravity alignment deviations
from one pixel to the next one. This then allows us to simplify Equation 
(\ref{eq_diff_sigma}) to 2 variables and compare the weight functions $w_B$
and $w_G$ in their relevance for $\Sigma_B$ (Figure \ref{figure_weight_simple}). 
This leads to some remarkable 
insight: (1) For any values in $\Delta_G$, small values in $\Delta_B$ (small
values in $\delta$) always mean that the magnetic field is irrelevant for a 
change in $\Sigma_B$. A change in $\Sigma_B$ is dominated by gravity and for
$d(\Delta_B)\approx d(\Delta_G)$ the field is not important and thus, $\Sigma_B<1$.
(2) For any value in $\Delta_G$, and $\Delta_B\le 0.618$ ($|\delta|\simlt 38^{\circ}$), the 
magnetic field is less important or at most equally important to a change in $\Sigma_B$.
(3) For values $\Delta_B > 0.618$ ($|\delta| \simgt 38^{\circ}$), the magnetic field can (but does 
not necessarily) dominate gravity. If $d(\Delta_G)$ and $d(\Delta_B)$ are known, 
the weight functions in Figure \ref{figure_weight_simple} can accordingly be shifted
vertically.

The above conclusions compare magnetic field and gravity force. We can additionally 
formulate 2 consequences from Figure \ref{figure_weight_simple} for the magnetic field 
separately: (1) Small values in $\Delta_B$ (small values in $|\delta|$) {\it always minimize}
the field contribution to $d\Sigma_B/\Sigma_B$. (2) Large values in $\Delta_B$ (large
values in $|\delta|$) {\it always maximize} its contribution. This is true for any values
in $\Delta_G$ and $d(\Delta_G)$. 

We have to acknowledge that most of the above conclusions can also be found from $\Sigma_B$
(Equation (\ref{eq_sigma_B})). What have we gained from further analyzing the above 
logarithmic differential? We recall that we are aiming at establishing $|\delta|$-maps
as a tracer for the change in the role of the magnetic field. Interestingly, the weight
function $w_B=\frac{\Delta_B}{1-\Delta_B^2}$ is close to $|\delta|$ ($\Delta_B=\sin|\delta|$), 
only with an additional factor $\frac{1}{1-\Delta_B^2}$. Since $1-\Delta_B^2\le 1$, $\sin|\delta|$
thus always provides a lower limit to a change in the field significance (blue dashed line in
Figure \ref{figure_weight_simple}). For values $\Delta_B \simlt 0.5$ ($|\delta|\simlt 30^{\circ}$), 
$\sin|\delta|$ is even accurate to within about 25\% of $w_B$. The logarithmic differential in 
Equation (\ref{eq_diff_sigma}) has, therefore, allowed us to explicitly put $|\delta|$ in 
relation  with the projection-free quantity $d\Sigma_B/\Sigma_B$, thus establishing $|\delta|$
as a valid approximation for it.

In summary, whatever the value for $\Xi_G$, $\Xi_B$ can always be approximated with $|\delta|$.
Therefore, $|\delta|$-maps by themselves -- despite possibly being affected by projection effects --
already reflect the change in the role of the magnetic
field in a system. An increase (decrease) in $|\delta|$ will indicate an increase (decrease)
in $\Sigma_B$, and thus, point toward a growing (diminishing) significance of the field as compared
to gravity.

\subsection{Comparing $|\delta|$-Map and $\Sigma_B$-Map}  \label{sub_section_comparison}

\begin{figure}
\begin{center}
\includegraphics[scale=0.35]{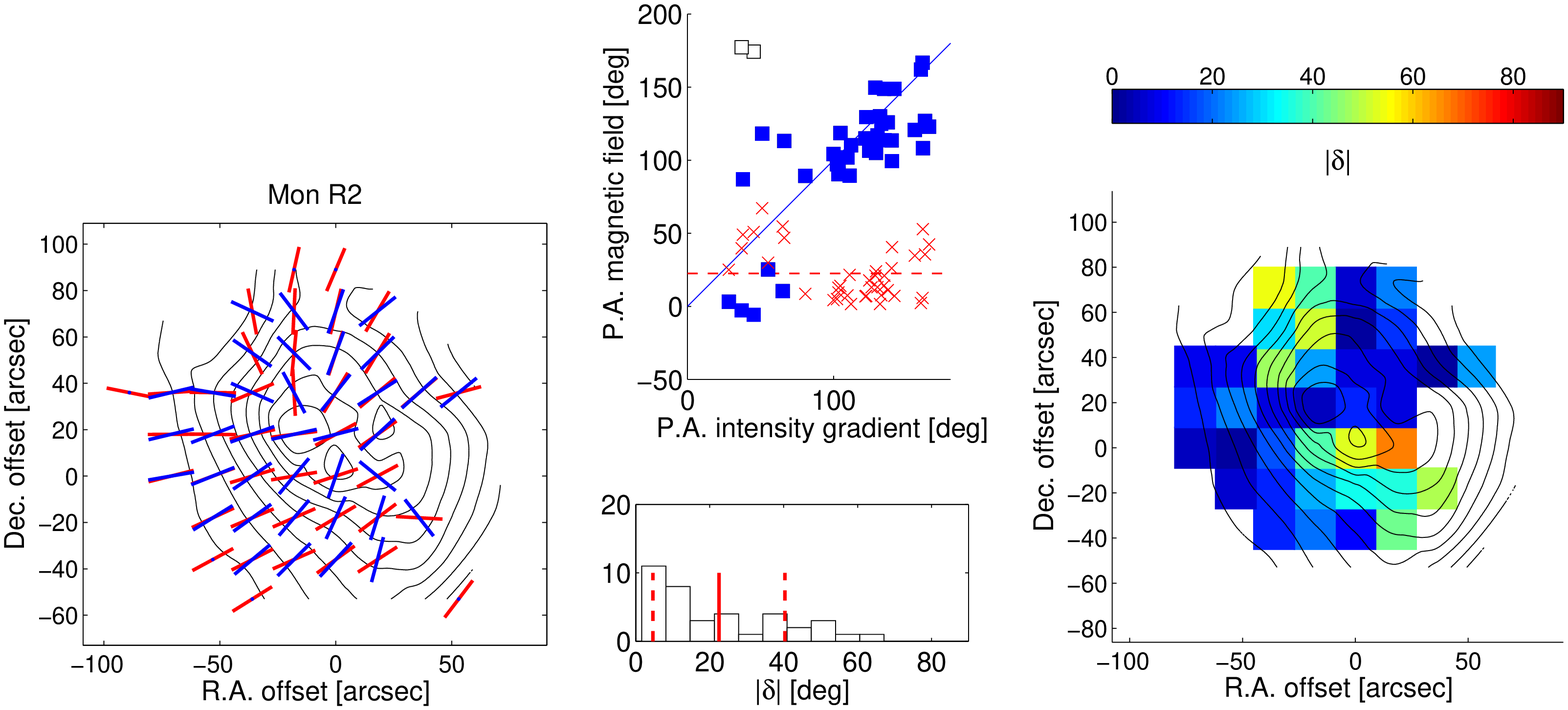}
\includegraphics[scale=0.35]{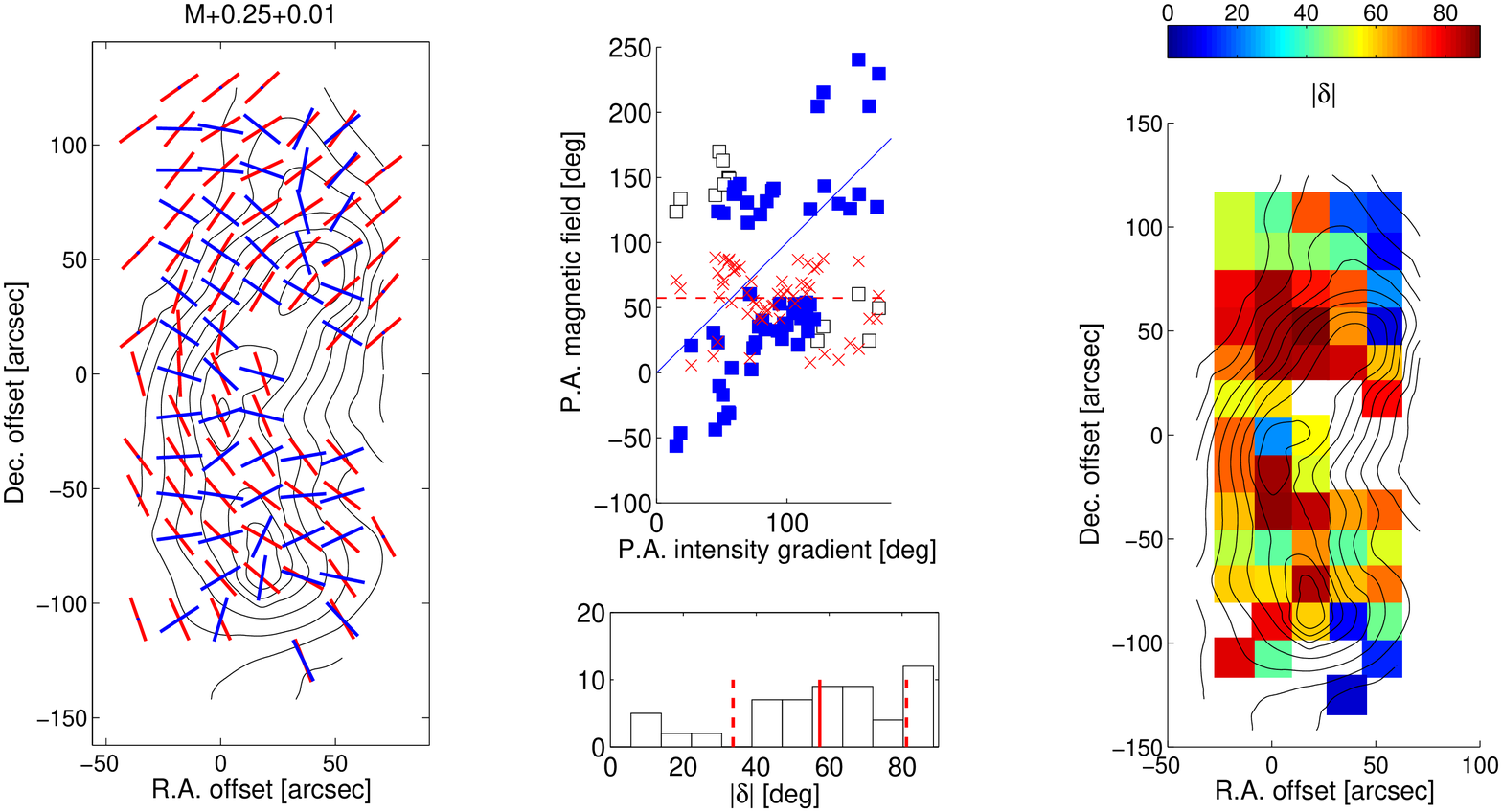}
\includegraphics[scale=0.35]{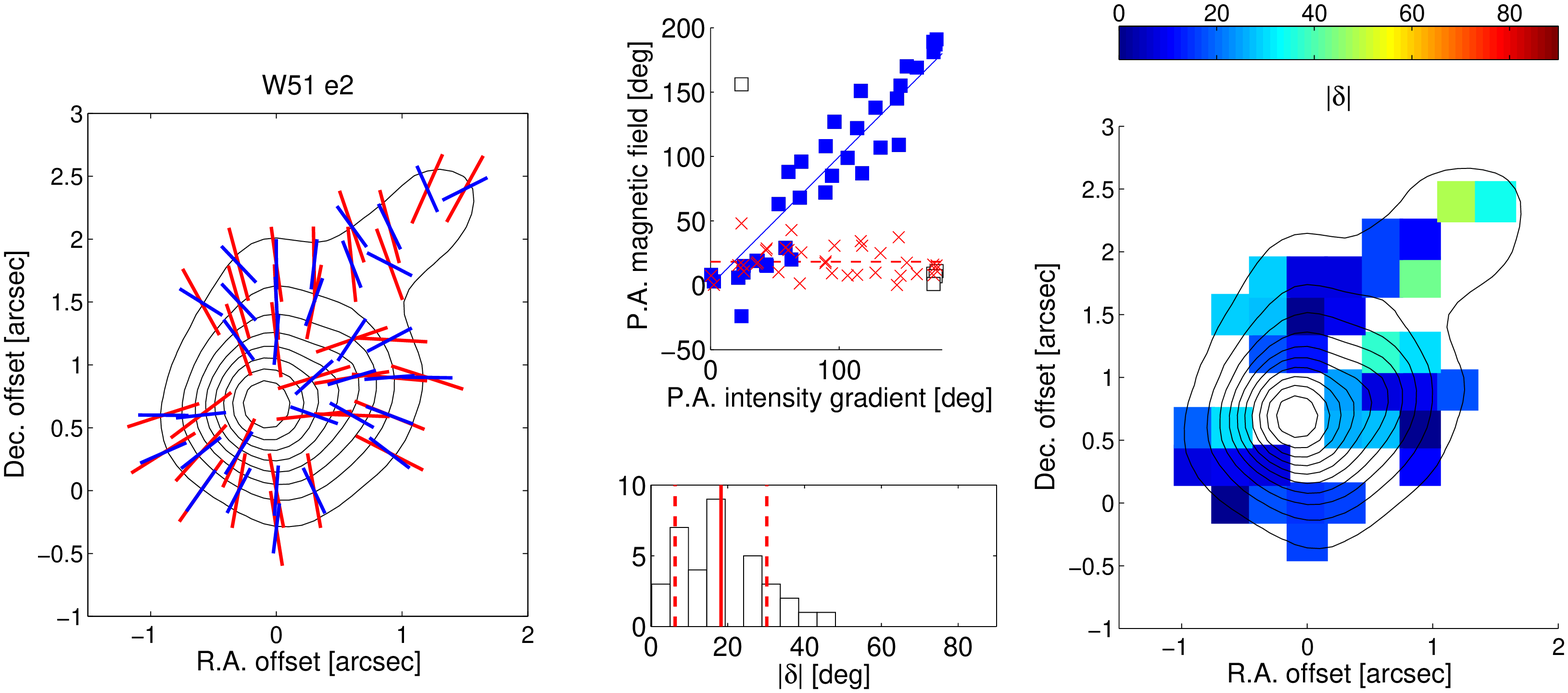} 
 \caption{\label{figure_delta_scales_cont} \footnotesize
The same as in Figure \ref{figure_delta_scales} but for 
the sources Mon R2 and M$+0.25+0.01$ (CSO, \citet{dotson10})
and W51 e2 \citep{tang09b} from top to bottom.
For completeness, left panel and histogram for W51 e2 are reproduced from 
\citet{koch12a}. The error estimate for $|\delta|$ in these sources is 
identical to the one for Figure \ref{figure_delta_scales}.
}
\end{center}
\end{figure}

\begin{figure}
\begin{center}
\includegraphics[scale=0.8]{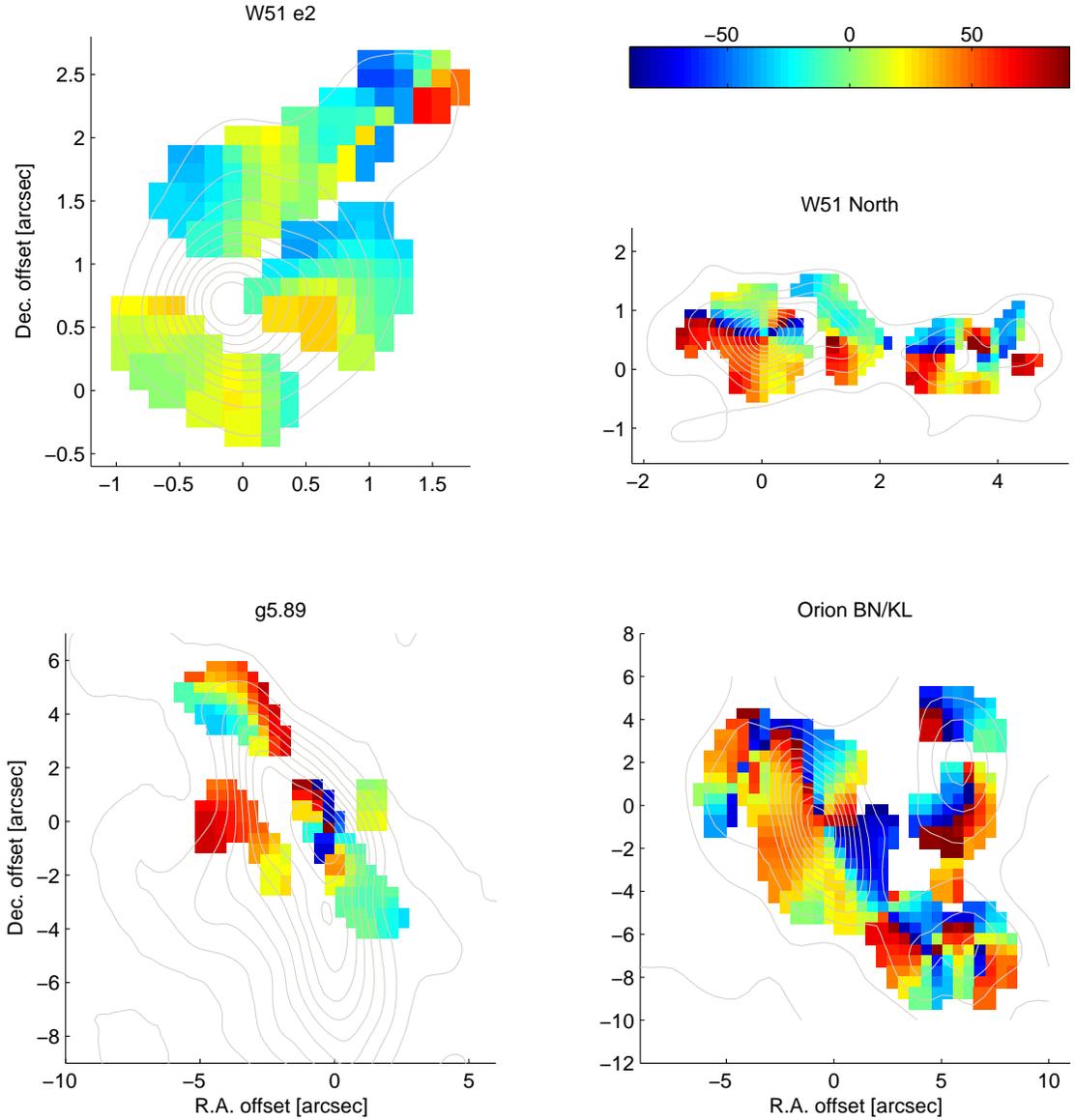}
 \caption{\label{figure_delta_cores}
Relative difference maps ($-90^{\circ} \le \delta \le 90^{\circ}$)
for the sources W51 e2, W51 North, g5.89 and Orion BN/KL.
Colors correspond to the color
wedge on the top with units in deg. 
For an enhanced visual impression, the data are over-gridded to
about $0.15\arcsec$ for W51 North and e2, $0.4\arcsec$ for g5.89
and  $0.5\arcsec$ for Orion BN/KL, which is about 5 times their synthesized
beam resolutions (Table \ref{table_quantities}).  
Overlaid are contours of the Stokes $I$ dust continuum emission.
Original maps with magnetic field segments are in \citet{tang12},
\citet{tang10} and \citet{tang09a} for W51 North, Orion BN/KL
and g5.89, respectively. 
}
\end{center}
\end{figure}

\begin{figure}
\begin{center}
\includegraphics[scale=0.35]{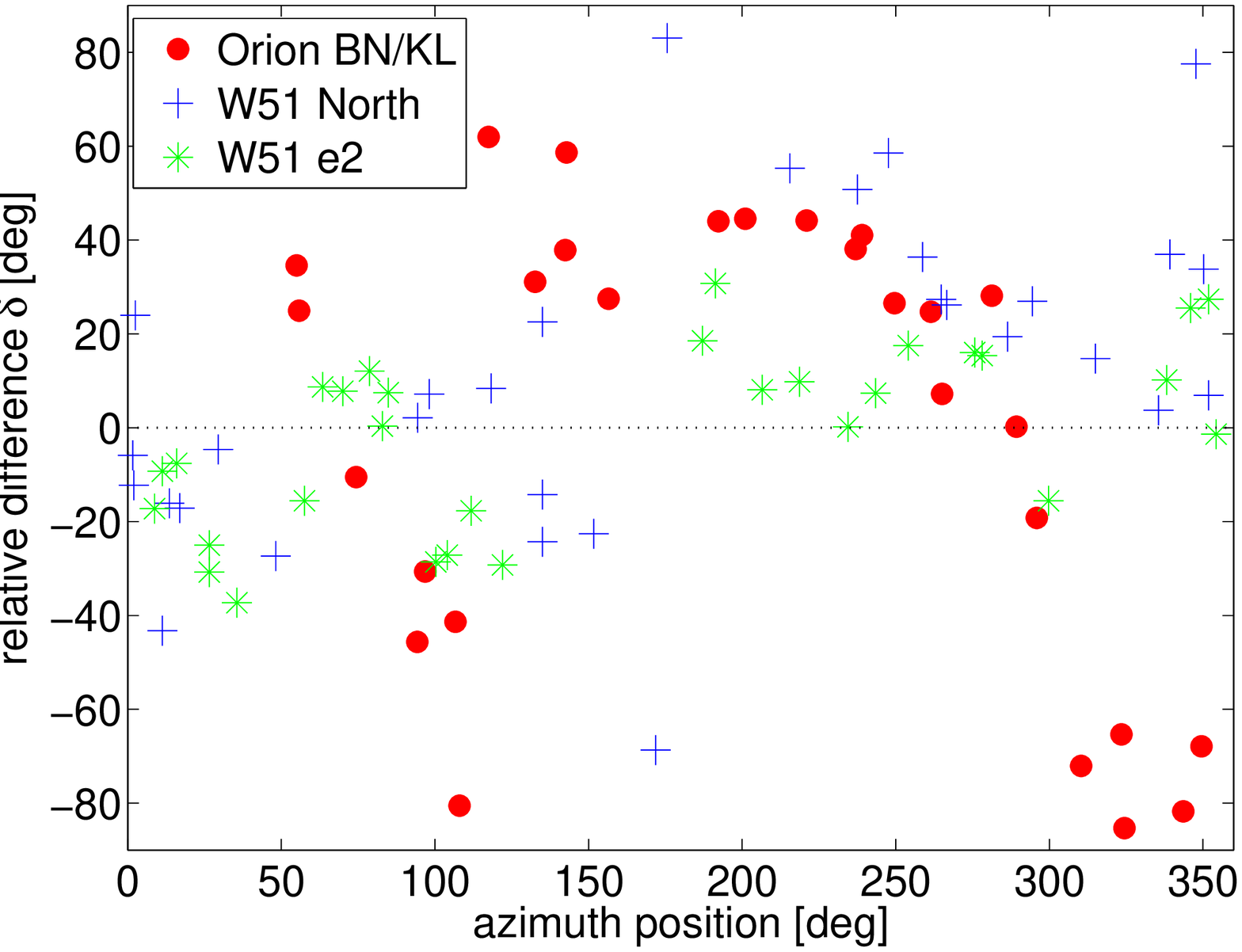}
\includegraphics[scale=0.35]{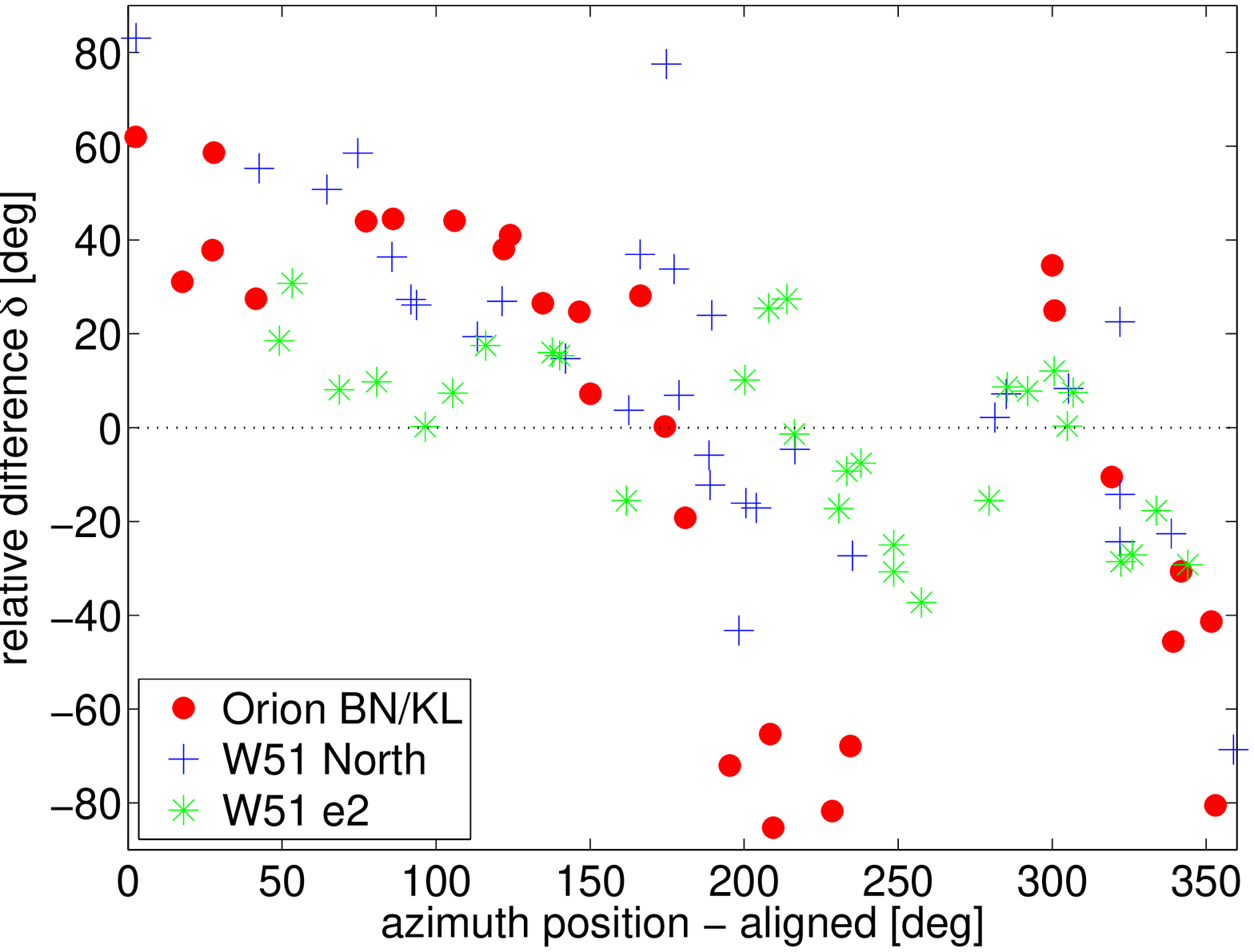}
\includegraphics[scale=0.35]{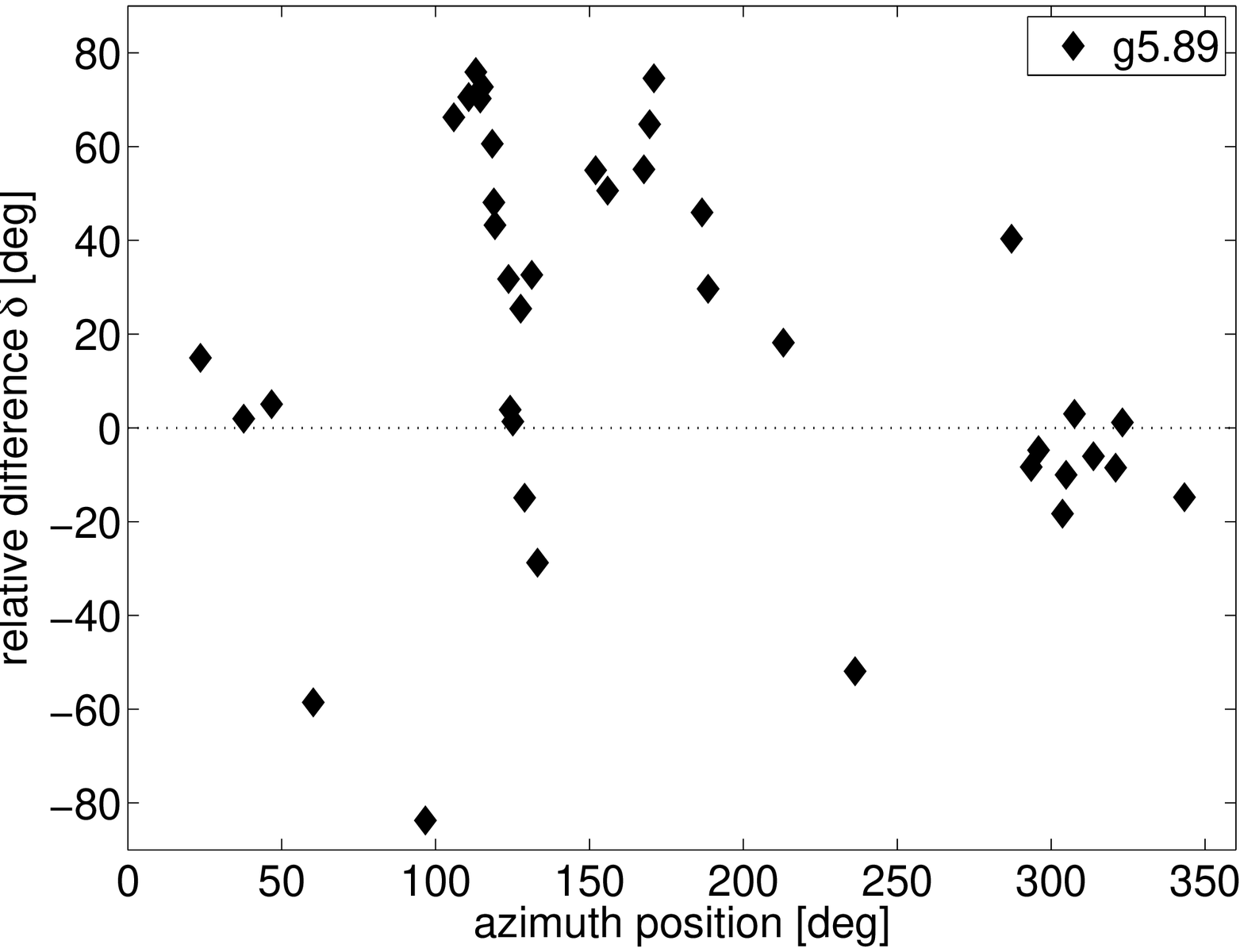}
 \caption{\label{figure_delta_azimuth}\footnotesize
Relative difference ($-90^{\circ} \le \delta \le 90^{\circ}$)
for the sources W51 e2, W51 North (main core in the East), 
Orion BN/KL (main core in the East) and g5.89 as a function of azimuth.
Top panel: azimuth is with respect to the original maps, i.e. measured 
counter-clockwise starting from west (W51 e2, W51 North and Orion BN/KL). 
For W51 e2, the 6 segments in the north-western extension 
where a possibly new core is forming
(bottom left panel in Figure \ref{figure_delta_scales_cont}), are excluded.
Middle panel: azimuth coordinates are shifted to maximally align features. 
With respect to the top panel, the shifts are -115$^{\circ}$, -173$^{\circ}$,
-138$^{\circ}$ for Orion BN/KL, W51 North and W51 e2, respectively.
Bottom panel: for comparison displayed is g5.89 which shows much less
pronounced azimuth features in $\delta$.
Unlike in Figure \ref{figure_delta_cores}, only the originally detected
non-over-gridded data are displayed here.  
}
\end{center}
\end{figure}

\begin{figure}
\begin{center}
\includegraphics[scale=0.5]{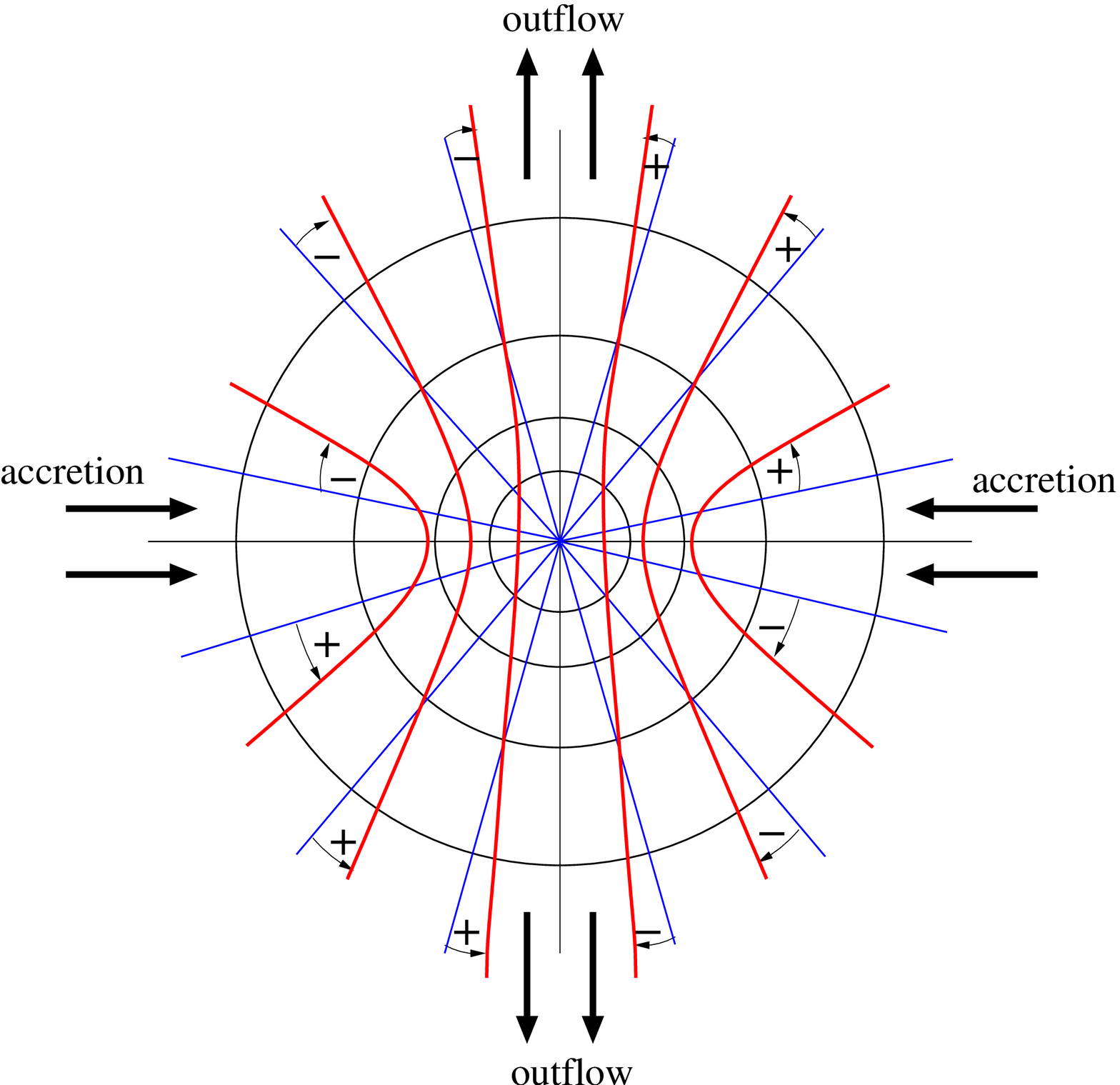}
\includegraphics[scale=0.5]{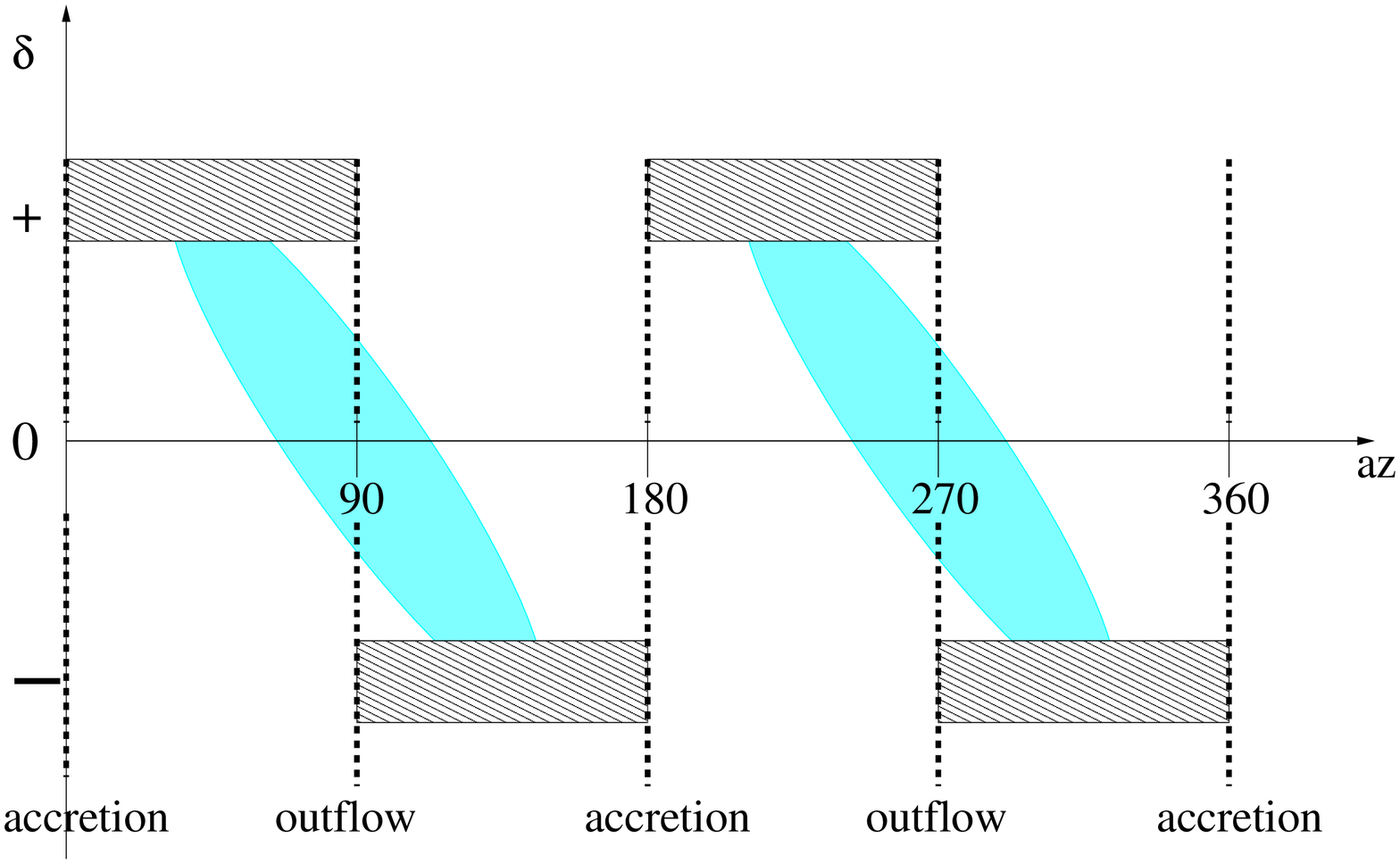}
 \caption{\label{figure_schematic_delta_azimuth}\footnotesize
Top panel: Schematic illustration of the relative difference 
($-90^{\circ} \le \delta \le 90^{\circ}$)
as a function of azimuth for a prototype collapsing core with pinched
magnetic field lines. Magnetic field lines are shown with red solid lines.
For simplicity circular intensity emission contours (black solid lines) with 
radial intensity emission gradients (solid blue lines) are assumed.
Expected accretion and outflow directions are indicated 
with arrows.
Bottom panel: $\delta$ as a function of azimuth, measured from the west
in the top panel.
The quadrants where $\delta$ is negative or positive are indicated with the 
hatched areas. The areas in cyan indicate the continuous change in $\delta$ 
across zero in the outflow areas (around azimuth 90 and 270), 
and the more abrupt change with a flip in sign
across the accretion zones (azimuth 0 and 180).
Note that a flattened ellipsoidal structure in the top panel will still
leave these characteristic zones unchanged.
The above schematics can also explain the bimodal distribution in $\delta$
around 0, which was first clearly observed for W51 e2 in Figure 2 in \citet{koch12a}.
$\delta$-distributions for all sources are shown in Figure \ref{figure_histogram}.
}
\end{center}
\end{figure}

\begin{figure}
\begin{center}
\includegraphics[scale=0.7]{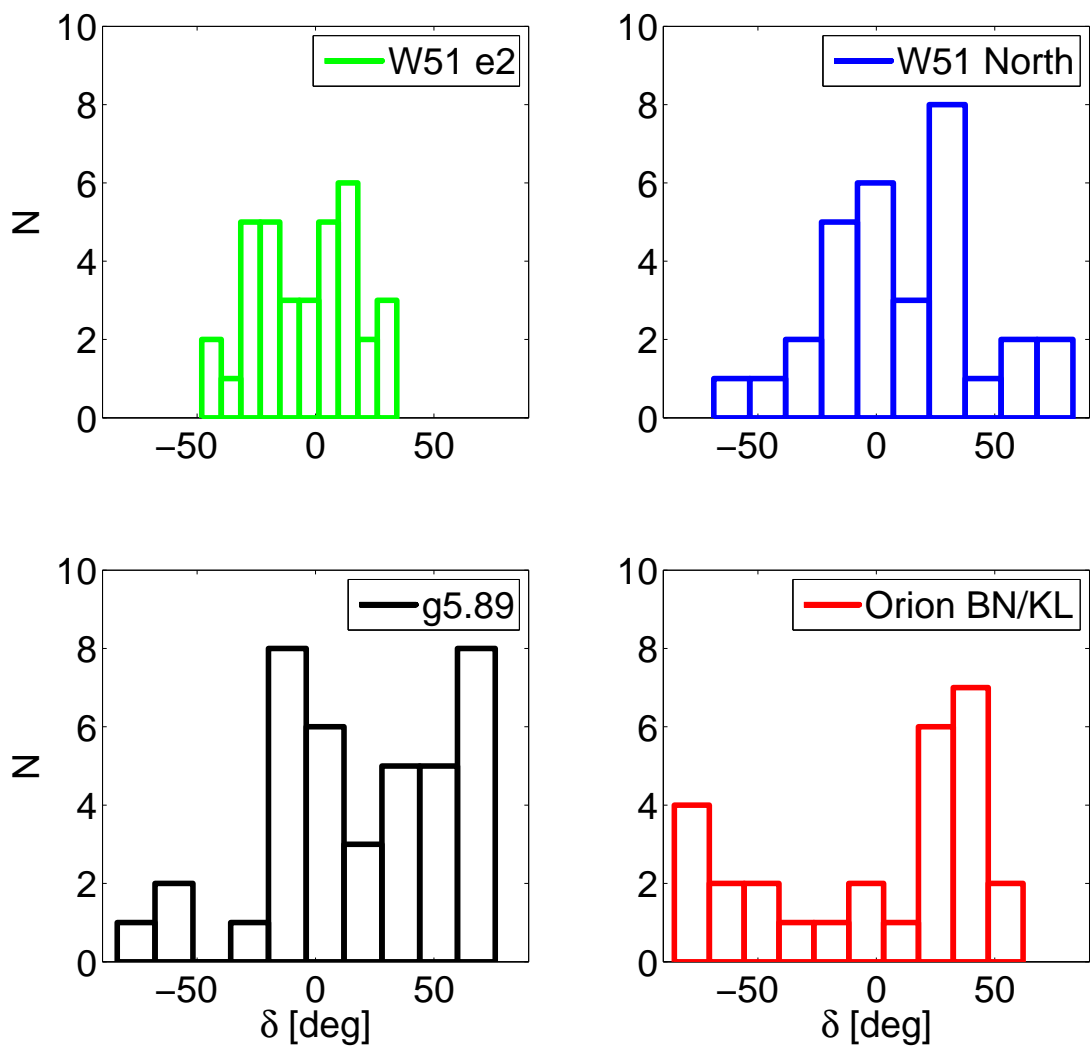}
 \caption{\label{figure_histogram}
Distributions of the relative-difference angles $\delta$ as displayed in 
Figure \ref{figure_delta_azimuth} for W51 e2, W51 North main core, g5.89
and Orion BN/KL main core. Distributions are clearly non-Gaussian.
Bimodal distributions are apparent even in the case of g5.89. }
\end{center}
\end{figure}

\begin{figure}
\begin{center}
\includegraphics[scale=0.45]{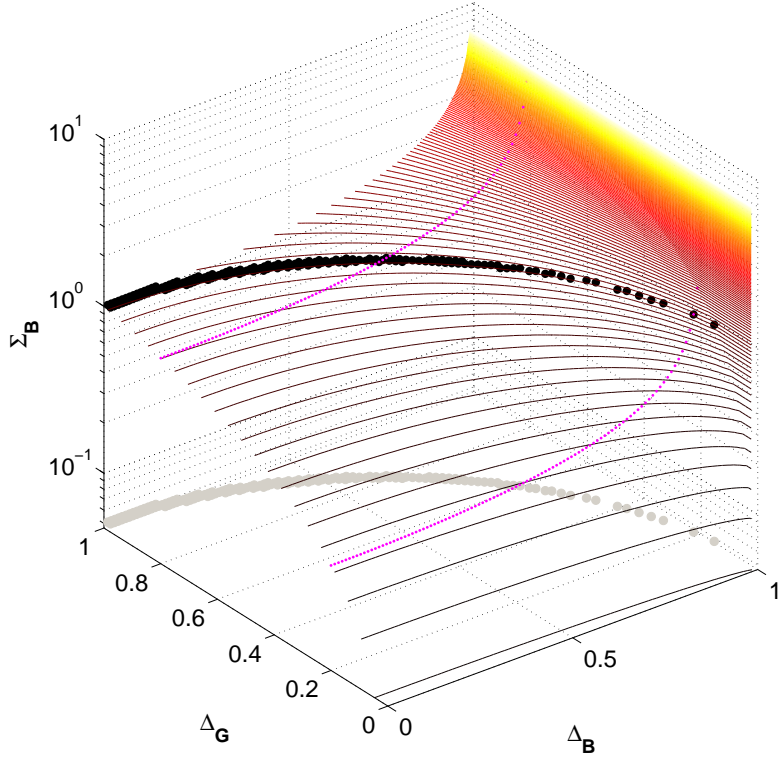}
 \caption{\label{sigma_b_inter}
The field significance $\Sigma_B$ as a function of the field and gravity
deviations $\Delta_B$ and $\Delta_G$. They gray-dotted line in the 
$(\Delta_G,\Delta_B)$-plane separates the parameter space into $\Sigma_B <1$
and $\Sigma_B >1$. The corresponding black-dotted contour line marks $\Sigma_B\equiv 1$.
The 2 magenta lines are illustrating $\Sigma_B$ as a function of $\Delta_B$ for
constant $\Delta_G$-values. At the boundary $\Delta_B\equiv 1$
(except for $\Delta_B=1$ with $\Delta_G=0$), 
$\Sigma_B$ is, strictly speaking, not defined as the polarization-intensity gradient
method \citep{koch12a} is failing here. 
}
\end{center}
\end{figure}

\begin{figure}
\begin{center}
\includegraphics[scale=1]{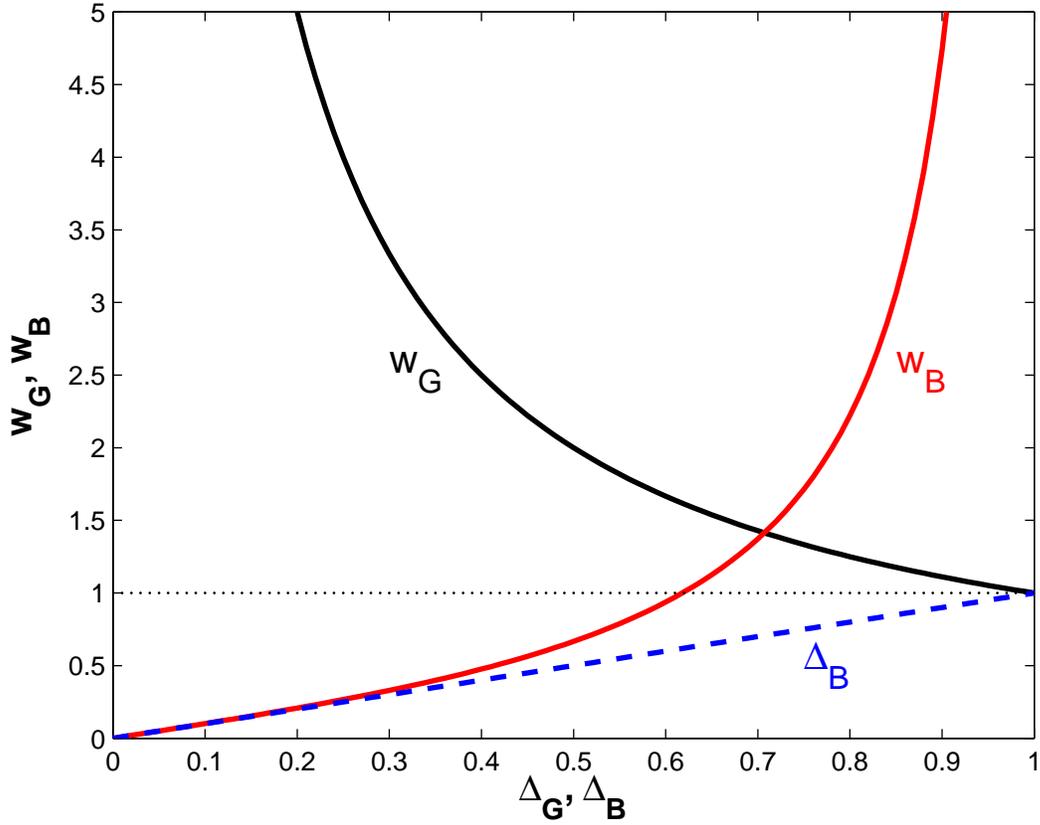}
 \caption{\label{figure_weight_simple}
The weight functions $w_B$ and $w_G$ for the magnetic field and gravity, respectively.
The lower limit, $\Delta_B=\sin|\delta|$, to the magnetic field weight $w_B$ is
indicated with the blue dashed line. 
This lower limit is directly obtainable from a $|\delta|-$map.
$w_B$ is unity at $\Delta_B\equiv\Delta_G=0.618$, which corresponds to a misalignment of 
about $38^{\circ}$. Magnetic field and gravity have equal weight at  
$\Delta_B\equiv\Delta_G=0.707$, i.e. $45^{\circ}$.
}
\end{center}
\end{figure}

\begin{figure}
\begin{center}
\includegraphics[scale=0.5]{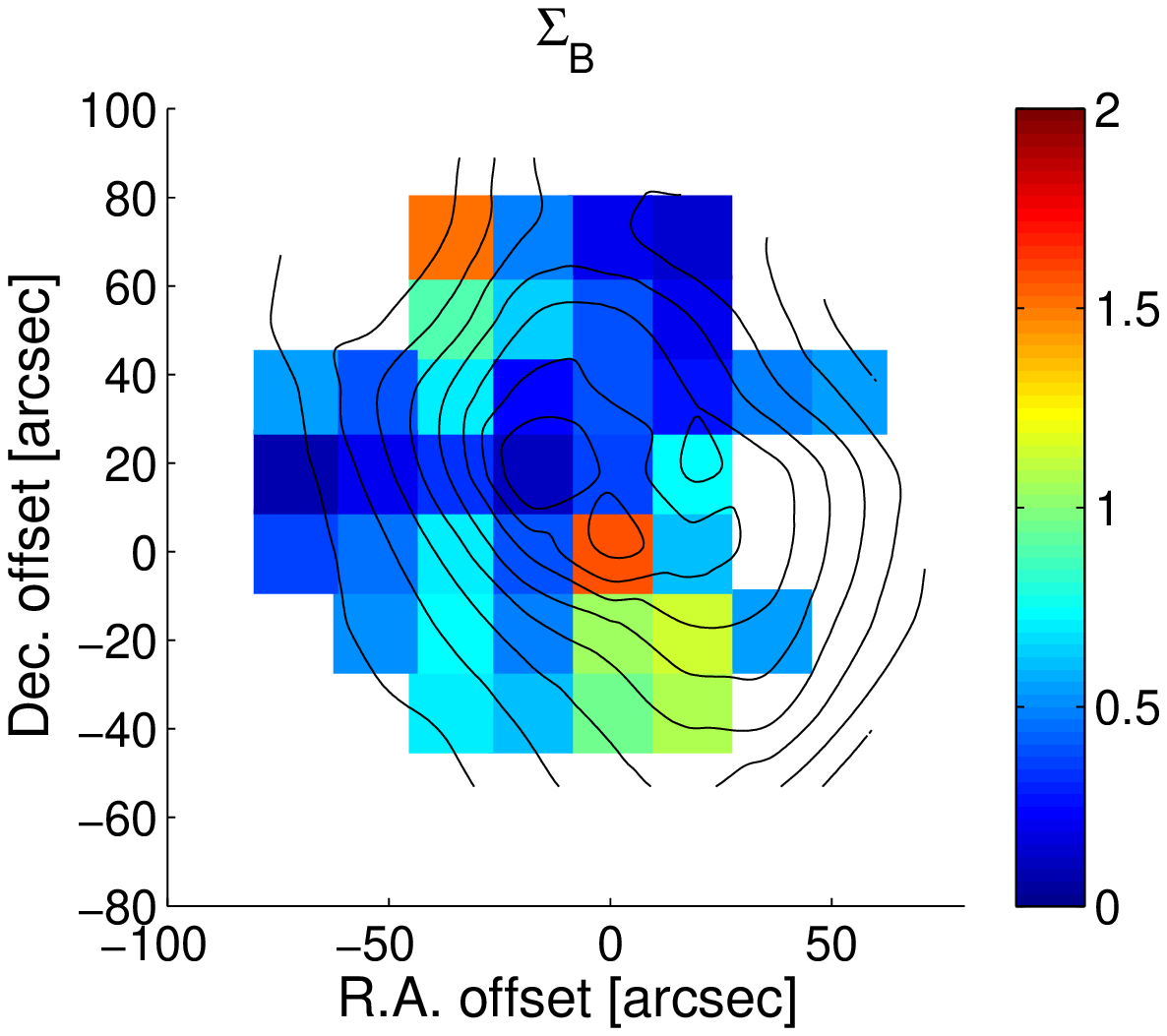}
\includegraphics[scale=0.45]{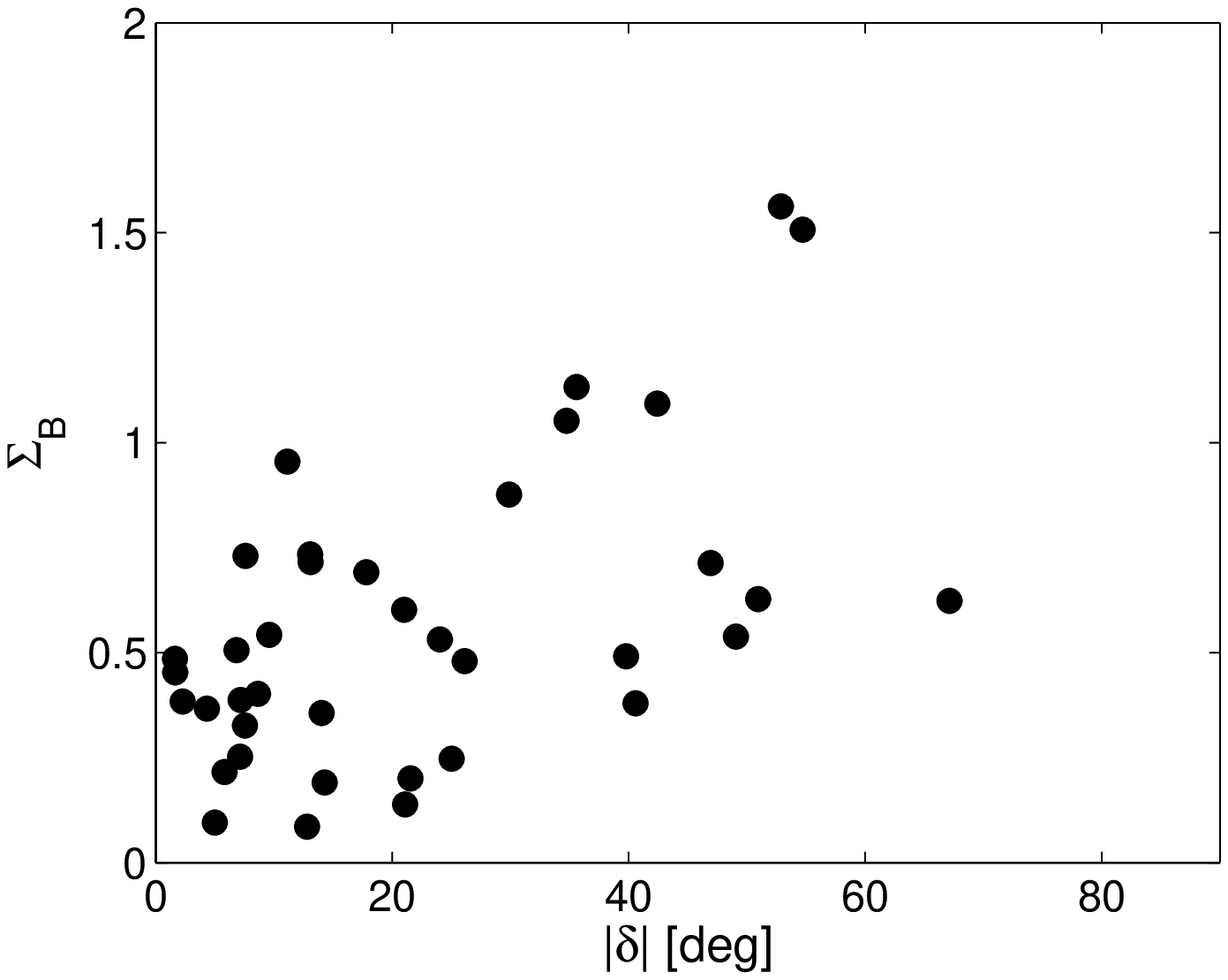}
\includegraphics[scale=0.5]{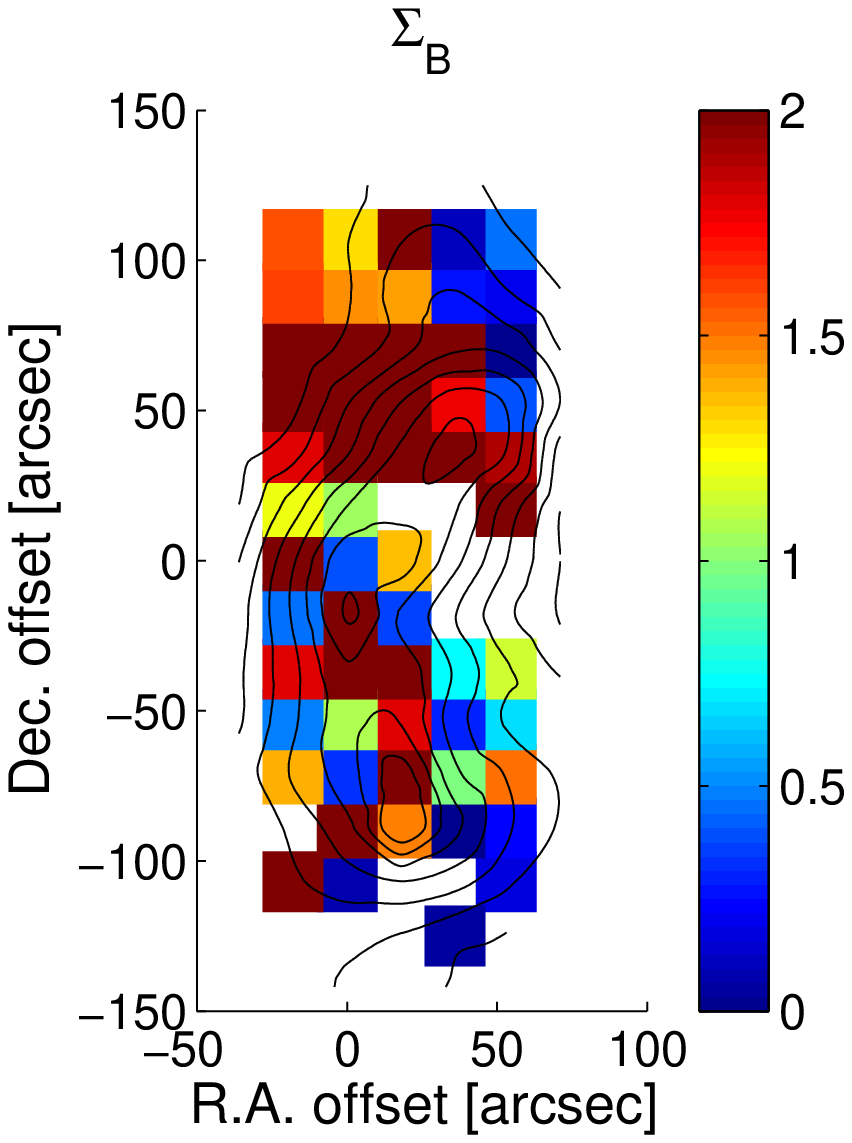}
\includegraphics[scale=0.45]{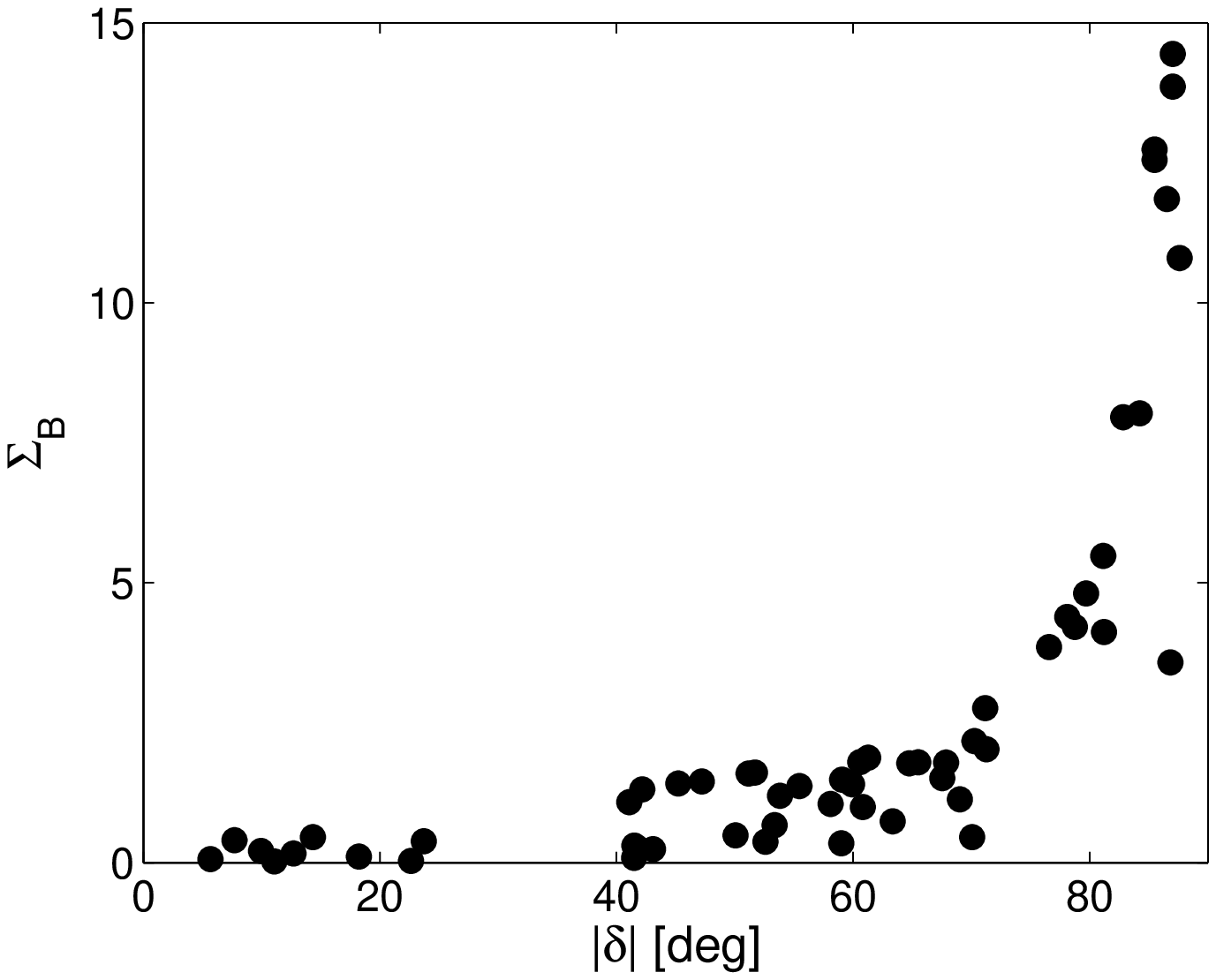}
\includegraphics[scale=0.5]{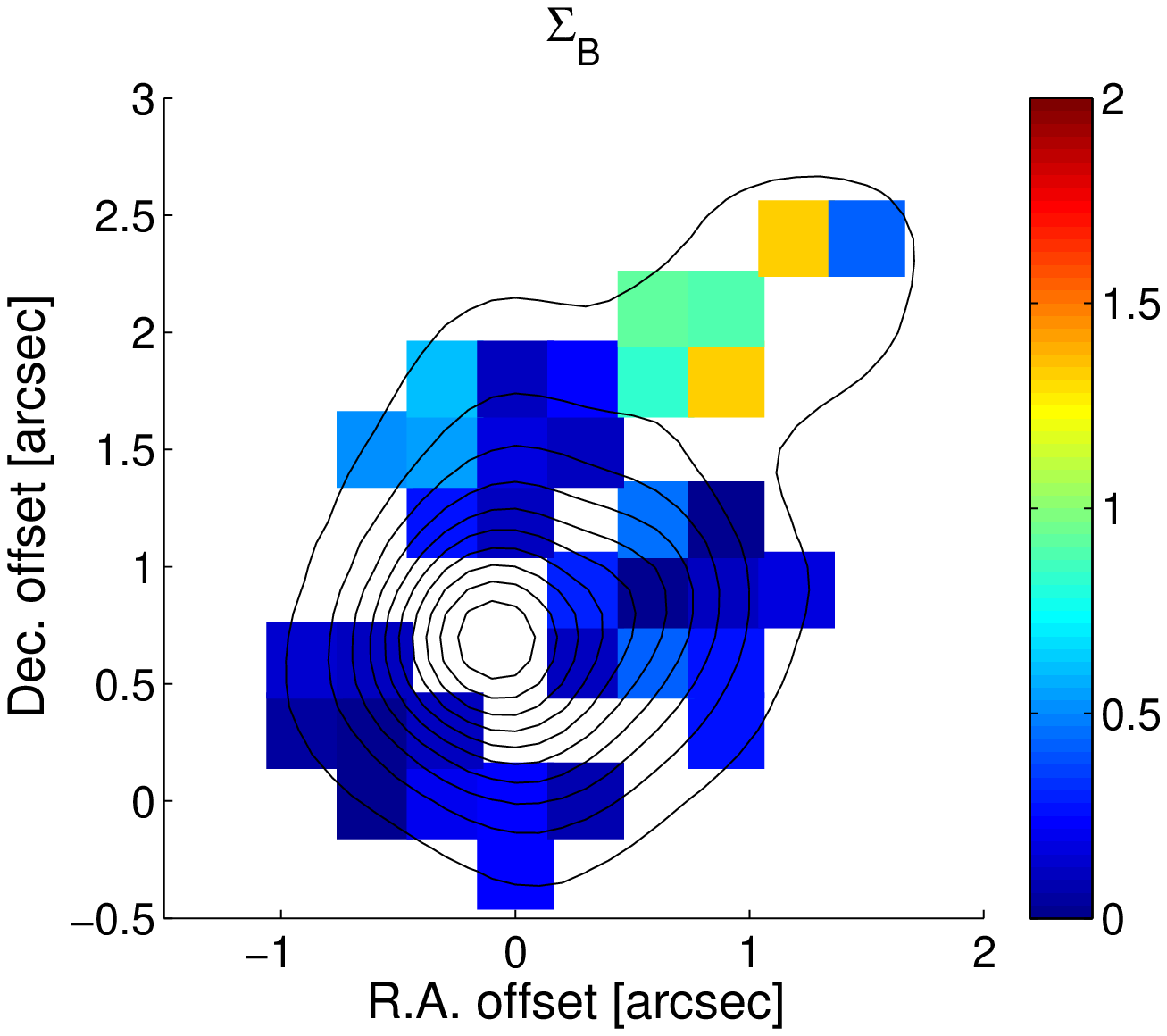}
\includegraphics[scale=0.45]{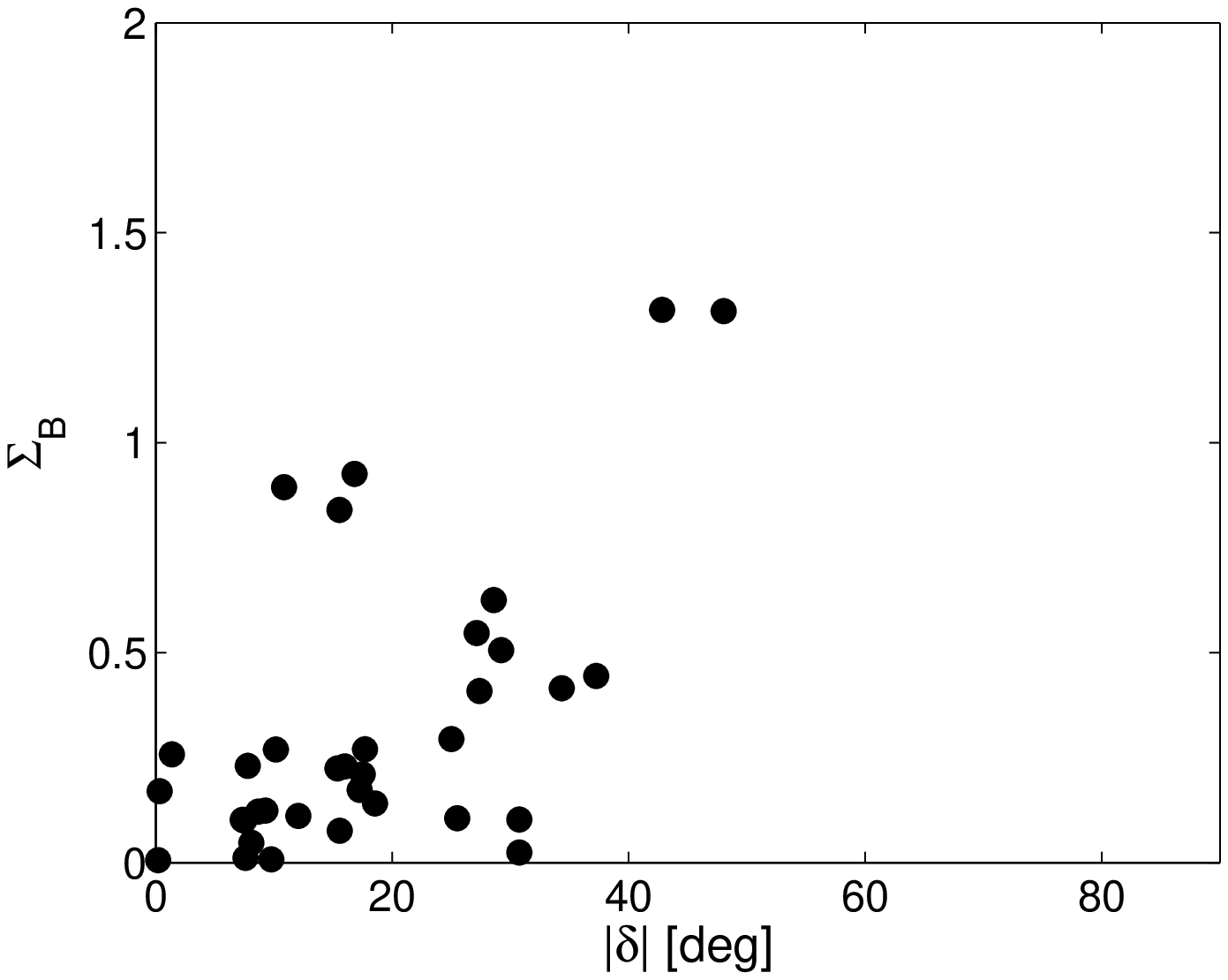} 
 \caption{\label{sigma_B_check}
Field significance $\Sigma_B$-maps (left panels) and $\Sigma_B$
versus $|\delta|$ correlation (right panels).
From top to bottom shown are the sources Mon R2 and 
M$+0.25+0.01$ (CSO, \citet{dotson10}) and W51 e2 \citep{tang09b}.
The exponential-like trend for M$+0.25+0.01$ results from large
$|\delta|$-values (small $\alpha$-values) which boost the contribution
of $1/\sin\alpha$ in $\Sigma_B$.
Uncertainties in $|\delta|$ are a few degrees up to a maximum of about 
10$^{\circ}$. Errors in $\Sigma_B$ are in the range of 10 to 20\%. 
They do, thus, not significantly alter the correlation.
}
\end{center}
\end{figure}

It remains to test in practice how reliably a $|\delta|$-map -- in the presence of (unknown)
projection effects -- reflects the role of the magnetic field. To that purpose, 
$\Sigma_B$-maps -- mostly free of projection effects -- are shown for comparison in 
Figure \ref{sigma_B_check}. All 3 sources, Mon R2, M$+0.25+0.01$ and W51 e2 show structures in 
their $\Sigma_B$-maps that seem to have analogous features in their $|\delta|$-maps
(Figure \ref{figure_delta_scales_cont}, right panels). Regions with small $|\delta|$-values, 
$\sim$0-20$^{\circ}$, in Mon R2 mostly coincide with $\Sigma_B$-values below 0.5. Maximum 
field-to-gravity ratios around 1 are found at the 2 ends of the source major axis in a 
northeast-southwest direction, which is also where the $|\delta|$-values peak around 60$^{\circ}$.
The right panel in Figure \ref{sigma_B_check} shows this positive correlation between $\Sigma_B$
and $|\delta|$ with a growing spread in $\Sigma_B$ for increasing $|\delta|$-values.
The smaller-scale e2 core (Figure \ref{sigma_B_check}, bottom panels) reveals 
a similar result. The immediate core area with $|\delta|\simlt 30^{\circ}$ 
(Figure \ref{figure_delta_scales_cont}) shows force ratios $\Sigma_B\simlt 0.5$. In the newly 
forming core in the northwest extension $\Sigma_B$ increases to about 1.5. Its maximum
values are found to coincide with the maximum $|\delta|$-values around $50^{\circ}$. 
Additionally, there is a hint of larger $\Sigma_B$-values toward the accretion plane
in the upper northern half of the core. The same tendency is again found in the $|\delta|$-map.
As for the case of Mon R2, the $\Sigma_B-|\delta|$-plot again reveals a positive correlation
with a smaller dispersion for small $|\delta|$.

The source M$+0.25+0.01$ -- with its magnetic field mostly aligned with the source major
axis -- reveals inverse patterns in its $|\delta|$-map (Figure \ref{figure_delta_scales_cont}, 
middle row), as compared to Mon R2 and W51 e2. Regions with small values in $|\delta|$ 
are found only in the north and south. In most areas, the magnetic field - intensity 
gradient deviations $|\delta|$ are large, up to 80-90$^{\circ}$. Strikingly, 
the $\Sigma_B$-map in Figure \ref{sigma_B_check}, middle panel, shows a tight 
correlation both for the zones of small and large values. Zones with $\Sigma_B \simlt 0.5$
are found only in the north and south and at a few locations between a Dec offset $\approx 0$
and $\approx 50$. In between these
areas, the magnetic field significance is growing to values where the field tension force
is dominating gravity by a factor of up to 10 or more. The $\Sigma_B - |\delta|$-plot 
shows a linear relation up to $|\delta|\approx 60^{\circ}$, with an exponentially growing
field significance $\Sigma_B$ for the largest deviations around 80-90$^{\circ}$. 
This is expected because large values in $|\delta|$ will boost $1/\sin\alpha$ in $\Sigma_B$.

Contrary to Mon R2 and W51 e2 where maximum values in $\Sigma_B$ are around 1 and 1.5, 
M$+0.25+0.01$ reveals a much more dominating field in most areas. Nevertheless, all 3 
sources manifestly show a positive correlation between $|\delta|$ and $\Sigma_B$. We,
therefore, conclude that $|\delta|$ indeed can trace the change in the magnetic field 
significance. We stress that we conservatively focus on the {\it change} in the magnetic
field role / significance because from a range of $|\delta|$-values we can not yet conclude
a range of $\Sigma_B$-values; e.g., the $\sim 40 - 60^{\circ}$ range in Mon R2 covers $\Sigma_B$
up to $\sim 1$, whereas the same range in M$+0.25+0.01$ is more likely to point toward
$\Sigma_B \sim 2$. 
This is a consequence of the missing information from the angle $\psi$ (and some possibly
related projection effect) when looking at $|\delta|$ only. This explains and defines
the dispersion in  $\Sigma_B$ for a fixed  $|\delta|$-value. Due to this scatter,
disconnected patches in $|\delta|$-values will be of limited
information. Changes of $|\delta|$ over an entire source, nevertheless, show 
an overall positive correlation with $\Sigma_B$ and, therefore, reliably describe
the varying role of the magnetic field with position. 

An identical analysis of a larger sample of 24 CSO sources \citep{dotson10} is presented
in Appendix B. In the large majority of these sources, features in $\Sigma_B$-maps are 
also reflected in $|\delta|$-maps. Occasional incomplete polarization coverages can limit
this resemblance. From the total sample of 31 sources -- 27 from the CSO and 4 from 
the SMA -- we arrive at a main conclusion: changes in the field significance $\Sigma_B$
are revealed by changes in the angle $|\delta|$. On average, larger angles $|\delta|$ are
pointing toward larger $\Sigma_B$-values, thus indicating a dynamically more significant
magnetic field. Basic statistical quantities are summarized in Table \ref{table_quantities}.

\section{Schematic Scenario}          \label{section_scenarios}

The previous sections have established the angle $|\delta|$ as a reliable tracer for the 
change in the role of the magnetic field. This is a consequence of 2 main results: (1) the 
angle $|\delta|$ can be interpreted as a field alignment deviation (efficiency) in the MHD
force equation; (2) the angle $|\delta|$ always yields an approximation (lower limit to 
the magnetic field weight function $w_B$) to 
a change in 
the field significance $\Sigma_B$. Figure \ref{sigma_B_check} 
and Appendix B
demonstrate and confirm the expected
connection between $|\delta|$ and $\Sigma_B$. In the following, 
we now propose an evolutionary scenario, across which the role of the magnetic field can 
be assessed with the angle $|\delta|$. 

\begin{figure}
\begin{center}
\includegraphics[scale=0.27]{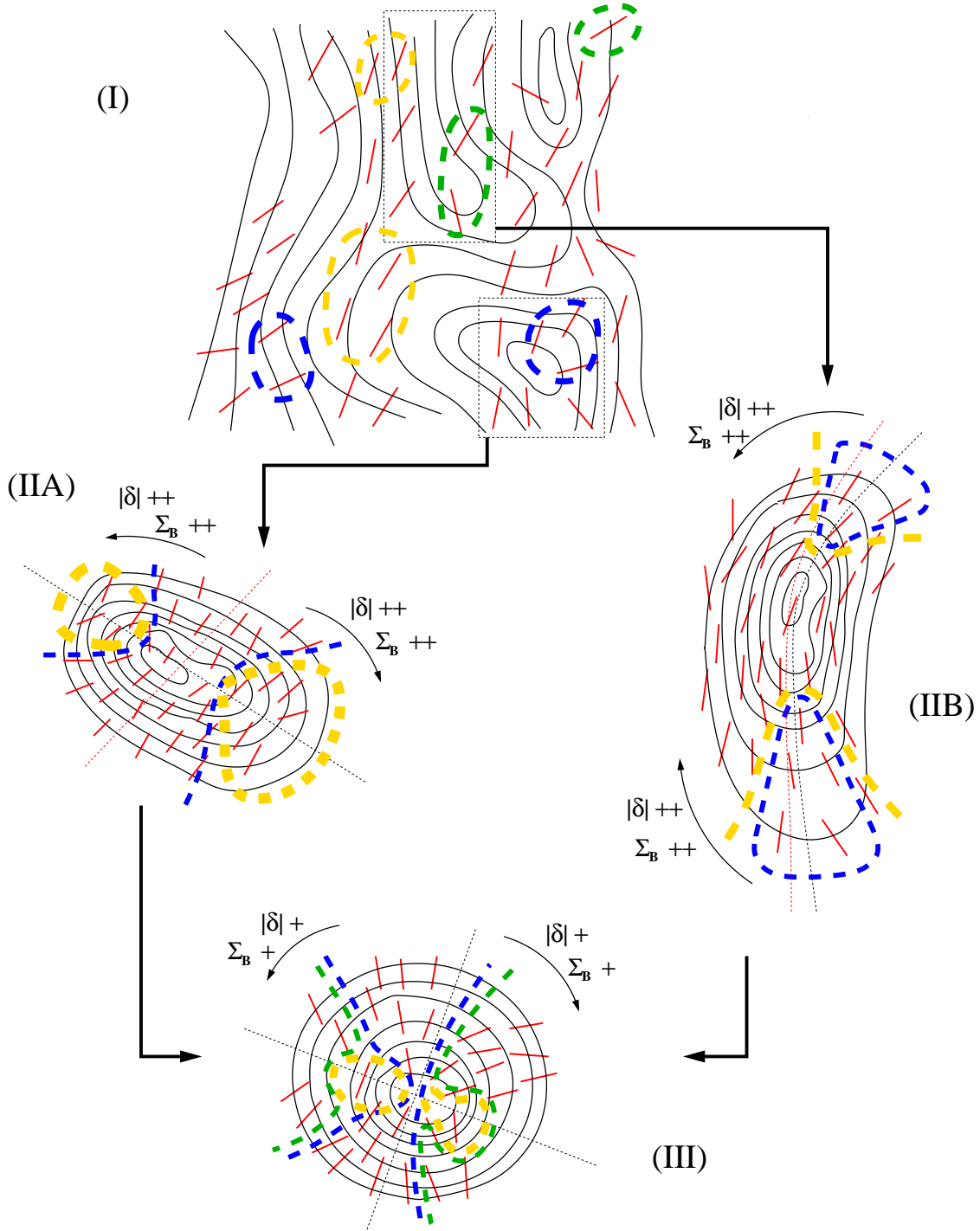}
 \caption{\label{figure_schematic_delta} \footnotesize
Schematic illustration of the angle $|\delta|$ across an evolutionary sequence in 
3 phases as described in Section \ref{section_scenarios} and summarized in 
Table \ref{table_evolution}. The red segments indicate the magnetic field orientations.
The black dotted squares in (I) mark the precursor
regions for (IIA) and (IIB). Following the color coding used in the 
Figures \ref{figure_delta_scales} and \ref{figure_delta_scales_cont} for $|\delta|$,  
blue dashed areas mark regions with close alignment between
field and intensity gradient orientations. Yellow areas mark zones with large deviations.
Green areas are in between. Mean field orientations (red dotted lines) and dust emission 
major axis (black dotted lines) are shown in (IIA) and (IIB). Accretion and outflow orientations
are shown with black dotted lines in (III), following the interpretation in 
Section \ref{section_delta_high_resolution}. Arrows with $|\delta| ++$ ($\Sigma_B ++$) and 
 $|\delta| +$ ($\Sigma_B +$) indicate directions of significant and moderate increase in 
$|\delta|$ ($\Sigma_B$).
}
\end{center}
\end{figure}

Similarities and analogous features are apparent between $|\delta|$-maps 
(Figure \ref{figure_delta_scales} and \ref{figure_delta_scales_cont})
and $\Sigma_B$-maps (Figure \ref{sigma_B_check}). In Figure \ref{figure_schematic_delta}
we adopt CO$+0.02-0.02$-, M$-0.02-0.07$-, M$+0.25+0.01$-, Mon R2- and W51 e2-type sources
as prototype sources with clear $|\delta|$- (and consequently $\Sigma_B$-) structures.
The competition between the magnetic field tension and other forces is reflected in 
distinct $|\delta|$-patches. In an early phase -- labeled as (I) in 
Figure \ref{figure_schematic_delta} -- 
larger scale dynamics in the surroundings
lead to initial
mass shuffling, leading to density contrasts but not yet well-defined shapes with clear
gravity centers. Patches with different $|\delta|$-values appear. The absence of clear 
gravity centers leads to more random-like $|\delta|$-structures that are less organized
without clear symmetry features. Nevertheless, areas with small (large) $|\delta|$ point
toward a minor (dominant) role of the magnetic field compared to other forces like gravity
and/or pressure gradients. On these largest scales in (I), some regions might already reveal
themselves as to what type of magnetic field - density configuration they will develop into.
The dotted squares in (I) illustrate areas that are likely to evolve into (IIA) and (IIB) in 
a later phase. Elongated structures of two types can be formed, with either a mean magnetic 
field orientation roughly perpendicular (IIA) or roughly parallel (IIB) to the density major
axis. One can speculate that certain initial configurations in (I) will either preferentially
lead to (IIA) or (IIB): gravity possibly has just started to shape the (IIA)-precursor region, 
whereas 
large-scale flows and/or turbulence 
seem more plausible to create the elongated structure in the 
(IIB)-precursor region with the field being compressed perpendicular to its mean orientation.
For any of these two regions, gravity is growing more important in (IIA) and (IIB) as the field
lines are being pulled in. Due to the different initial configurations, $|\delta|$-maps will 
reveal opposite trends here. The largest deviations in (IIA) are found at the two ends of the
dust emission major axis where the field maximally resists gravity. The field significance
$\Sigma_B$ is maximum here. On the other hand, the same areas in (IIB) show field orientations
that are closely aligned with the dust emission gradients, minimizing $|\delta|$ and 
$\Sigma_B$. Outside of these areas, $\Sigma_B$ is growing and diminishing in (IIB) and (IIA), 
respectively. As gravity leads to further contraction, these elongated structures are more
symmetrized in (III). Areas with large and small $|\delta|$ move closer together. Zones with 
maximum $|\delta|$- and $\Sigma_B$-values are likely shrunk and moved closer to the center
(yellow dashed patches in (III)), 
leaving more field lines closer to radial directions (green dashed patches in (III)). 
In this way, the initially different 
configurations (IIA) and (IIB) might evolve into one and the same system (III). This 
(smaller-scale) system can then be further analyzed using the symmetry properties of 
$\delta$ (including its sign) as introduced in Section \ref{section_delta_high_resolution}.
Thus, in this sequence of evolution, symmetries, features and changes in $|\delta|$ (and $\delta$)
reveal the role of the magnetic field. A minor or dominant role of the field can be classified
via $|\delta|$. A more quantitative estimate is achieved via $\Sigma_B$.

Many of the additional sources presented in Appendix B can be 
categorized as (IIA) or (IIB). Essentially no source is found that qualifies as (III) because
the CSO ~ 20\arcsec~ resolution typically does not fully reveal the densest core structures.
We note that a thorough analysis of mean polarization orientations versus source shapes is 
presented in \citet{tassis09}. Whereas in their work differences in mean orientations are
compared with the source aspect ratios, we are making use of the $|\delta|$-distributions.
By looking at broad features in these distributions -- i.e., peaking toward small values
versus peaking toward large values -- a simple bimodal categorization is possible. Offsets of 
roughly perpendicular or roughly parallel can still be identified with a peak and can, 
therefore, still be categorized as (IIA) and (IIB). Table \ref{table_quantities} summarizes 
our results. Sources are tentatively categorized as (I), (IIA), (IIB) or (III) by visual
inspection of $|\delta|$-maps and distributions. A more thorough statistical analysis
of Table \ref{table_quantities} will be presented in a forthcoming work.

\section{Summary and Conclusion}   \label{section_conclusion}

Based on the observed angle $\delta$ between magnetic field and dust emission gradient
orientations, we propose a generally valid interpretation for the role of the magnetic
field. This then further leads to a schematic scenario for the magnetic field in the 
evolutionary phases from large-scale systems to collapsing cores. The key points 
are summarized in the following.

\begin{enumerate}

\item
{\it Observed angle $|\delta|$:} Magnetic field and dust emission gradient orientations
are observed to show distinct patterns and symmetry features. This is both found in lower-
resolution CSO and higher-resolution (sub-)arcsecond SMA data. The angle $|\delta|$ between
these two orientations is straightforward to observe, does not rely on any modeling and can
often already be estimated by eye. $|\delta|$-histograms and  correlation coefficients 
reflect ensemble-averaged properties and differences. 

\item
{\it Observed high-resolution cores:} SMA high-resolution (collapsing) cores reveal 
additional symmetry properties in $\delta$ when taking into account the relative 
orientation between magnetic field and dust emission gradient. These features can be 
explained assuming a core with accretion and outflow zones. These findings seem to be
preserved even in the presence of possibly unknown projection effects. 

\item
{\it $|\delta|$ in the MHD force equation:} When  identifying an observed intensity gradient
orientation with the density gradient orientation in the MHD force equation, the magnetic 
field tension force term can be written with an additional factor $\sin|\delta|$. The angle
$|\delta|$ then measures to what extent the field tension force contributes to an observed
density gradient. Thus, $\sin|\delta|$ can be interpreted as a field alignment deviation or
a field deviation efficiency. 

\item
{\it $|\delta|$-map tracer:} Taking into account projection effects, it can be demonstrated
that a $|\delta|$-map always yields an approximation (lower limit to the magnetic field 
weight function $w_B$) to the more elaborate 
$\Sigma_B$-map which fully quantifies the magnetic field significance. Moreover, we show in 
practice, that positional changes in a $|\delta|$-map closely reflect changes in a $\Sigma_B$-map.
Therefore, we generally propose $|\delta|$-maps as reliable tracers for the change in the role
of the magnetic field over an observed map. 

\item
{\it Schematic evolutionary scenario:} Symmetries and features in $|\delta|$-maps (and 
consequently in $\Sigma_B$-maps) evolve from larger-scale cloud complexes to  collapsing 
cores. On the largest scales, apparently random $|\delta|$-patches are found which point toward
a variable magnetic field influence. In a next phase, elongated structures appear with a 
mean field orientation either roughly parallel or orthogonal to the major axis of the structure.
The field is here more dominant toward either the center or toward the two 
ends of the major axis. In a later third phase, gravity leads to less elongated but more
symmetrized (collapsing) cores. Here, $\delta$ can reveal zones of accretion and outflows. 
In many areas, the field is increasingly radially aligned, illustrating its diminishing 
role in this phase.

\end{enumerate}

The authors acknowledge the referees for very careful and thorough comments
that led to further insight in this work.
P.T.P.H. is supported by NSC Grant NSC99-2119-M-001-002- MY4.


\section*{Appendix}

\subsection*{Appendix A -- About the Intensity Gradient}       \label{appendix_gradient}

The derivation of the angle $\delta$ in Equation (\ref{mhd_delta_vect}) with its interpretation
as field alignment deviation (Equation (\ref{eq_delta_align})) is based on identifying an 
observed intensity gradient direction with the density gradient direction. This might seem at 
odds with the assumption in \citet{koch12a}, where the intensity gradient is identified with the 
orientation of the inertial term on the left-hand side in the MHD force equation (\ref{mhd_momentum}).
The assumption in \citet{koch12a} is phenomenologically motivated, by picturing a gradient in 
emission as a consequence of accumulating and compressing material. Such a process is the 
result of all the combined forces acting in a molecular cloud and is, therefore, measured with 
the inertial term. The assumption then states that the emission gradient must have some of this
information of the inertial term encrypted in its direction. We stress that -- in both the 
derivation of the force ratio $\Sigma_B$ and the local field strength -- only the direction of the emission
gradient but not its magnitude is relevant.

In the following we compare these two assumptions --
(a) emission gradient direction measures density gradient direction and (b) emission gradient
direction is a measure for inertial term direction --
by further analyzing the inertial term in 
Equation (\ref{mhd_momentum}). We assume stationarity, i.e. $\partial / \partial t \equiv 0$
\footnote{If necessary, this condition of strict stationarity can be relaxed by considering 
time scales $\Delta t$ where changes in velocity are small compared to typical changes per 
resolution element $\Delta R$, i.e. $\Delta \mathbf{v}/\Delta t \ll \Delta \mathbf{v}/\Delta R$.
In such situations, $\Delta \mathbf{v}/\Delta t$ will then not significantly alter the direction
of the inertial term.
}.
Using a vector identity, the inertial term can then be written as
$\rho\left(\mathbf{v}\cdot\mathbf{\nabla}\right)\mathbf{v}=-\rho\mathbf{v}\times \left(\nabla\times\mathbf{v}\right)
+ \rho \nabla \left(1/2 \mathbf{v}^2\right)$. Since we are interested in its
connection to the density gradient $\nabla \rho$, we are making use of 
$\nabla \left ( \rho v^2\right)=\rho \nabla v^2 + v^2 \nabla \rho$. This then leads to:
\begin{equation}
\rho\left(\mathbf{v}\cdot\mathbf{\nabla}\right)\mathbf{v}=-\rho\mathbf{v}\times \left(\nabla\times\mathbf{v}\right)
+ \frac{1}{2}\nabla \left(\rho v^2\right) - \frac{1}{2} v^2 \nabla \rho,         \label{eq_inertial}
\end{equation}
where $v$ is the absolute value of the velocity $\mathbf{v}$. The first term on the right-hand side depends
on the curl of the velocity field. For a curl-free flow or for a curl small compared to a density
change over the resolution of an observation, this term can be neglected. Otherwise, the curl of 
a velocity field being along the rotation axis that is normal to the plane where the velocities 
change, the direction of the term $\mathbf{v}\times \left(\nabla\times\mathbf{v}\right)$ is in the 
plane of the velocity flow. The second term 
describes a change in ram pressure. The change along a density gradient
direction is measured with the third term. 

\begin{figure}
\begin{center}
\includegraphics[scale=1]{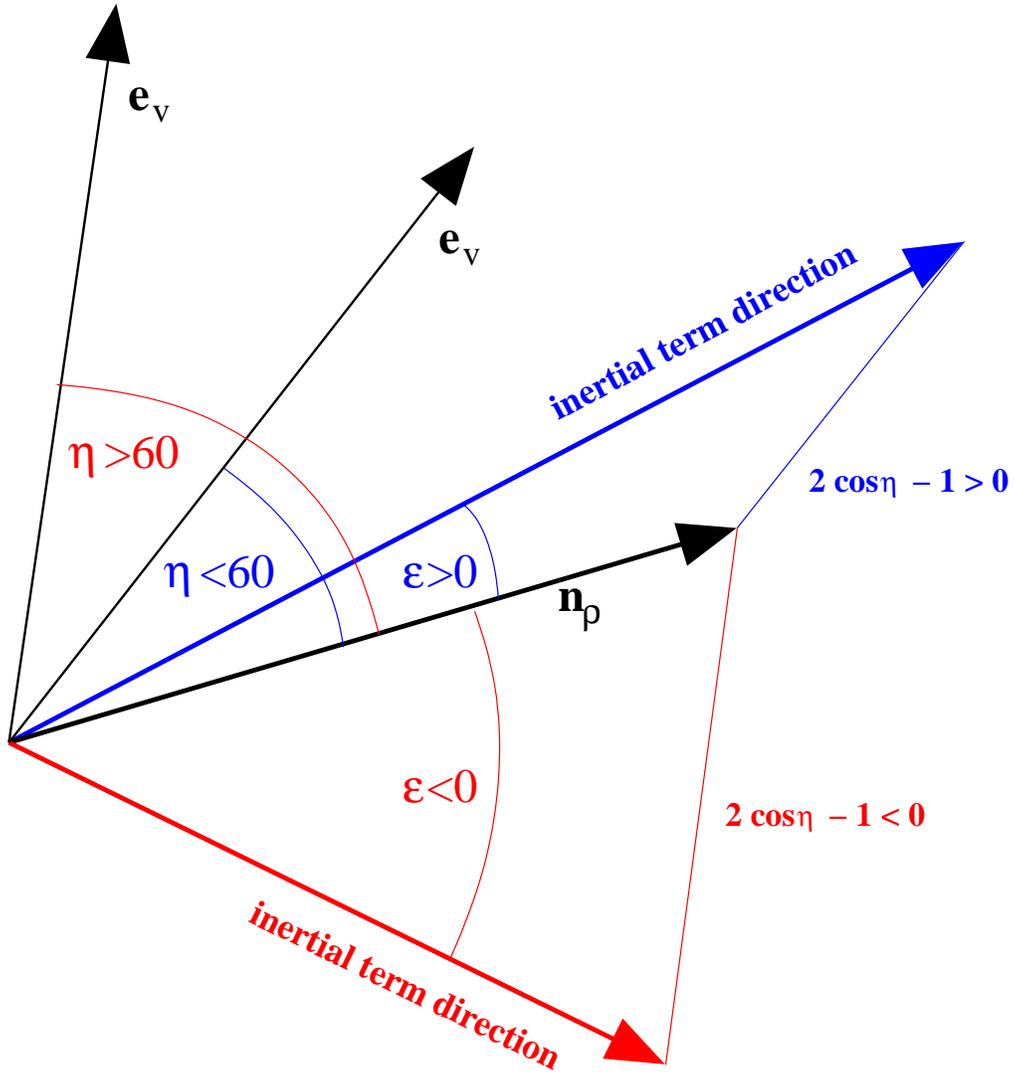}
 \caption{\label{schematic_inertial}
Schematic illustration of the inertial term deviation $\epsilon$. The angle between the 
density gradient direction $\mathbf{n}_{\rho}$ and the velocity direction $\mathbf{e}_v$
is labeled with $\eta$. Following Equation (\ref{eq_eta}), two cases need to be distinguished
where $\eta \le 60^{\circ}$ (illustrated in blue) or $\eta > 60^{\circ}$ (in red). 
The resulting inertial term direction deviates from $\mathbf{n}_{\rho}$ by $\epsilon >0$
(blue) or $\epsilon <0$ (red).
}
\end{center}
\end{figure}

\begin{figure}
\begin{center}
\includegraphics[scale=1]{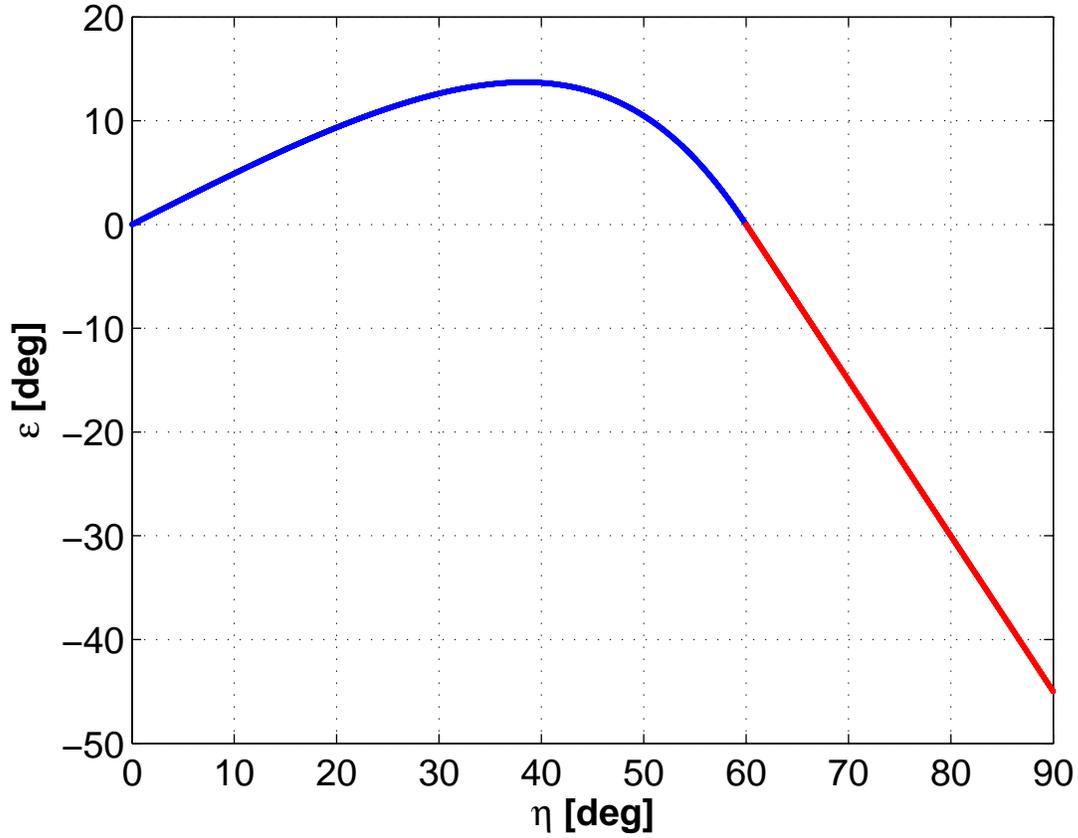}
 \caption{\label{fig_eps_eta}
Inertial term deviation $\epsilon$ as a function of the angle $\eta$ between the density 
gradient and the velocity direction (Figure \ref{schematic_inertial}). 
$\epsilon =0$ indicates that the inertial term is aligned with the density gradient. 
For angles $\eta<60^{\circ}$ the velocity contribution along $\mathbf{e}_v$ is added to 
$\mathbf{n}_{\rho}$ and the deviation $\epsilon$ is positive (blue regime)
with a maximum around $14^{\circ}$ and an average deviation of about $8.8^{\circ}$. 
For angles $\eta$ larger than $60^{\circ}$ the deviation becomes negative because 
$\mathbf{e}_v$ is subtracted (red regime). 
}
\end{center}
\end{figure}

Based on Equation (\ref{eq_inertial}) we can now distinguish two main situations first:
(i) {\it The velocity field has a negligible curl and the velocity is 
aligned with the density gradient:} A density change occurs along a velocity direction. The 
direction of the inertial term  follows the density gradient\footnote{A hydrodynamical collapse, 
or a close-to-hydrodynamical collapse is an example for this situation. In a spherically 
symmetrical collapse model, density gradients and accelerations only occur in radial direction.
Independent of any symmetries, if compression (roughly uniform over a resolution element) 
happens along streamlines, inertial term and 
density gradients will be aligned.
}.
(ii) {\it The velocity field has a negligible curl, but the velocity direction is different
from the density gradient direction:} In order to analyze this general case, we approximate the 
differential operators on the right-hand side in Equation (\ref{eq_inertial}) as changes per
resolution element $\Delta R$: 
$1/2\nabla \left(\rho v^2\right) - 1/2 v^2 \nabla \rho \sim 
-1/2 v^2 \Delta\rho/\Delta R \mathbf{n}_{\rho} + 1/2 \left(v^2 \Delta \rho/\Delta R + 
2\rho v \Delta v / \Delta R \right) \mathbf{e}_v $, where $\mathbf{e}_v $ is a unity vector
tangential to a streamline\footnote{
In writing so, we assume that a dominating change in ram pressure (over a resolution element)
is directed along a (bulk) streamline, i.e., we are omitting a possible component 
$\nabla \left(\rho v^2\right )_{\bot}$ orthogonal to a bulk flow. Note that this does not strictly exclude
the existence of turbulence. Turbulence can exist up to a level such that a dominating organized bulk flow 
(in a resolution element) can still be identified.
}. 
We additionally make use of the continuity equation with 
$\partial \rho /\partial t \approx 0$, thus giving $\mathbf{v}\cdot \nabla \rho 
+ \rho (\nabla \cdot \mathbf{v}) = 0$. Introducing the angle $\eta$ between the velocity and 
the density gradient directions (Figure \ref{schematic_inertial}) together with again 
approximating the differential operators in the continuity equation with the resolution 
elements $\Delta R$, leads to $(\Delta v / \Delta R)/v \sim -\cos \eta (\Delta\rho / \Delta R)/\rho$.
This relation states that any relative change in velocity (relative to the absolute 
velocity) is balanced by a relative change in density modulo the alignment factor $\cos\eta$.
This then allows us to express the above approximation in terms of $\Delta \rho/\Delta R$ only:
\begin{equation}
-\frac{1}{2}v^2 \frac{\Delta \rho}{\Delta R}\mathbf{n}_{\rho} + 
\frac{1}{2}\left(v^2 \frac{\Delta \rho}{\Delta R} + 2\rho v \frac{\Delta v}{\Delta R} \right) \mathbf{e}_v
= -\frac{1}{2}v^2 \frac{\Delta \rho}{\Delta R}\left [\mathbf{n}_{\rho} + \left(2\cos\eta -1\right)\mathbf{e}_v \right].
                                                                        \label{eq_eta}
\end{equation}
Equation (\ref{eq_eta}) is an estimate of the inertial term, neglecting the curl and 
approximating differential operators with changes per resolution element $\Delta R$. 
In particular, the direction of the inertial term -- which is relevant for the polarization-
intensity gradient method -- is now linked to the density gradient and the velocity direction 
with the expression in square brackets. We now investigate this term further, with the goal
of quantifying the resulting deviation $\epsilon$ from the direction of the 
density gradient $\mathbf{n}_{\rho}$ (Figure \ref{schematic_inertial}). The alignment angle $\eta$ between 
$\mathbf{n}_{\rho}$ and $\mathbf{e}_v$ is in the range of 0 to 90$^{\circ}$. Therefore, the 
factor $(2\cos\eta -1)$ changes sign at $\eta=60^{\circ}$, defining two different regimes where
$\mathbf{e}_v$ is either added to or subtracted from $\mathbf{n}_{\rho}$. Figure \ref{fig_eps_eta}
displays the inertial term deviation $\epsilon$ from the density gradient as a function of $\eta$.
For $\eta$ within $60^{\circ}$, the maximum deviation is about $14^{\circ}$ with an average of 
$8.8^{\circ}$. These relatively small deviations are a consequence of the continuity equation
which controls the magnitude $(2\cos\eta -1)$ along the velocity direction $\mathbf{e}_v$.
For angles $\eta>60^{\circ}$, the deviation $\epsilon$ grows linearly to a maximum of $-45^{\circ}$
where $\mathbf{e}_v$ is orthogonal to $\mathbf{n}_{\rho}$ (red regime in Figure \ref{fig_eps_eta}).
The overall average deviation, including extreme values $\eta>60^{\circ}$, is 
$\langle \epsilon \rangle \approx 17^{\circ}$.

The analysis here makes clear that the possible difference between velocity and density 
gradient directions plays a major role in how closely the two assumptions (a) and (b) agree. The presence
of a significant velocity curl in Equation (\ref{eq_inertial}) might lead to an additional more complicated dependence of 
$\epsilon$ on $\eta$. We are not attempting to quantify this here. Further investigations
via numerical simulations will be more appropriate to tackle down the detailed influence
of a curl term. 
We are, thus, left with average deviations of about $8.8^{\circ}$ to about $17^{\circ}$ 
between a measured emission gradient direction and the inertial term direction. These are 
conservative estimates while assuming no significant velocity curl over a beam-averaged
resolution element. 
We now recall that uncertainties in polarization position angles from observations are typically 
in the range of about 5 to $10^{\circ}$. Errors in the emission gradient orientations are
slightly smaller, $\sim 3$ to $5^{\circ}$, due to averaging in the interpolation when calculating
the gradient. Additional statistical uncertainties in the field orientation resulting from turbulent
dispersion are also around 5 to $10^{\circ}$ \citep{koch10}. We, therefore, conclude that adopting 
the emission gradient orientation is a reasonably good approximation for the orientation of the 
inertial term in the polarization--intensity gradient method. Furthermore, in the presence of 
other (statistical) uncertainties, the overall error budget is not significantly affected. 

\begin{figure}
\begin{center}
\includegraphics[scale=0.7]{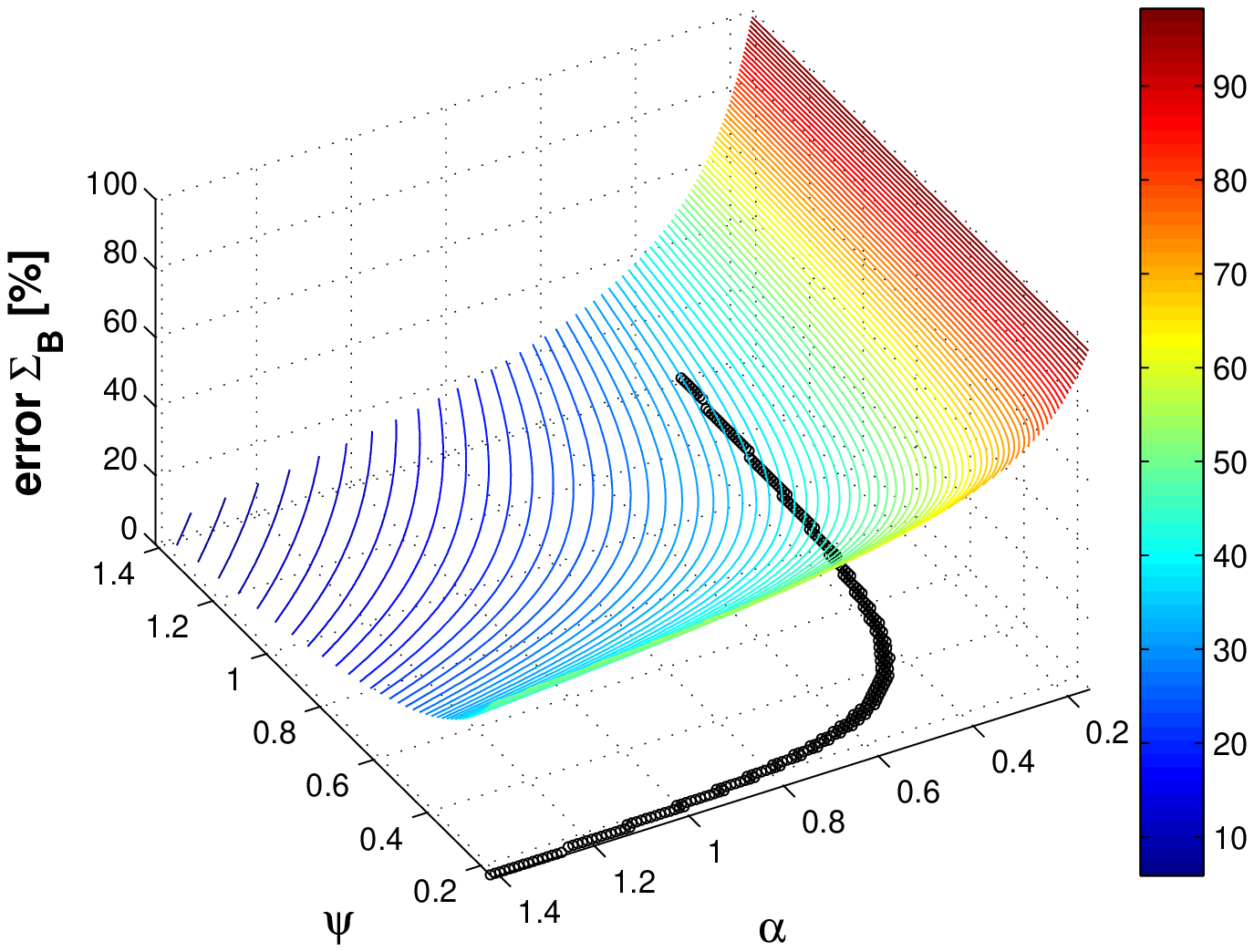}
\includegraphics[scale=0.7]{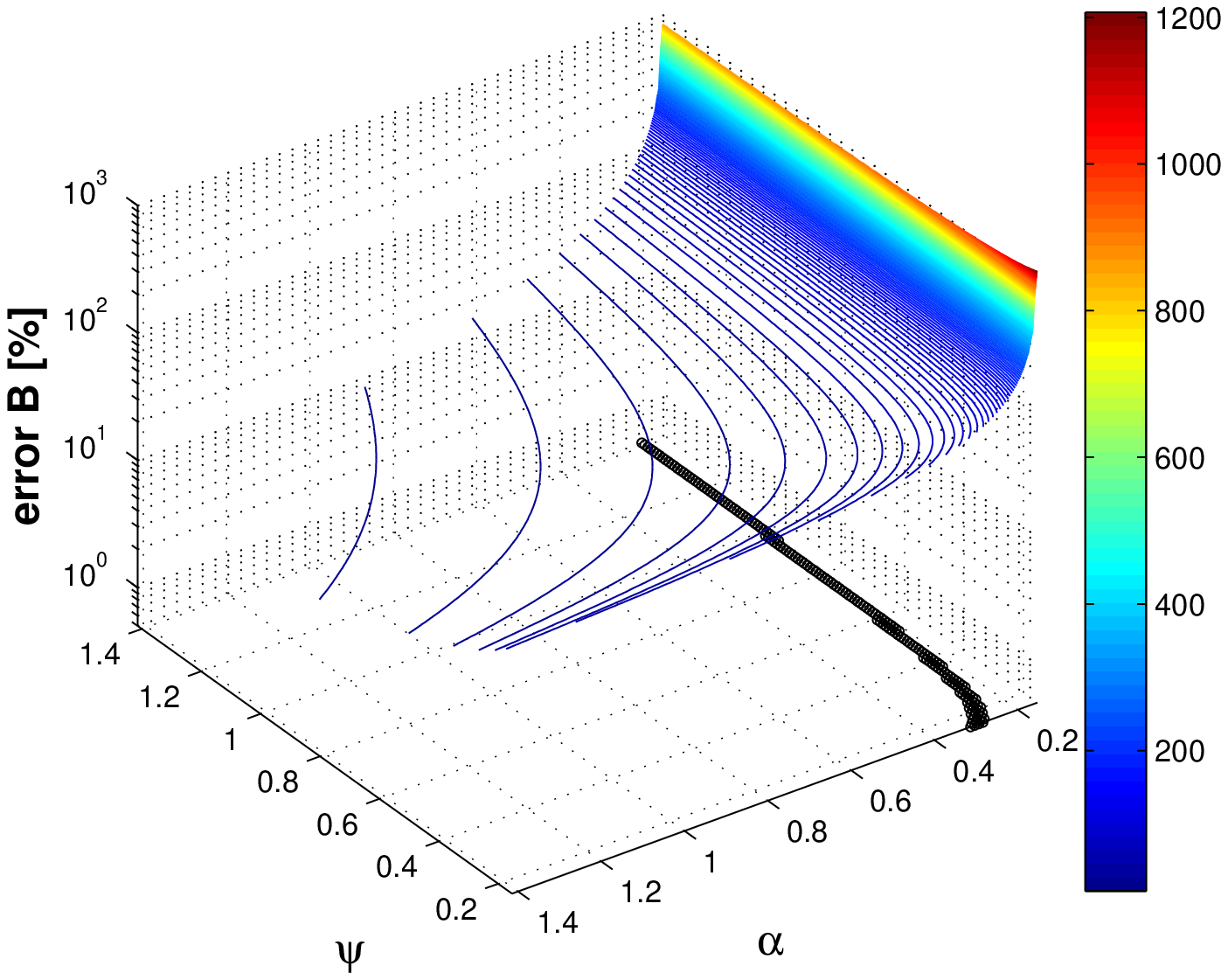}
 \caption{\label{error_sigma_B_epsilon}
Top panel: Errors in the force ratio $\Sigma_B$ for an average inertial term deviation of $8.8^{\circ}$
(Figure \ref{fig_eps_eta}) as a function of the angles $\alpha$ and $\psi$. 
The color coding displays the errors in percentage.
The black-dotted line in the $(\alpha,\psi)$-plane marks the 50 \% error boundary. 
Bottom panel: Errors in the magnetic field strength $B$ for an average inertial term deviation of $8.8^{\circ}$.
The black-dotted line again marks the 50\% error boundary.
}
\end{center}
\end{figure}

Finally, we propagate the uncertainty in the orientation of the inertial term, i.e., the 
deviation $\epsilon$, through to the magnetic field significance $\Sigma_B$ and the field strength
$B$. To that purpose, we replace $\sin\psi / \sin\alpha$ with $\sin(\psi+\epsilon)/\sin(\alpha-\epsilon)$
in the Equations (\ref{eq_sigma_B}) and (\ref{eq_B}). Figure \ref{error_sigma_B_epsilon}
displays the expected errors in $\Sigma_B$ and $B$ for $\langle \epsilon \rangle \approx 8.8^{\circ}$
as a function of the angles $\alpha$ and $\psi$. For most of the $(\alpha, \psi)-$parameter space, 
errors are less than 50\%. Larger errors are typically limited to small angles $\alpha$. This is, 
however, less of a consequence of the deviation $\epsilon$ than the result of the factor $1/\sin\alpha$
that amplifies any error for small values of $\alpha$. We, thus, finally conclude that within the 
framework of the polarization--intensity gradient method where the inertial term orientation is 
approximated with the emission gradient orientation, both $\Sigma_B$ and $B$ remain robust estimates.

\subsection*{Appendix B -- $|\delta|$ and $\Sigma_B$ across a Larger Sample}  \label{appendix_sample}

With the goal of adding further evidence to the similarities found in the $|\delta|$- and $\Sigma_B$-maps
(Section \ref{sub_section_comparison}) and the proposed evolutionary scenario (Section \ref{section_scenarios}), 
we analyze an additional 24 sources from the CSO catalog \citep{dotson10}. Sources are
chosen based on their coverage of detected polarized emission, together with the condition
that the Stokes $I$ continuum emission need to be strong enough to define a gradient. In order
to have at least some connected patches, we request a minimum of 10 measured polarization --
intensity gradient pairs. 

\begin{figure}
\begin{center}
\includegraphics[scale=0.55]{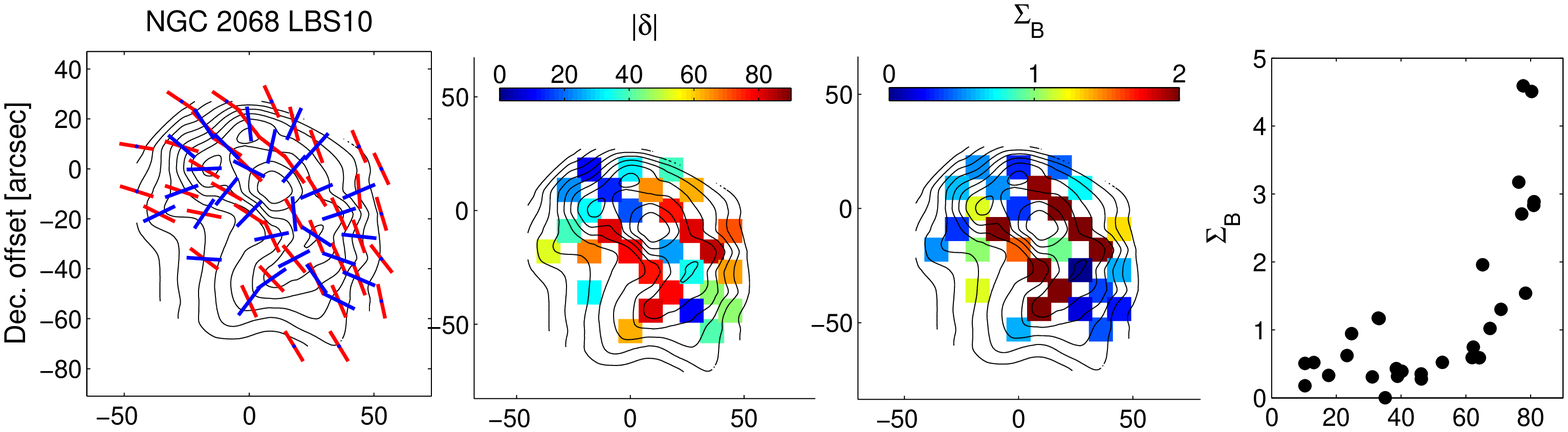}
\includegraphics[scale=0.55]{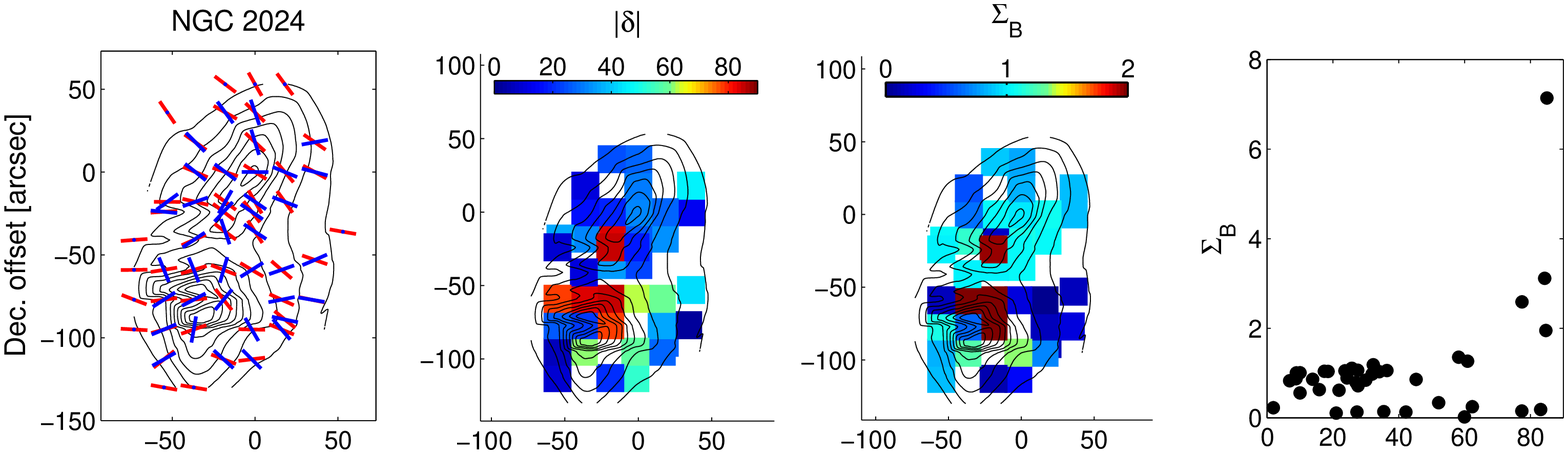}
\includegraphics[scale=0.55]{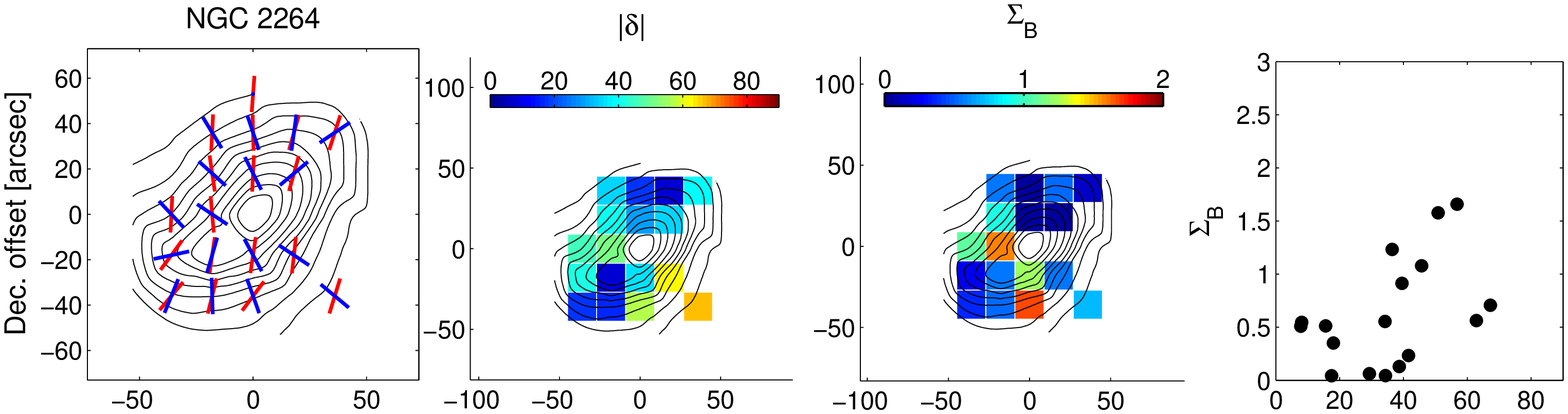}
\includegraphics[scale=0.55]{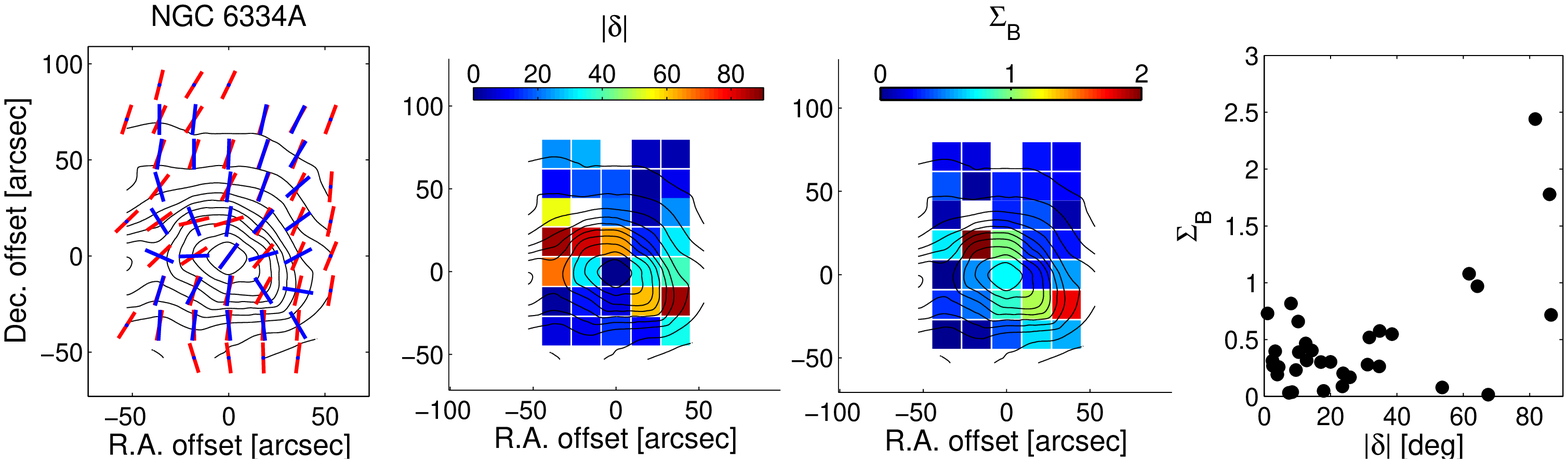}
 \caption{\label{sample_summary}
Additional sources from the CSO sample \citep{dotson10}. The panels from 
left to right display the Stokes $I$ dust continuum overlaid with 
magnetic field segments (red) and and dust intensity gradients (blue), 
$|\delta|$-map, $\Sigma_B$-map and the connection between $\Sigma_B$
and $|\delta|$. Note that, for a better uniform display across the sample, 
the color-coding for $\Sigma_B$ in the third-column
panels is saturated at 2.  
}
\end{center}
\end{figure}

\begin{figure}
\begin{center}
\includegraphics[scale=0.55]{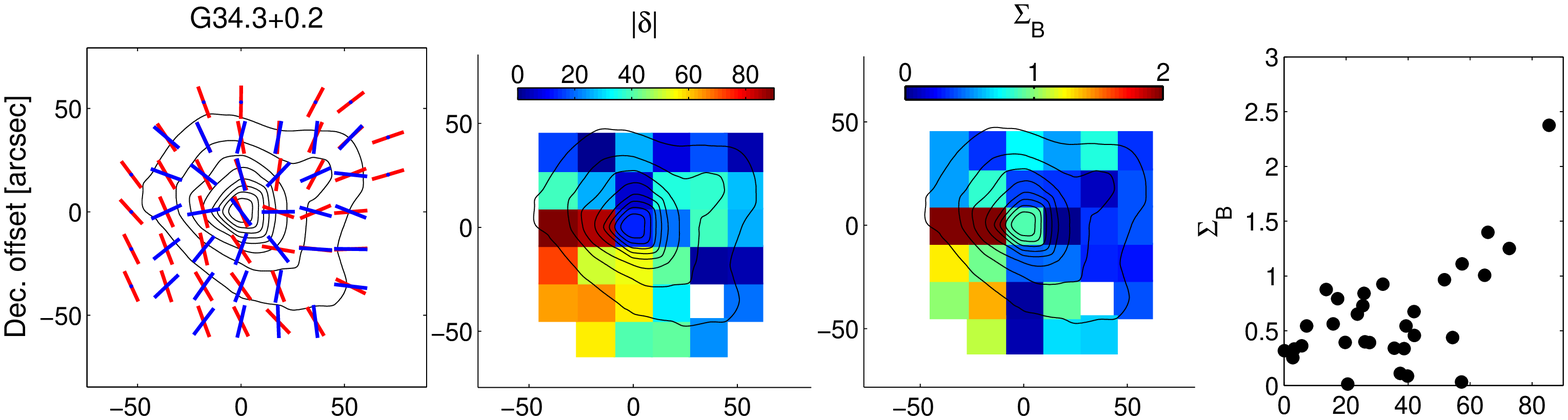}
\includegraphics[scale=0.55]{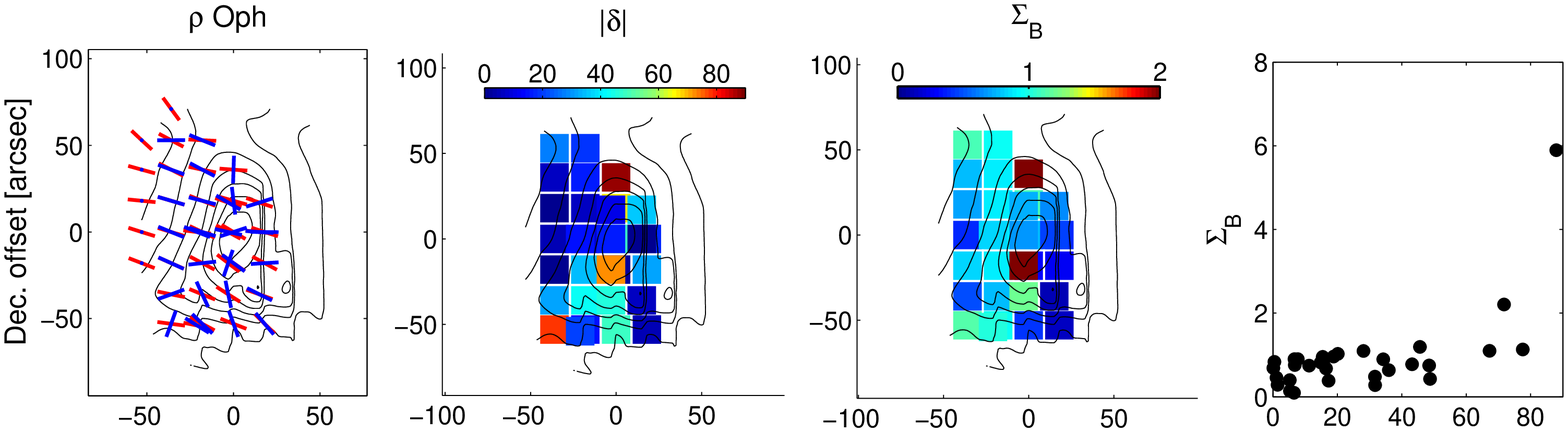}
\includegraphics[scale=0.55]{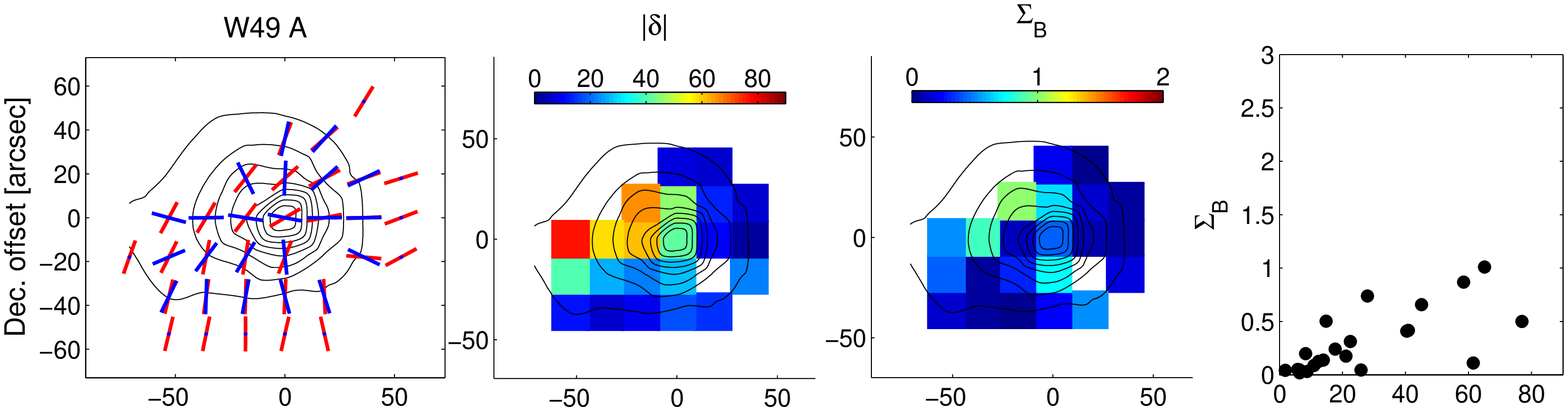}
\includegraphics[scale=0.55]{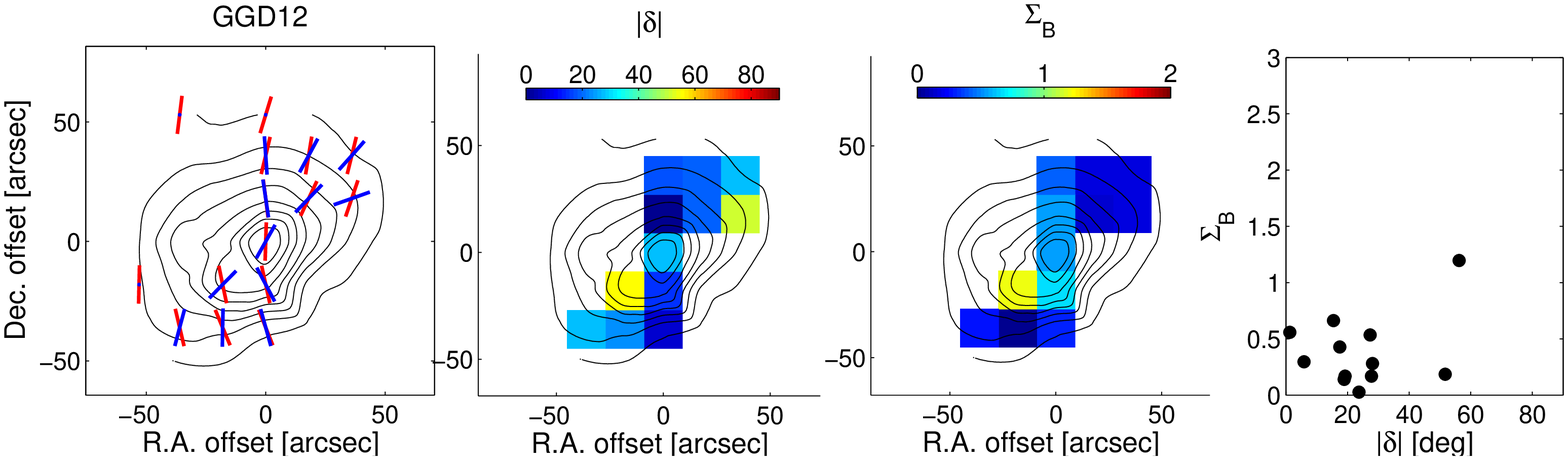}
 \caption{\label{sample_summary_2}
Same as in Figure \ref{sample_summary}.
}
\end{center}
\end{figure}

\begin{figure}
\begin{center}
\includegraphics[scale=0.55]{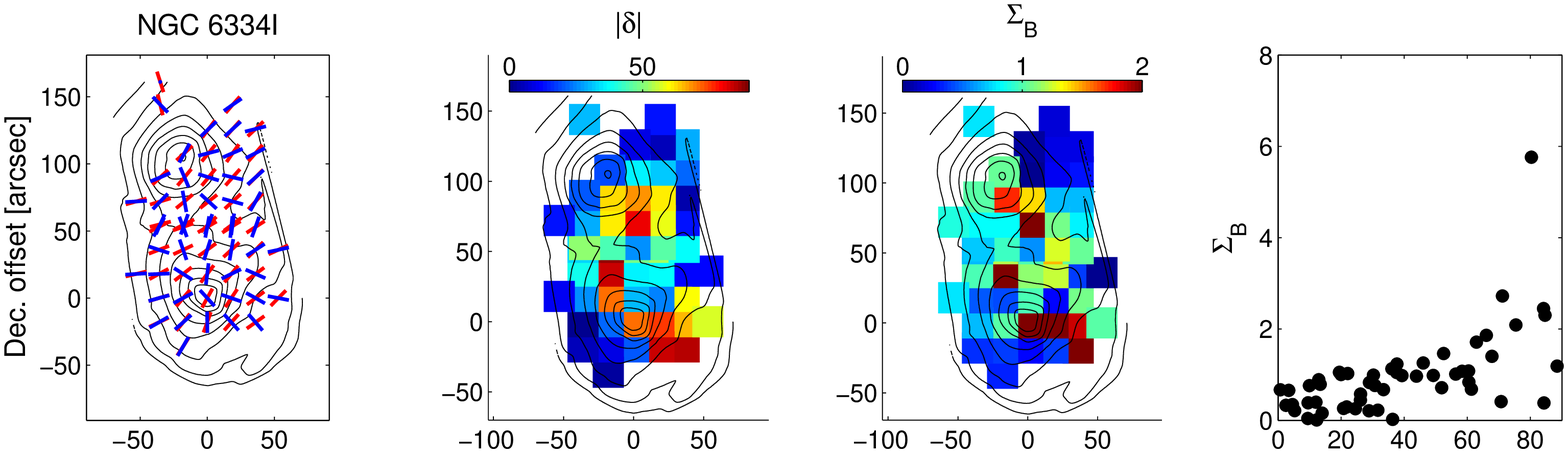}
\includegraphics[scale=0.55]{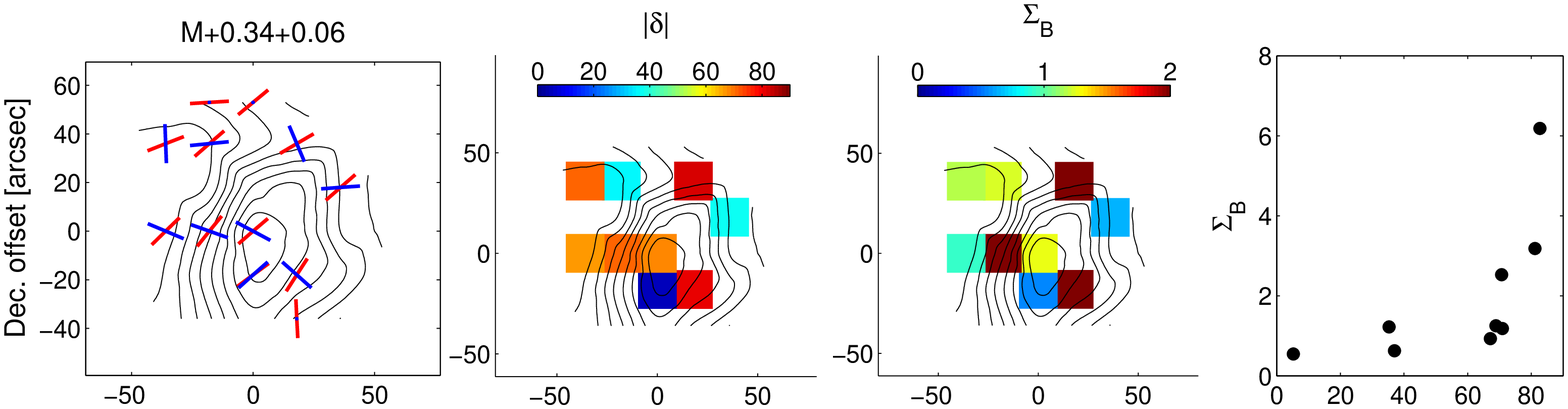}
\includegraphics[scale=0.55]{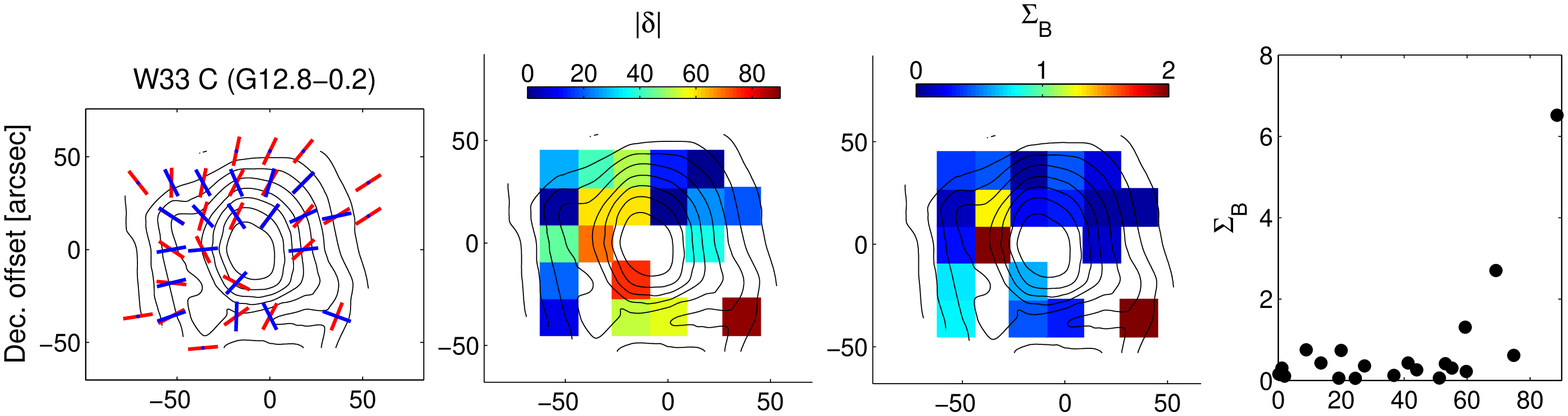}
\includegraphics[scale=0.55]{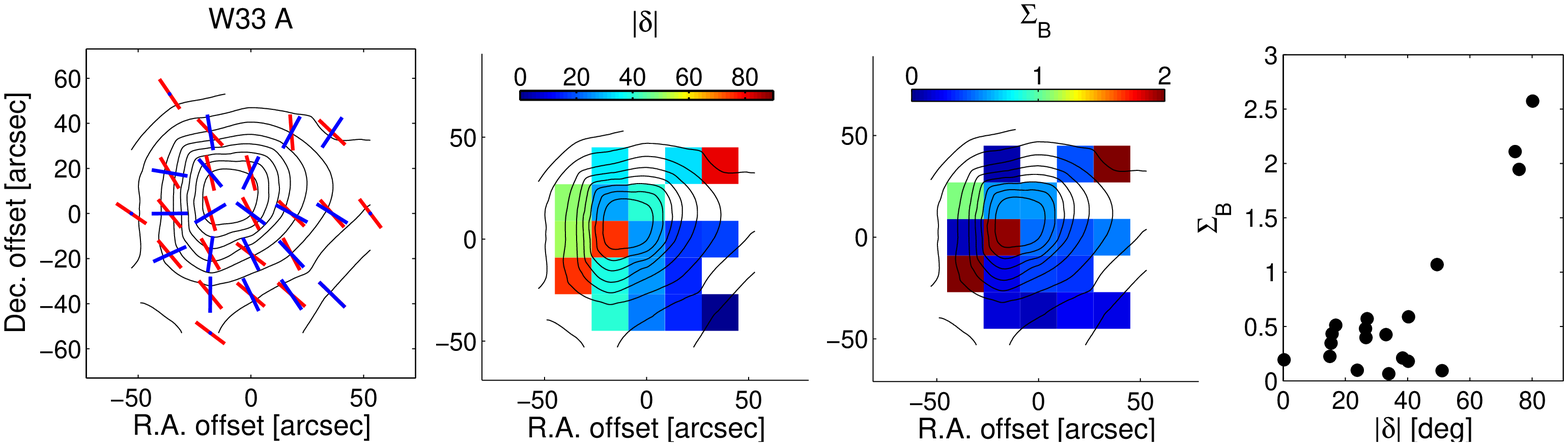}
 \caption{\label{sample_summary_3}
Same as in Figure \ref{sample_summary}.
}
\end{center}
\end{figure}

\begin{figure}
\begin{center}
\includegraphics[scale=0.55]{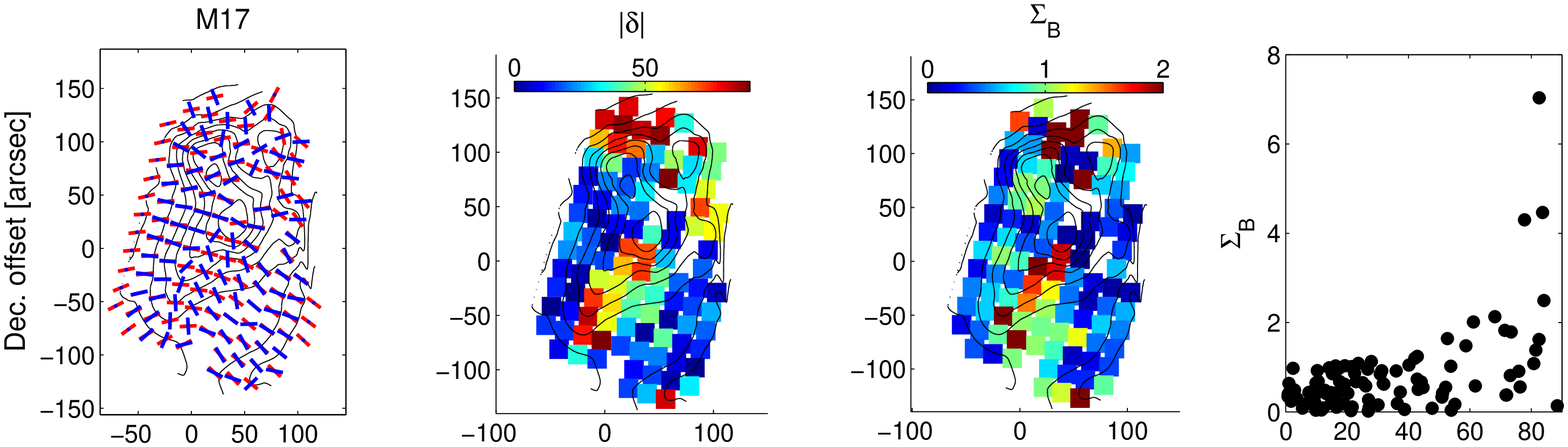}
\includegraphics[scale=0.55]{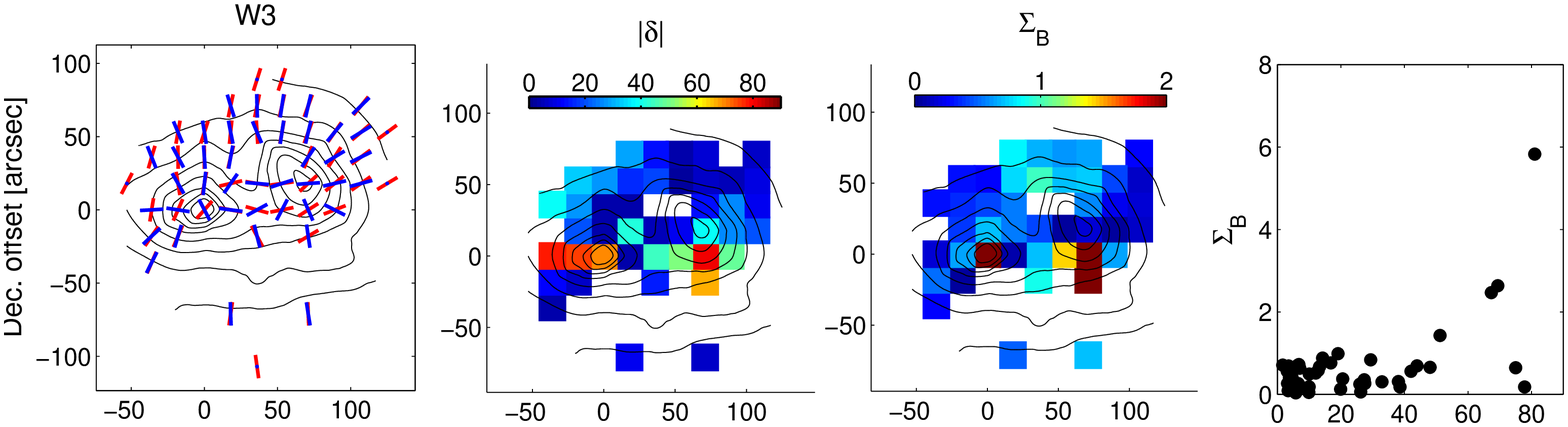}
\includegraphics[scale=0.55]{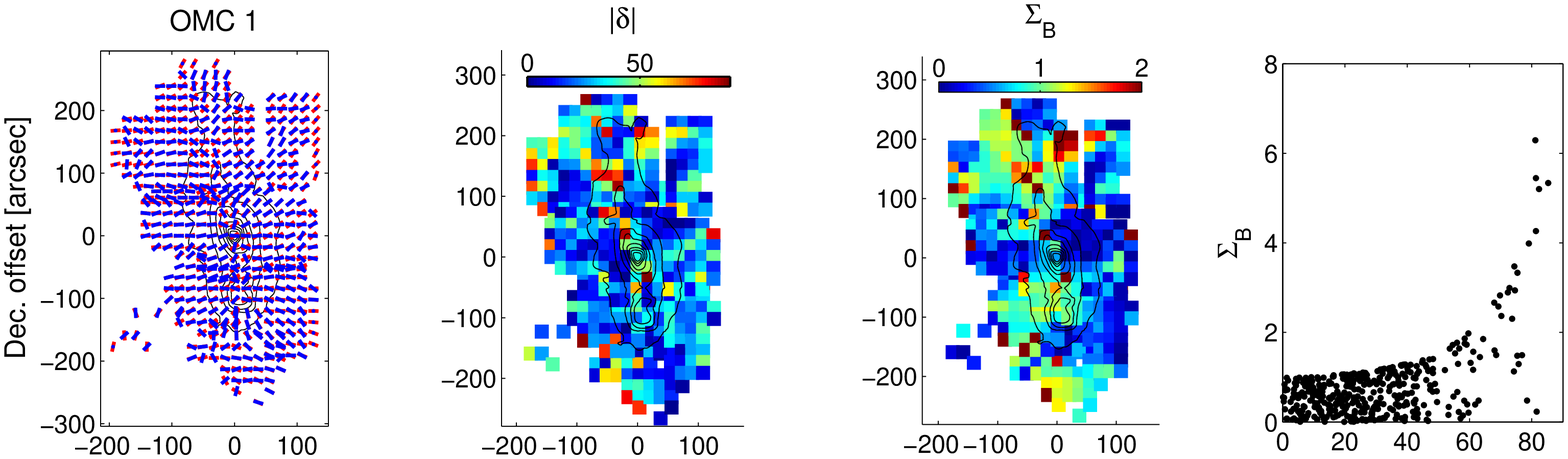}
\includegraphics[scale=0.55]{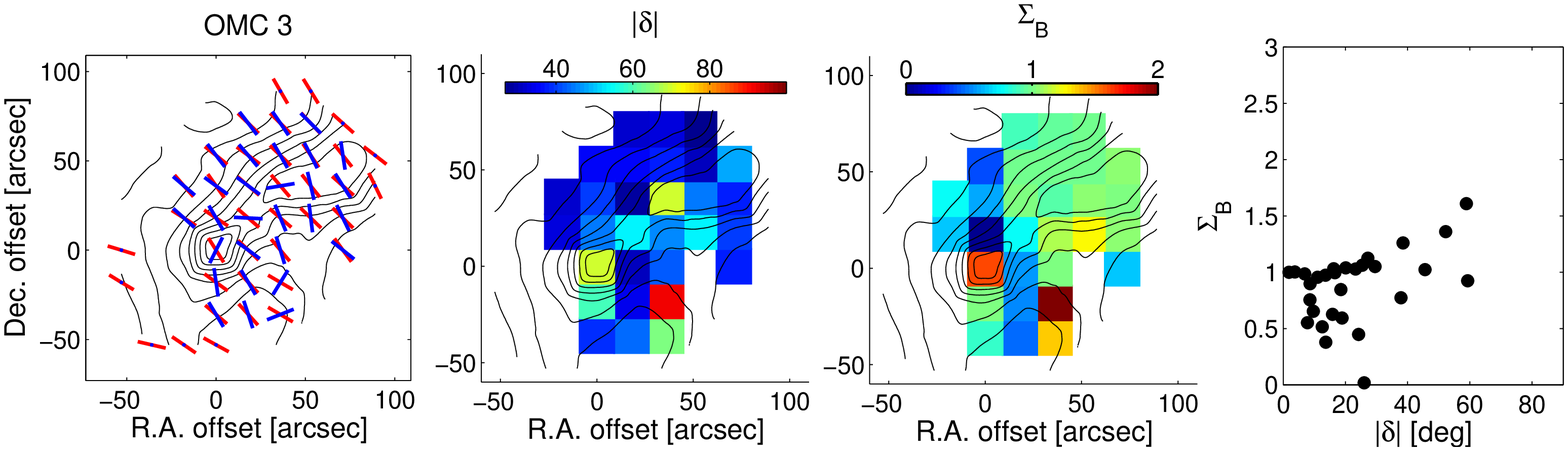}
 \caption{\label{sample_summary_4}
Same as in Figure \ref{sample_summary}.
}
\end{center}
\end{figure}

\begin{figure}
\begin{center}
\includegraphics[scale=0.55]{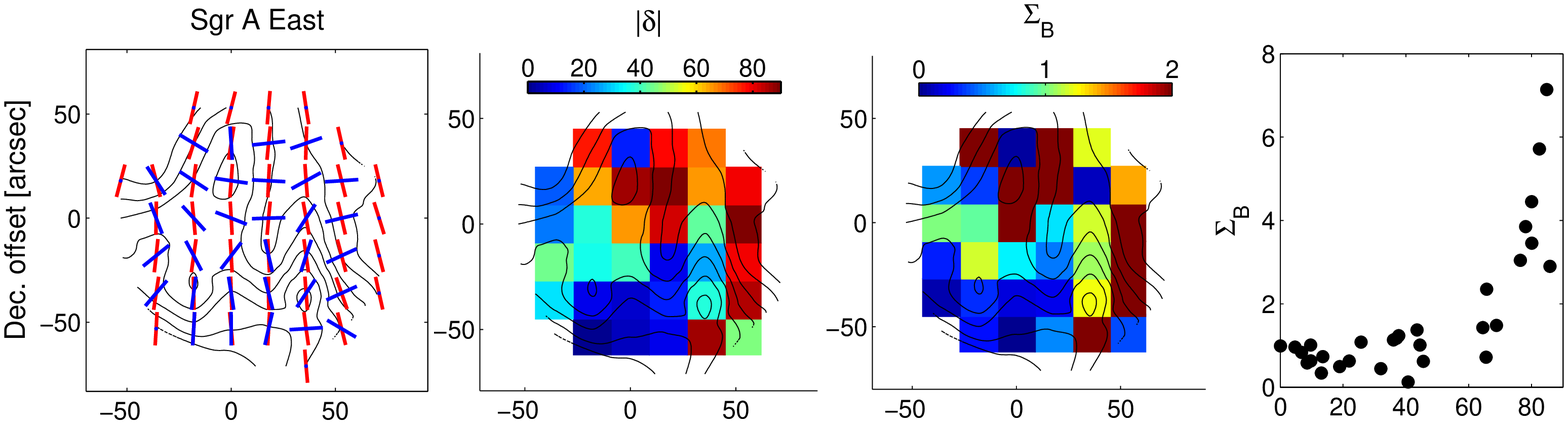}
\includegraphics[scale=0.55]{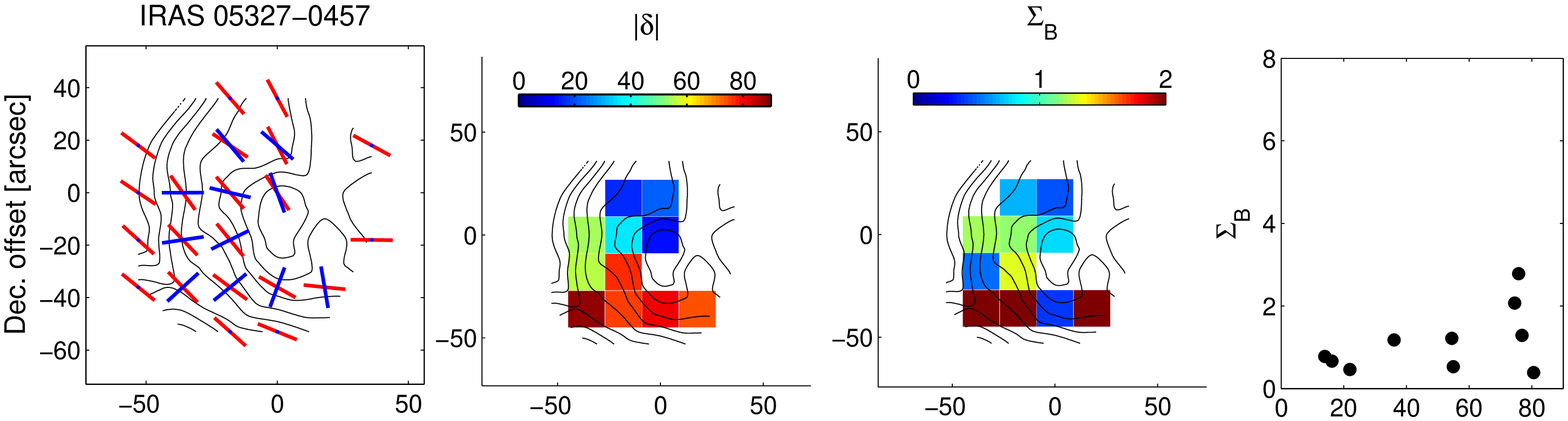}
\includegraphics[scale=0.55]{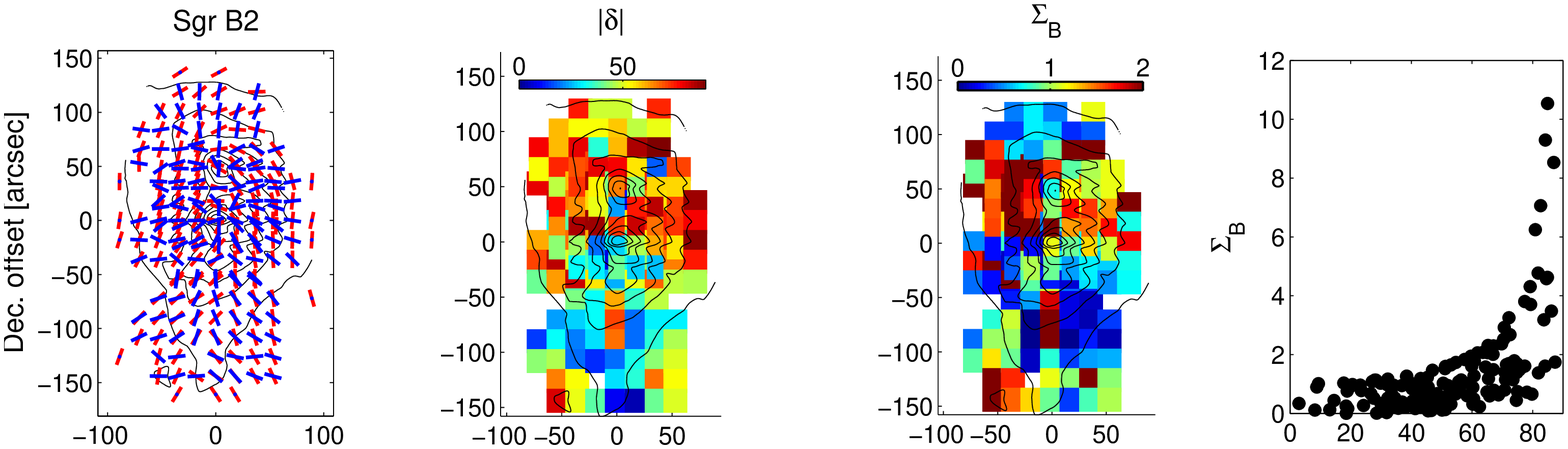}
\includegraphics[scale=0.55]{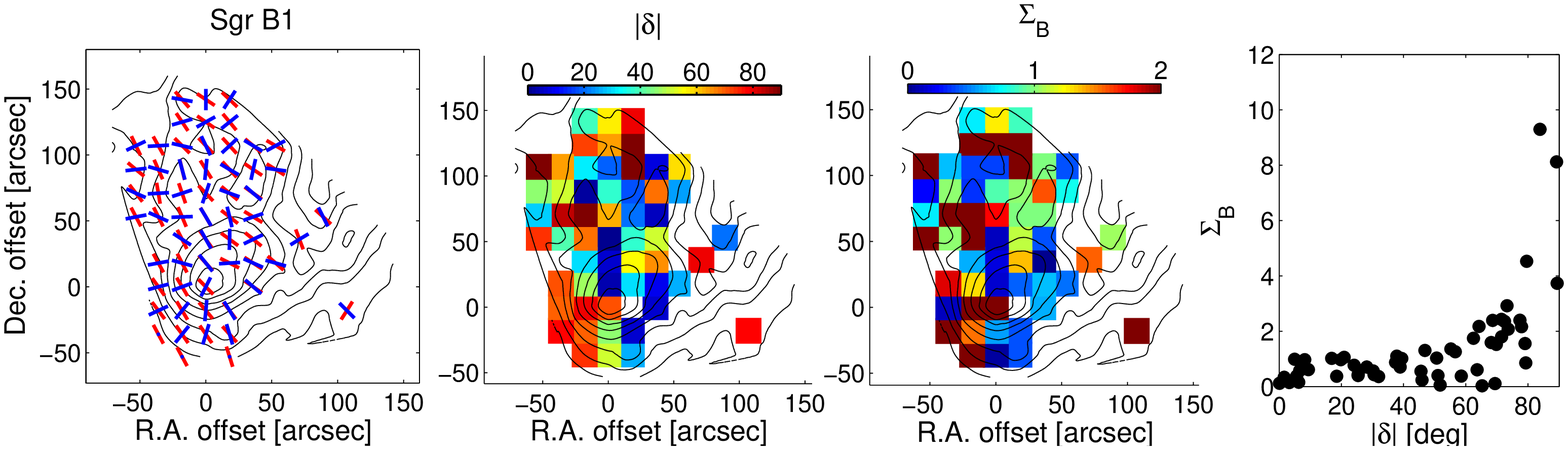}
 \caption{\label{sample_summary_5}
Same as in Figure \ref{sample_summary}.
}
\end{center}
\end{figure}

\begin{figure}
\begin{center}
\includegraphics[scale=0.55]{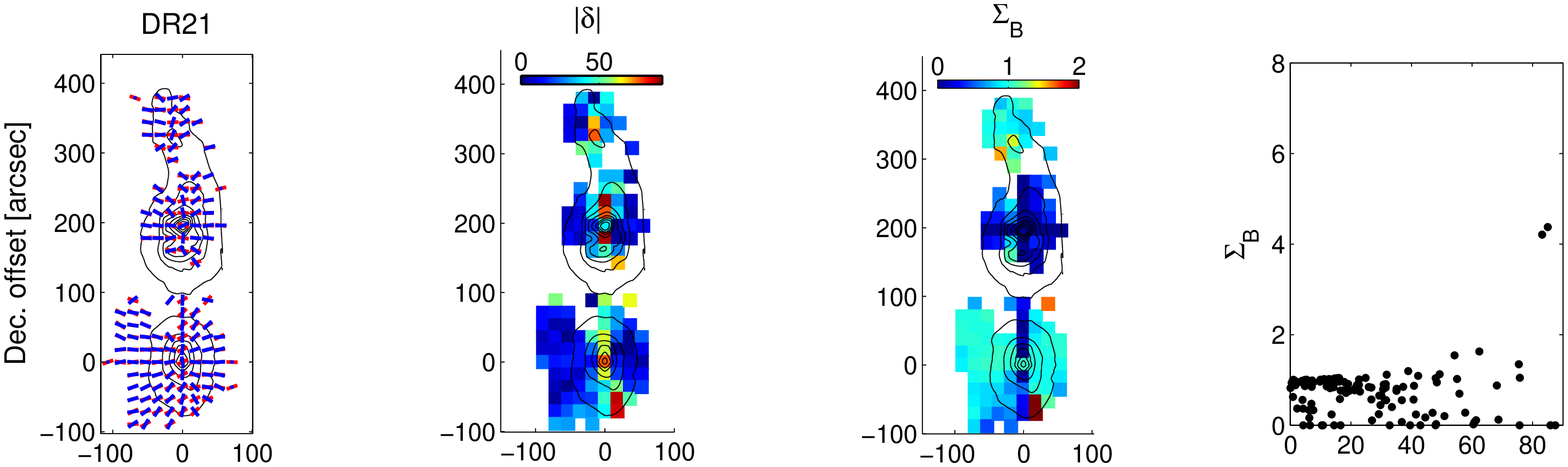}
\includegraphics[scale=0.55]{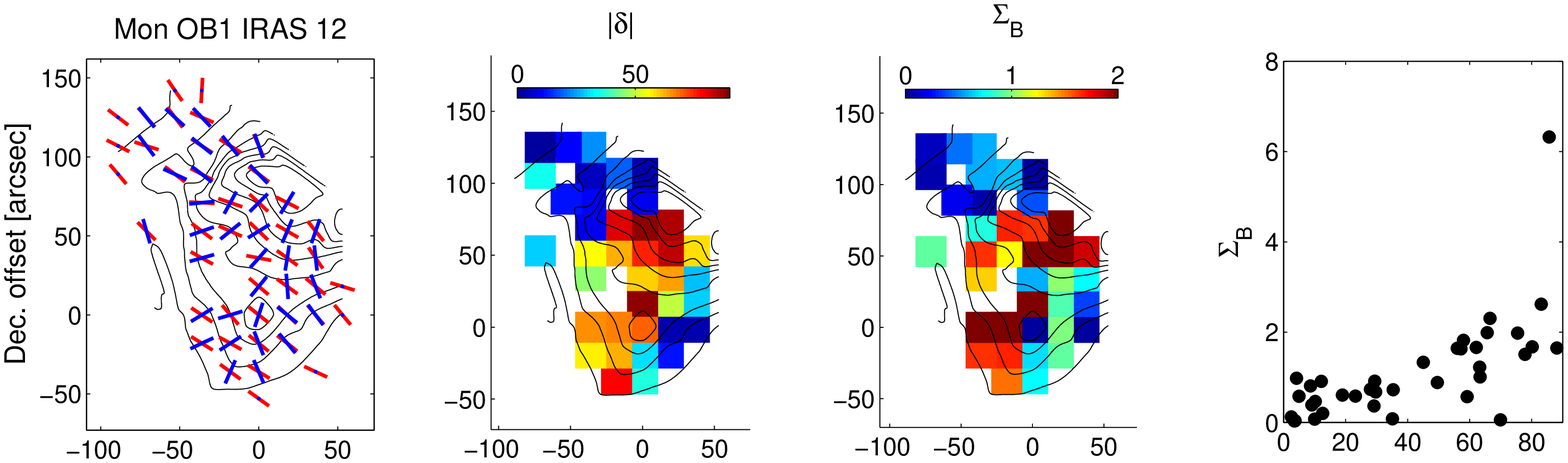}
\includegraphics[scale=0.55]{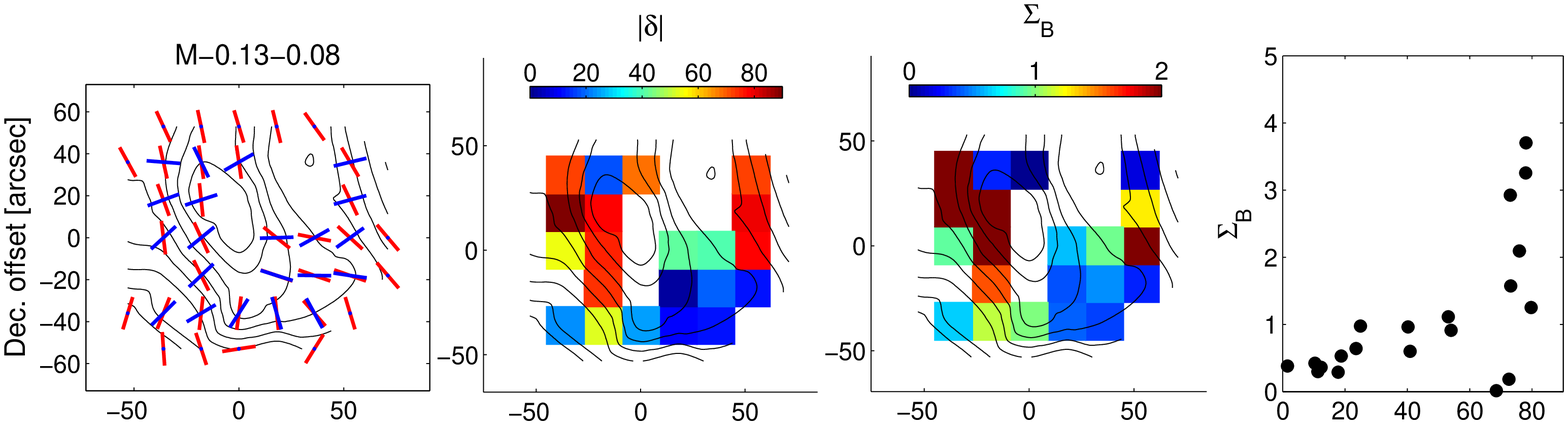}
\includegraphics[scale=0.55]{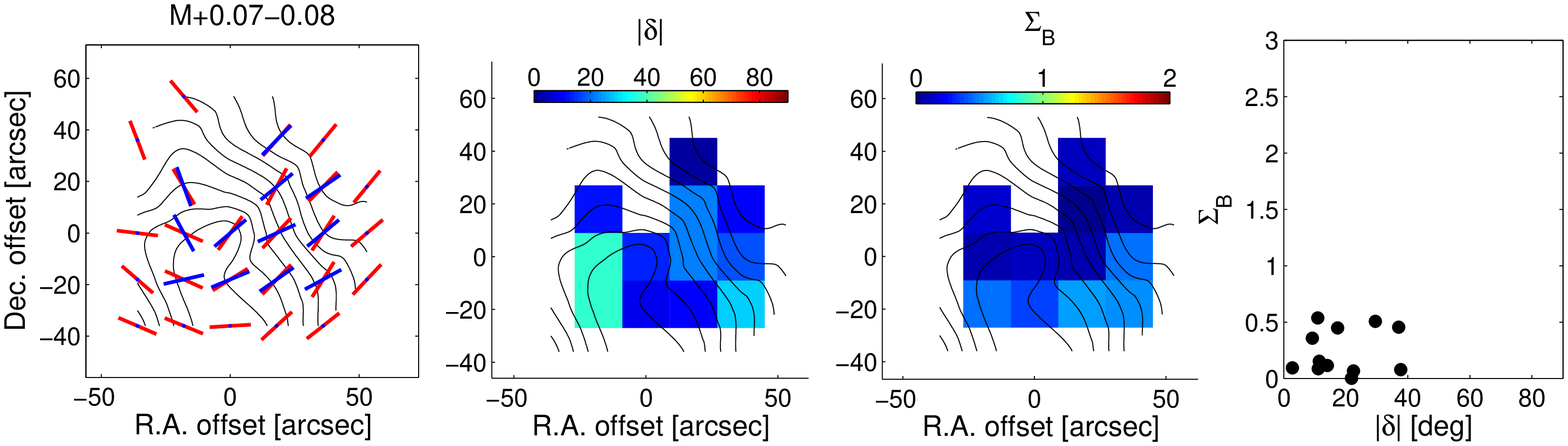}
 \caption{\label{sample_summary_6}
Same as in Figure \ref{sample_summary}.
}
\end{center}
\end{figure}

\begin{figure}
\begin{center}
\includegraphics[scale=0.5]{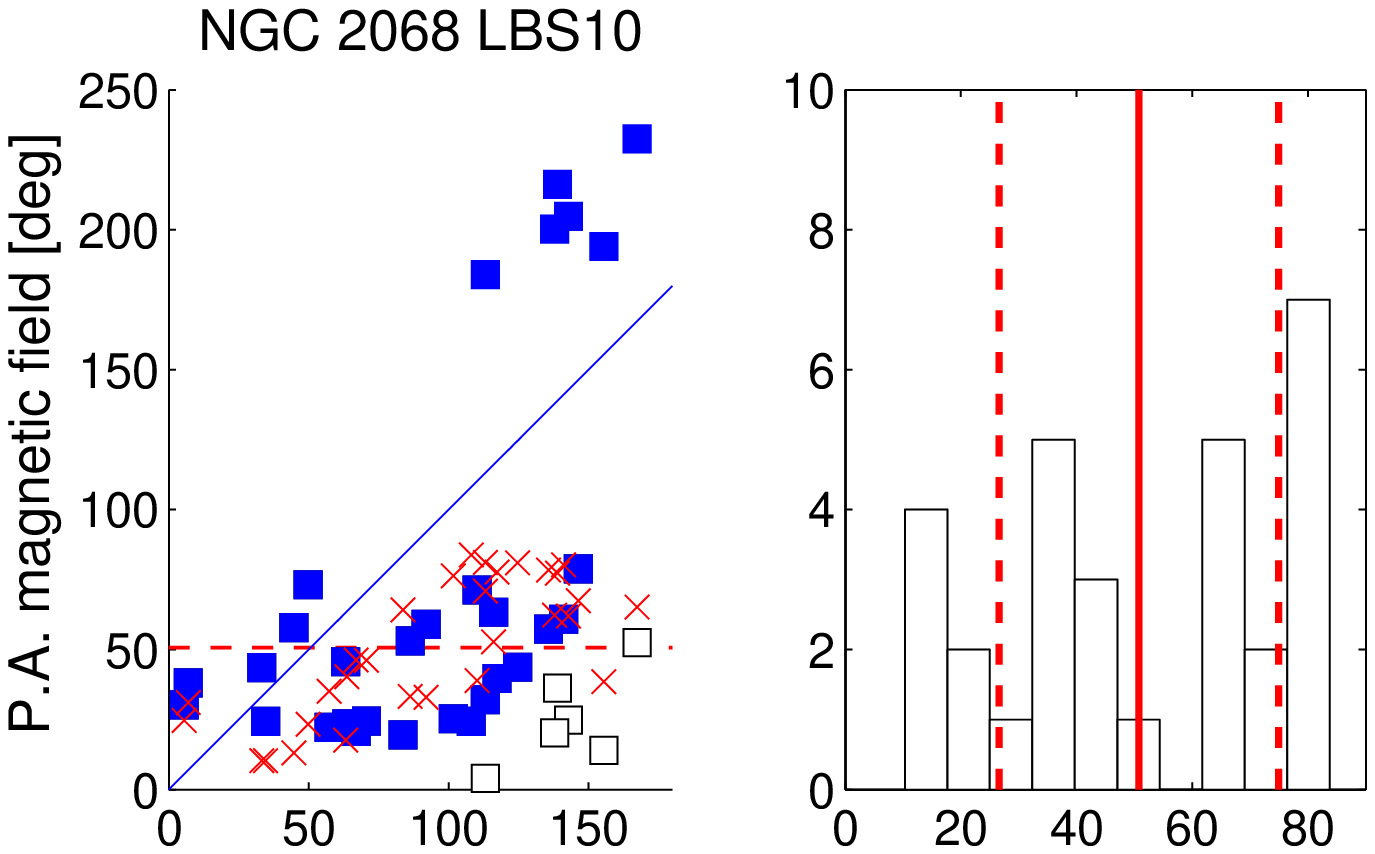}
\includegraphics[scale=0.5]{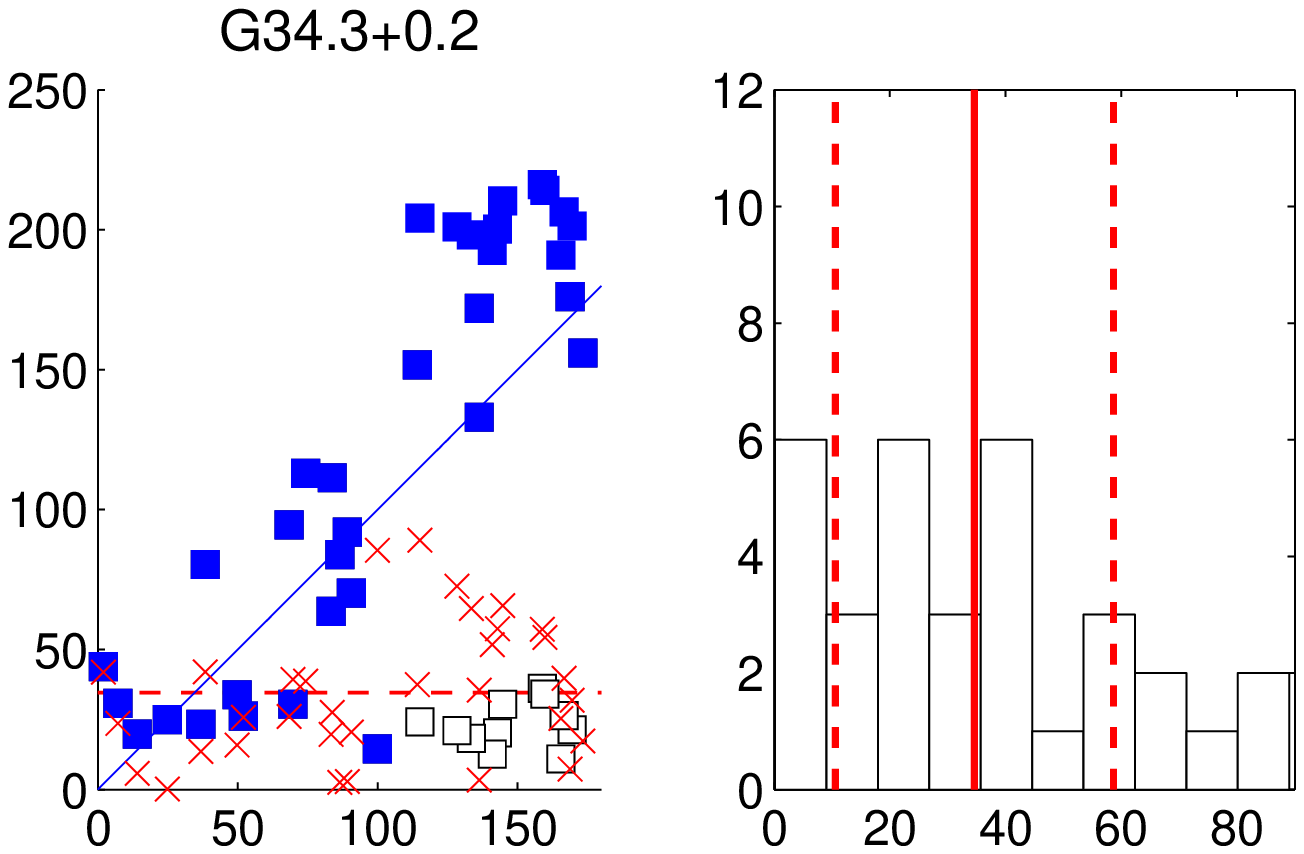}
\includegraphics[scale=0.5]{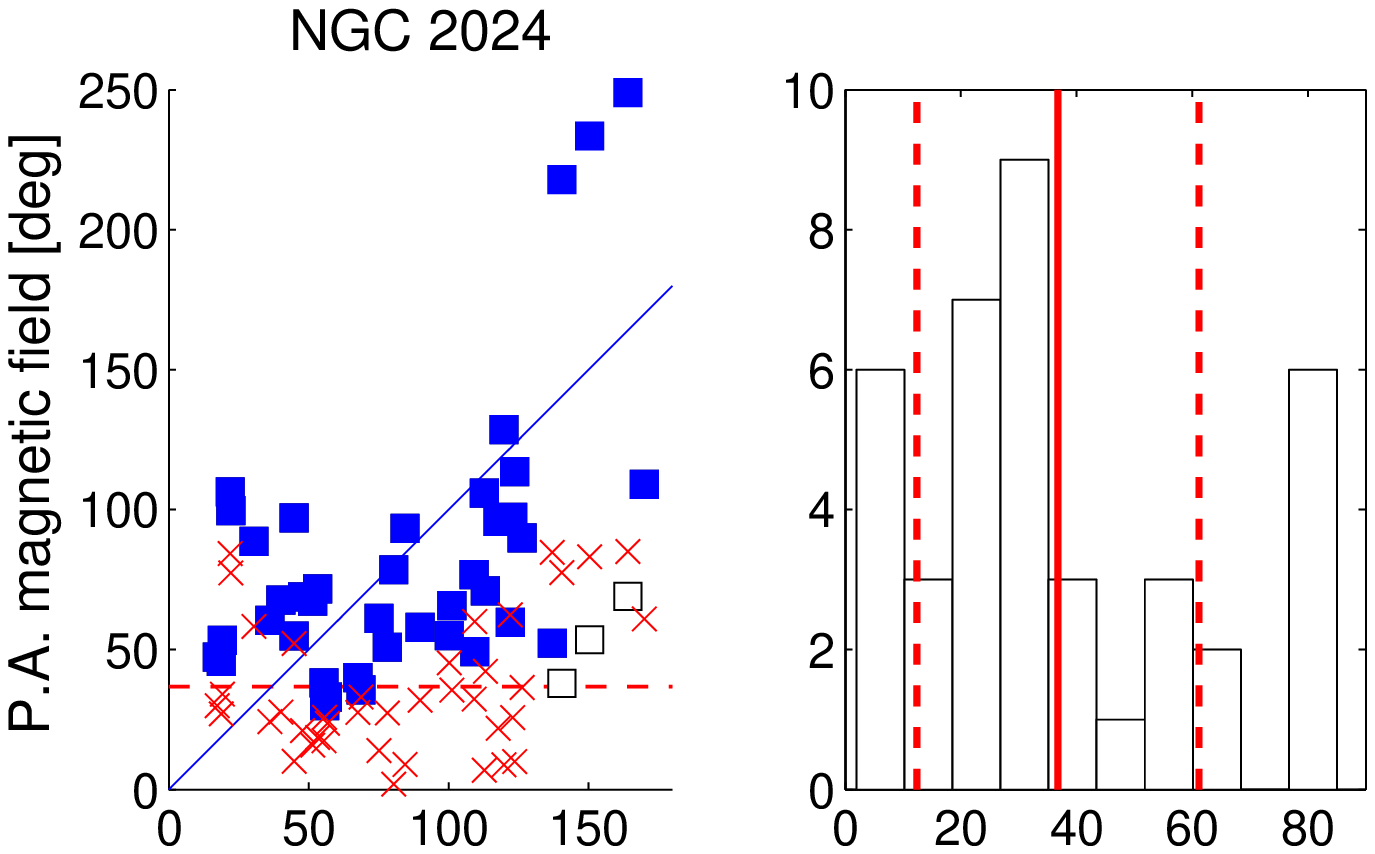}
\includegraphics[scale=0.5]{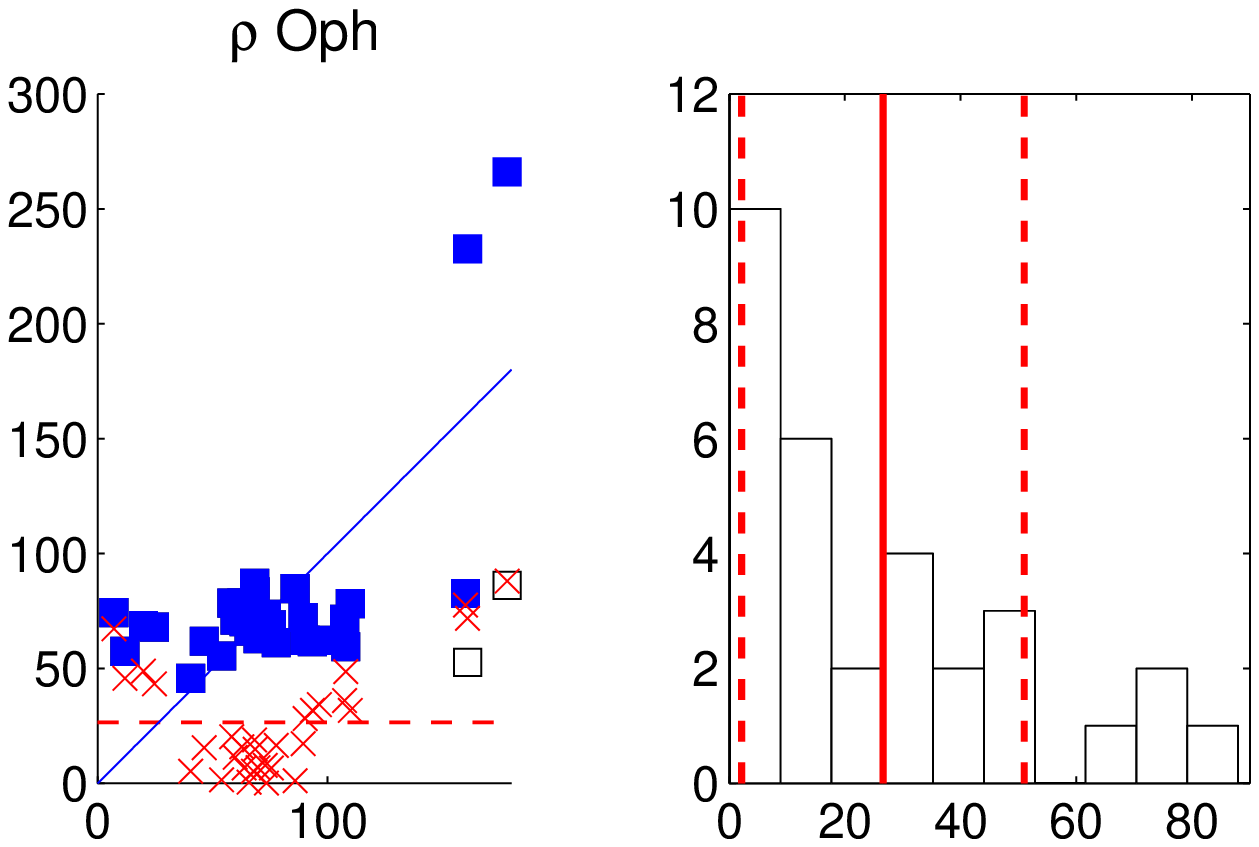}
\includegraphics[scale=0.5]{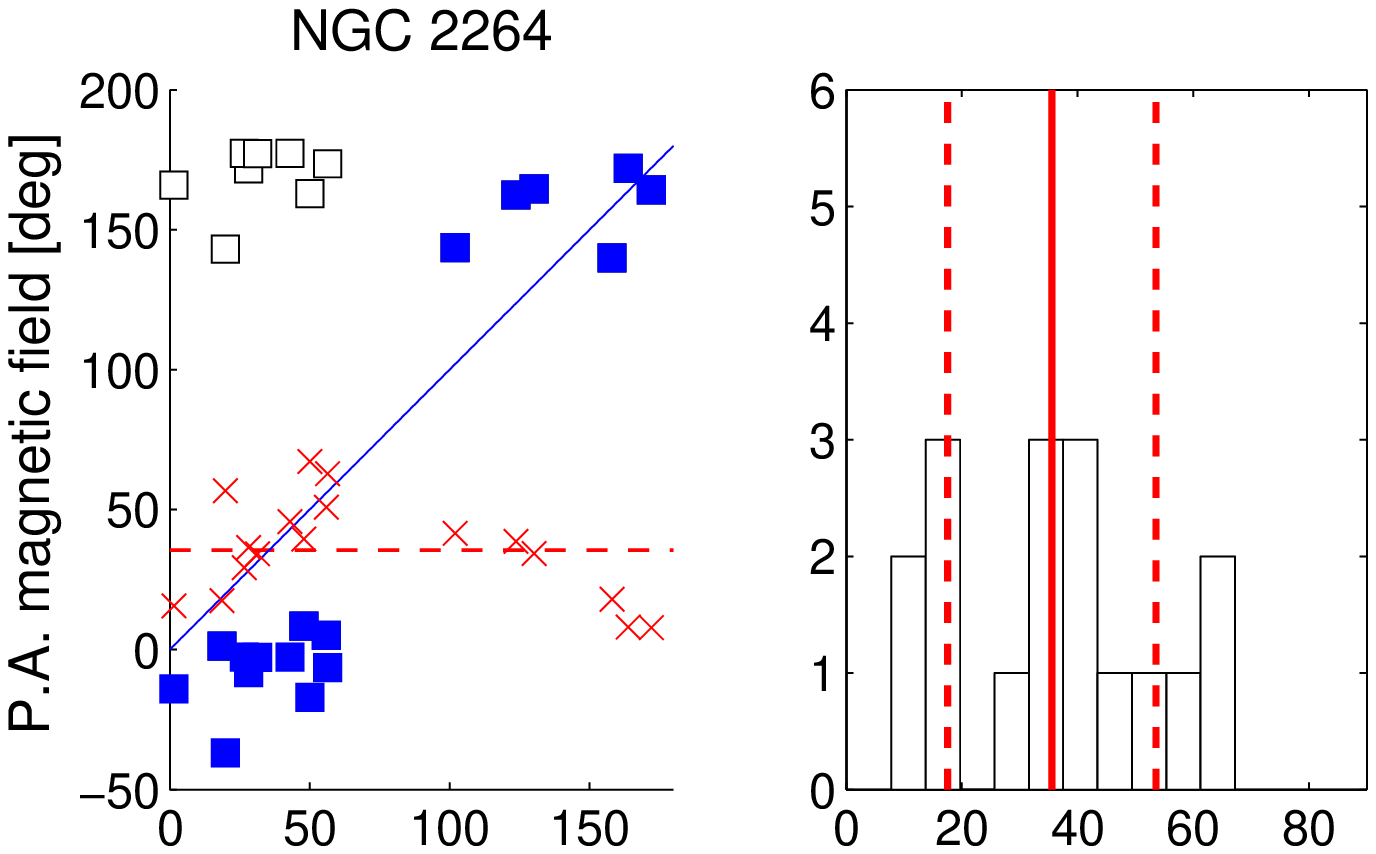}
\includegraphics[scale=0.5]{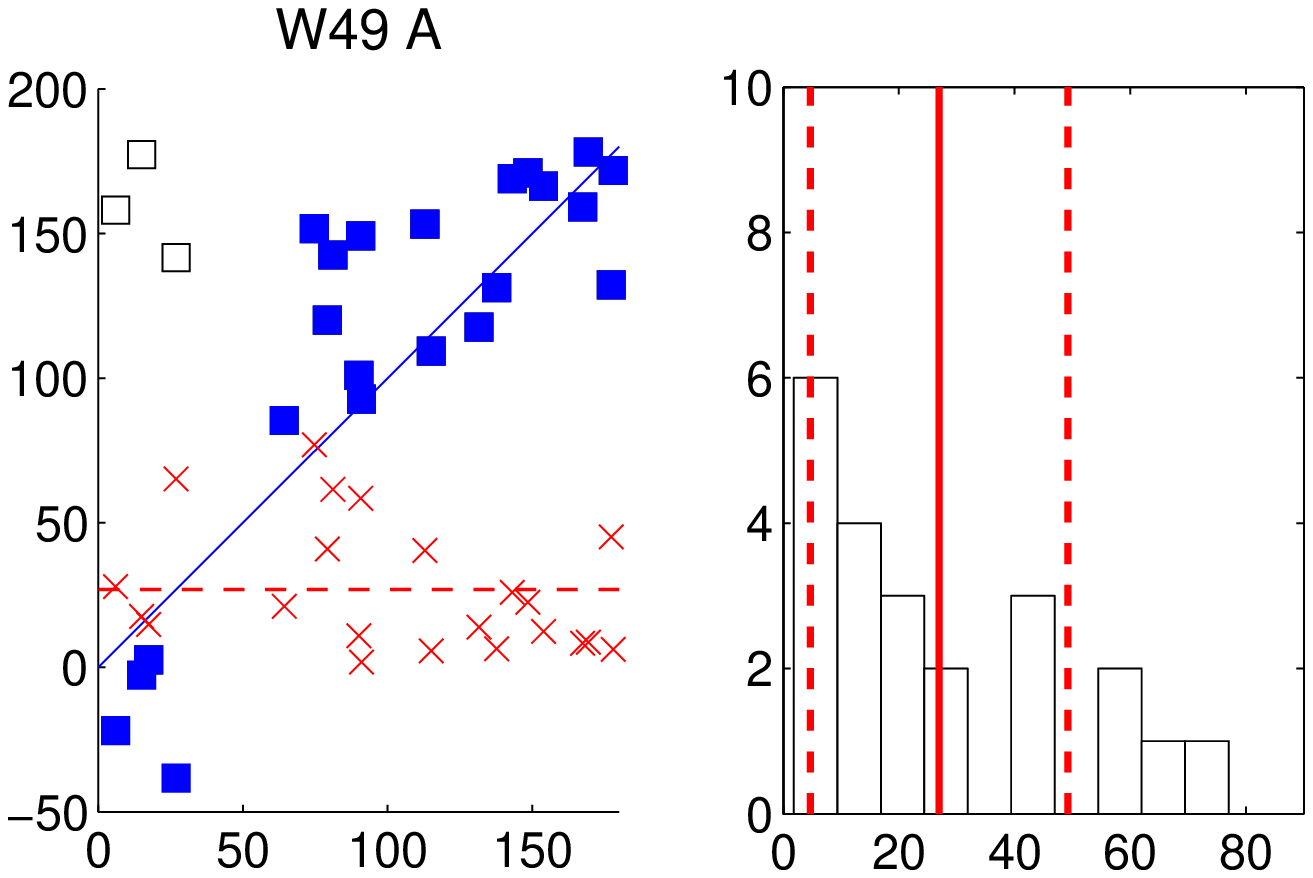}
\includegraphics[scale=0.5]{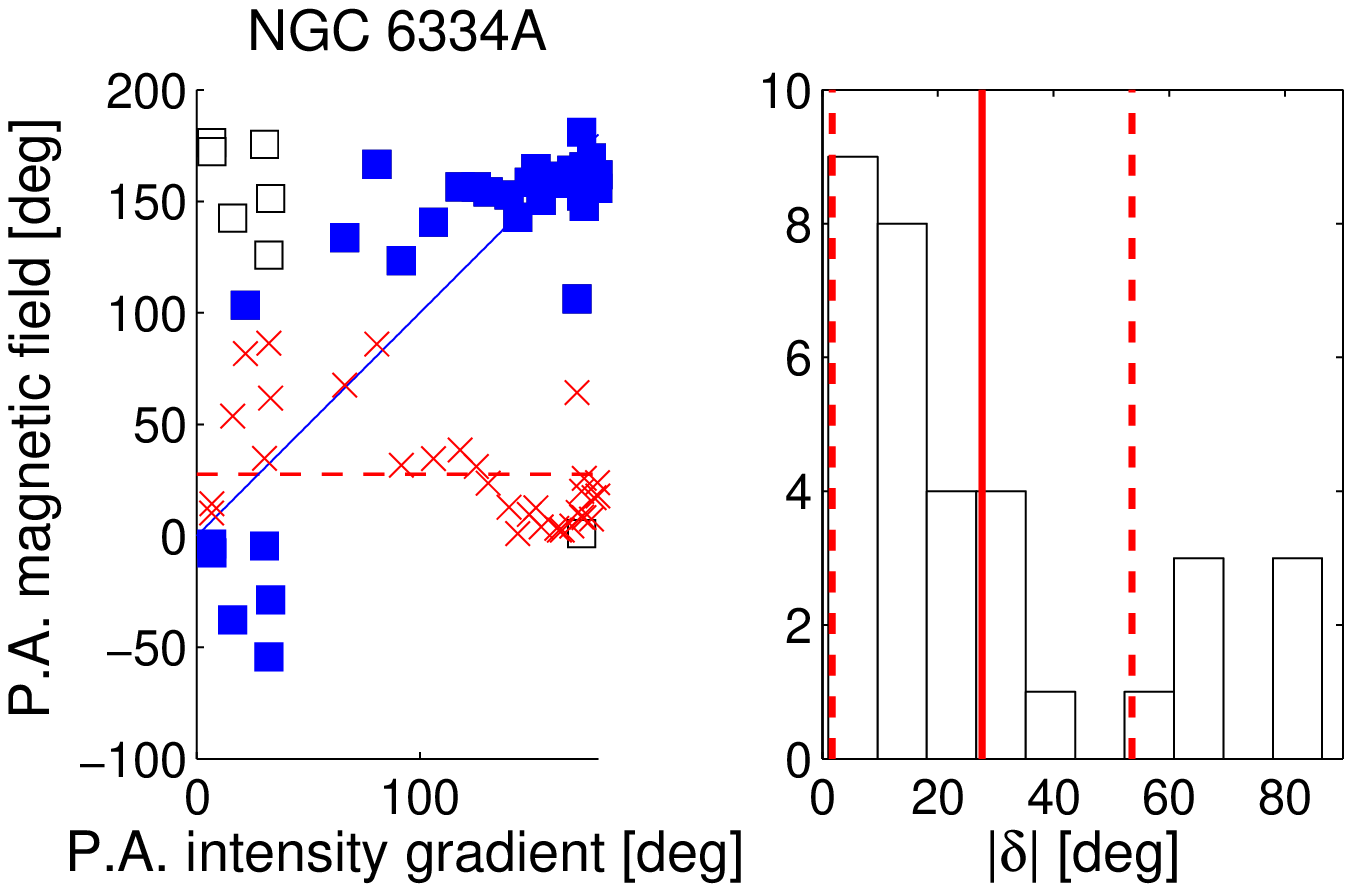}
\includegraphics[scale=0.5]{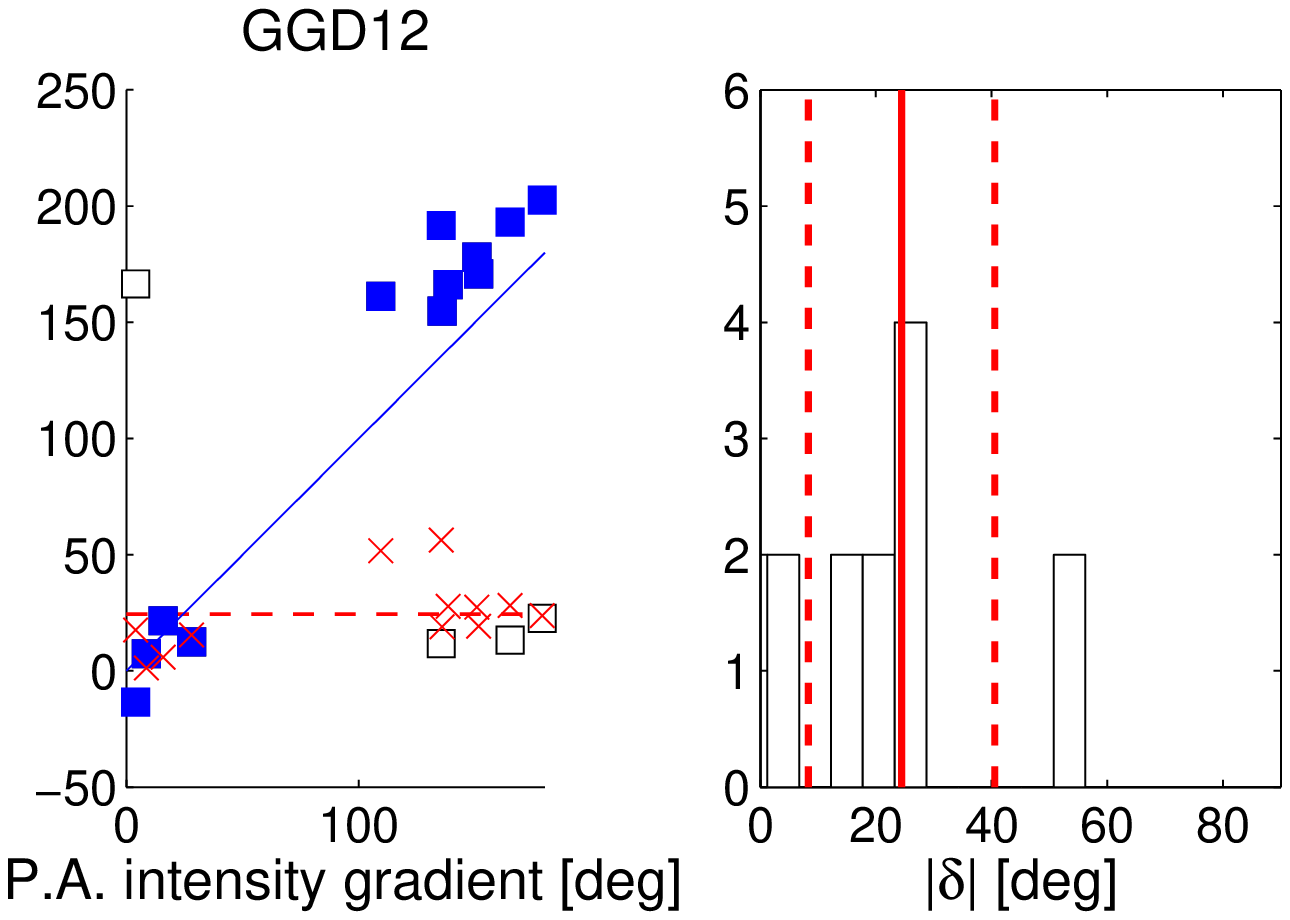}
 \caption{\label{sample_hist}
The correlation between the magnetic field and intensity gradient
orientations together with the histogram for the absolute angle 
$|\delta|$ in between the two orientations for the sources in 
the Figures \ref{sample_summary} and \ref{sample_summary_2}.
}
\end{center}
\end{figure}

\begin{figure}
\begin{center}
\includegraphics[scale=0.5]{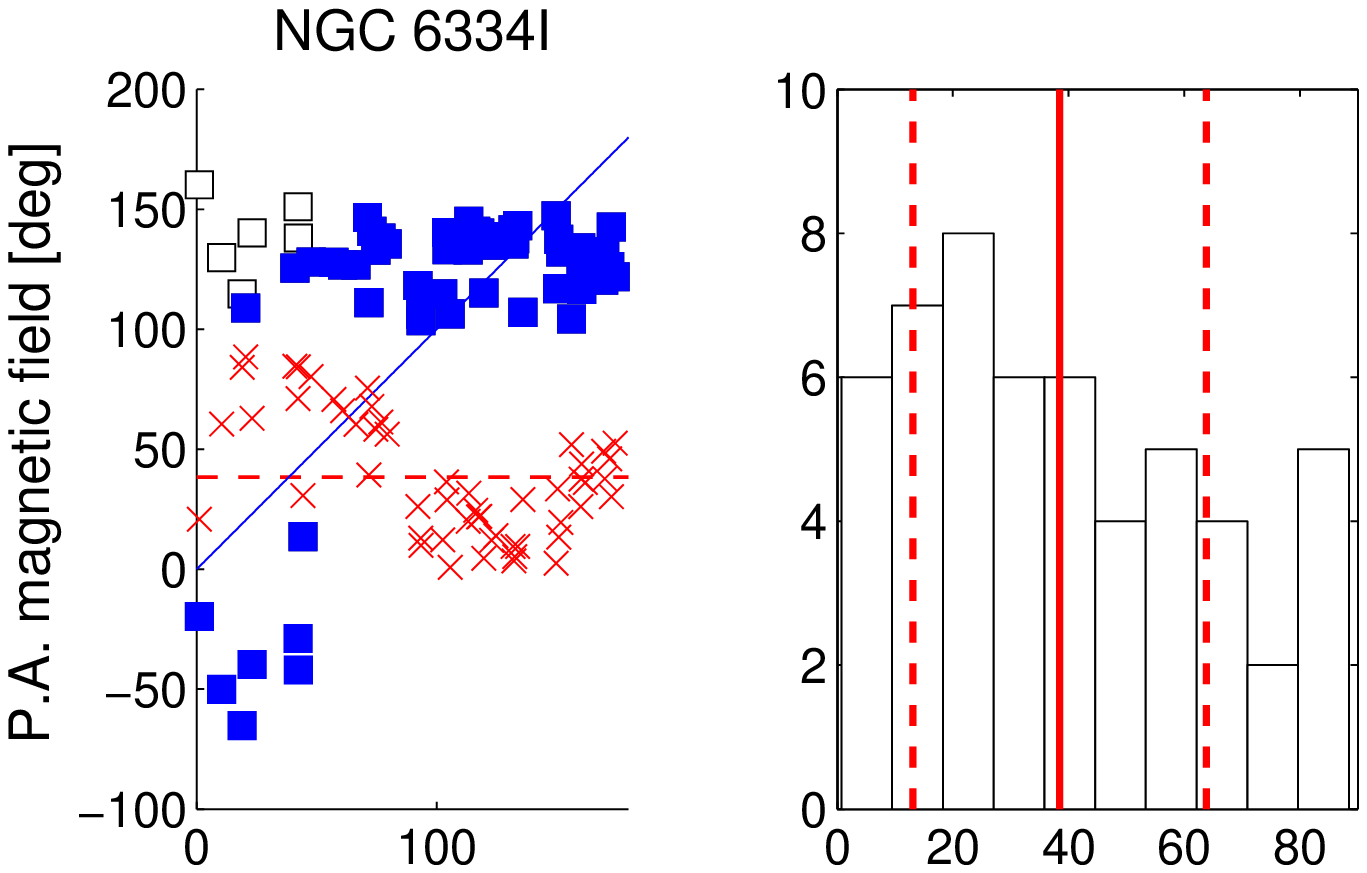}
\includegraphics[scale=0.5]{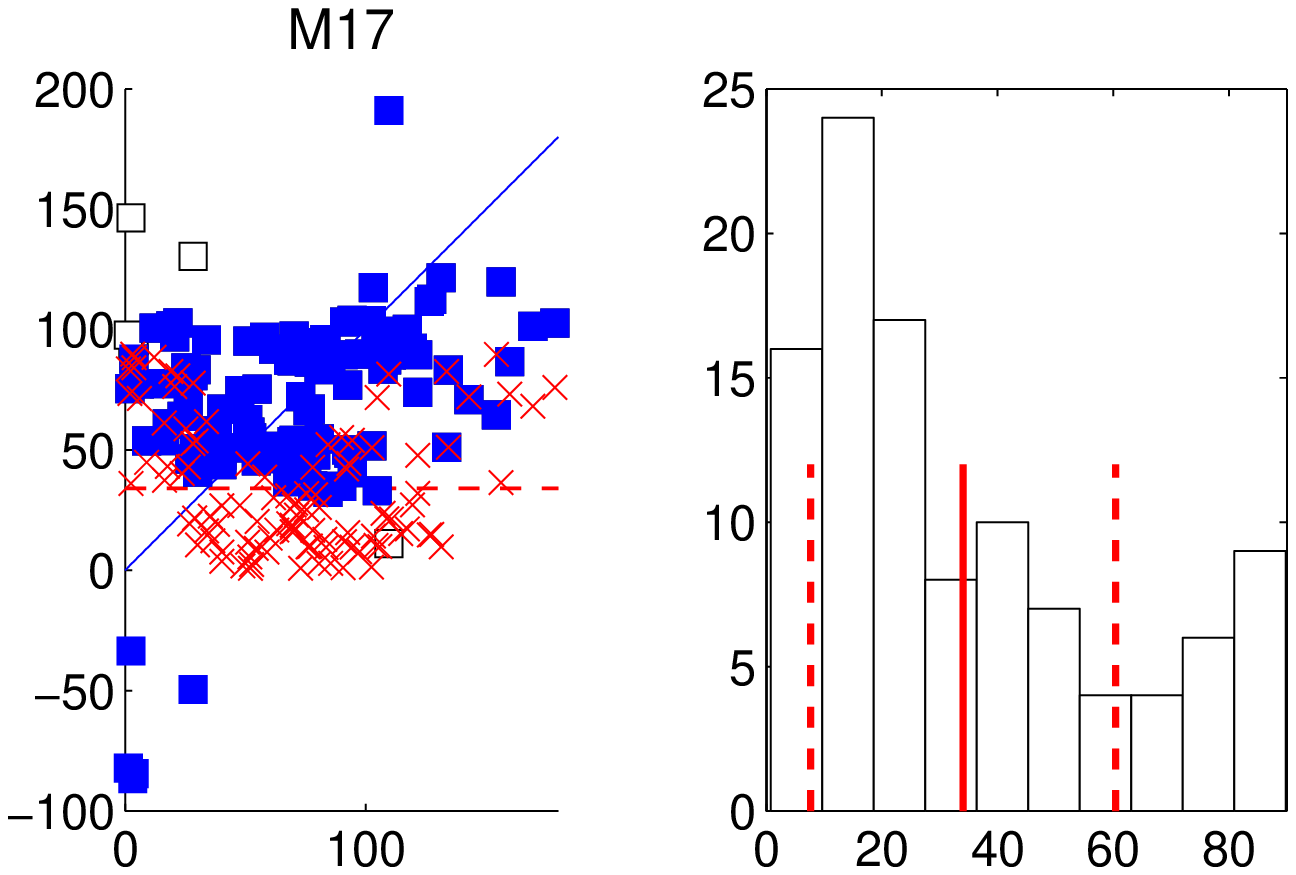}
\includegraphics[scale=0.5]{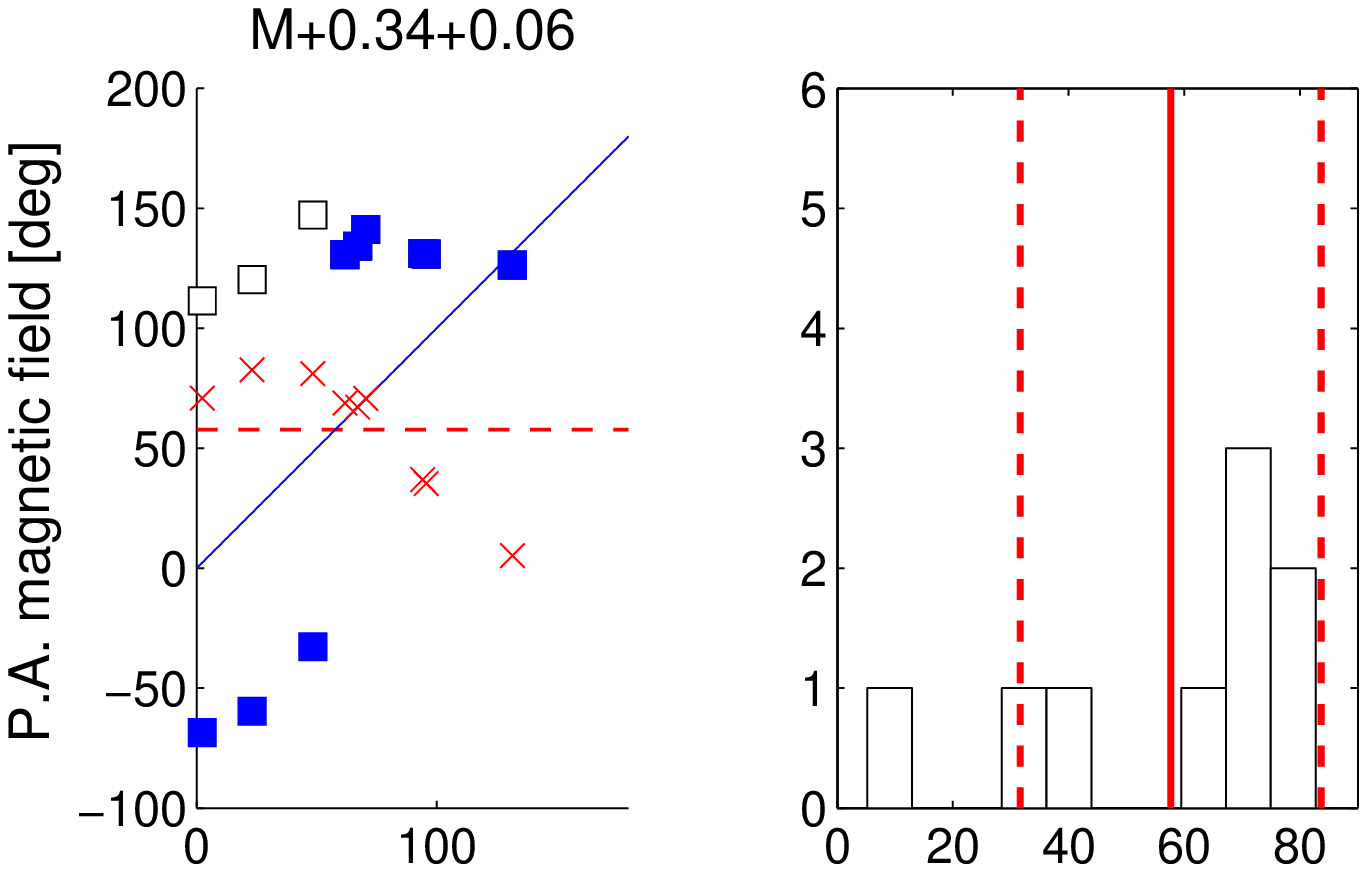}
\includegraphics[scale=0.5]{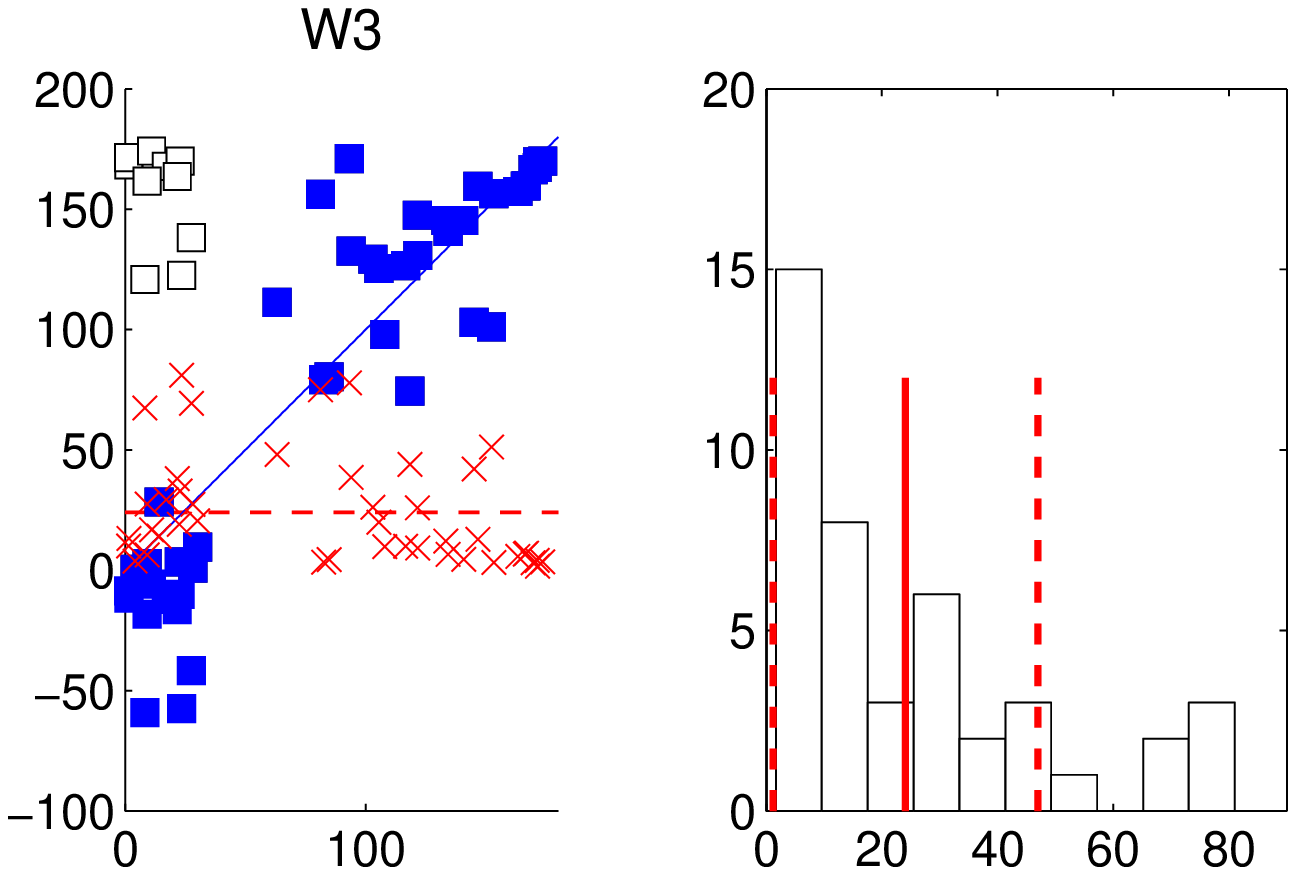}
\includegraphics[scale=0.5]{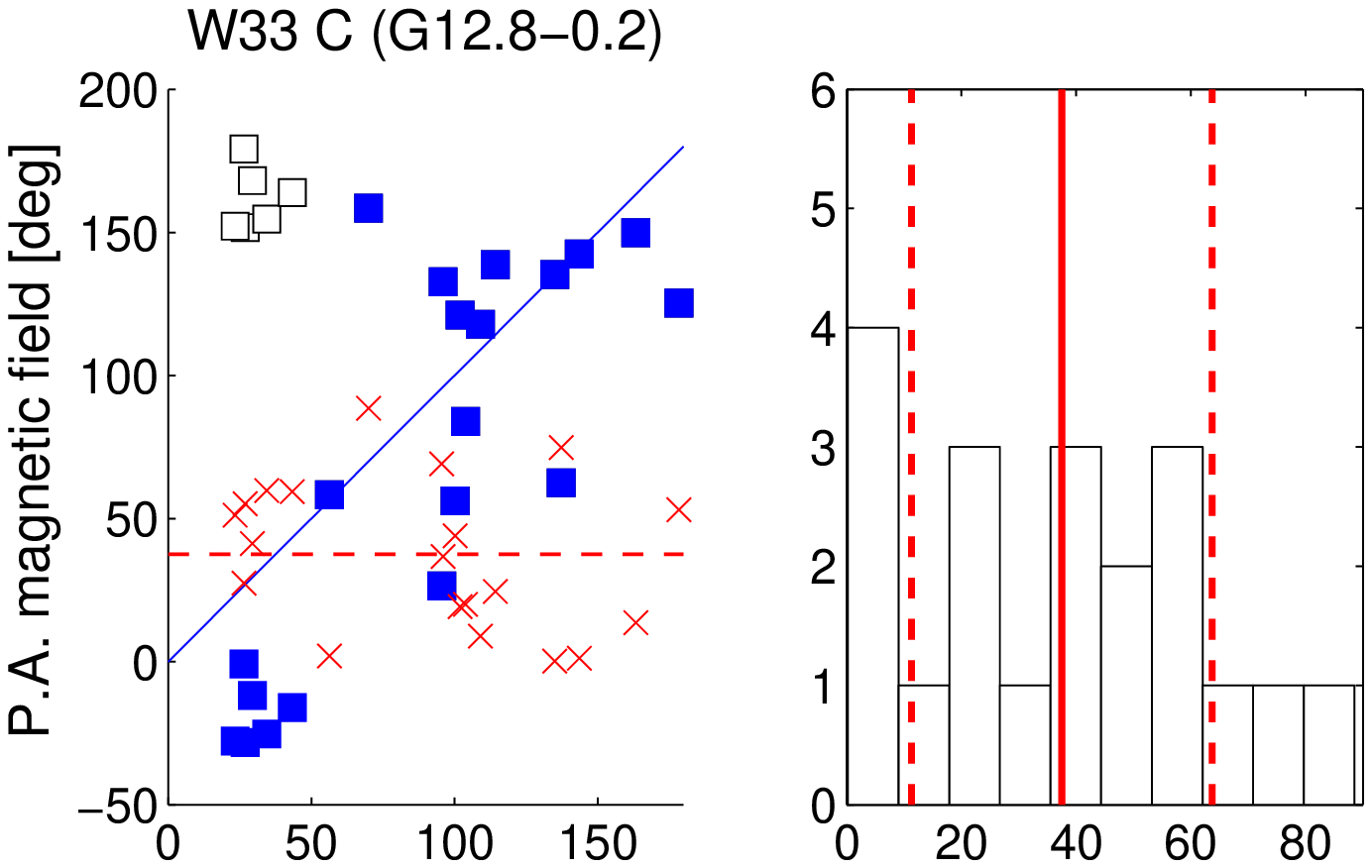}
\includegraphics[scale=0.5]{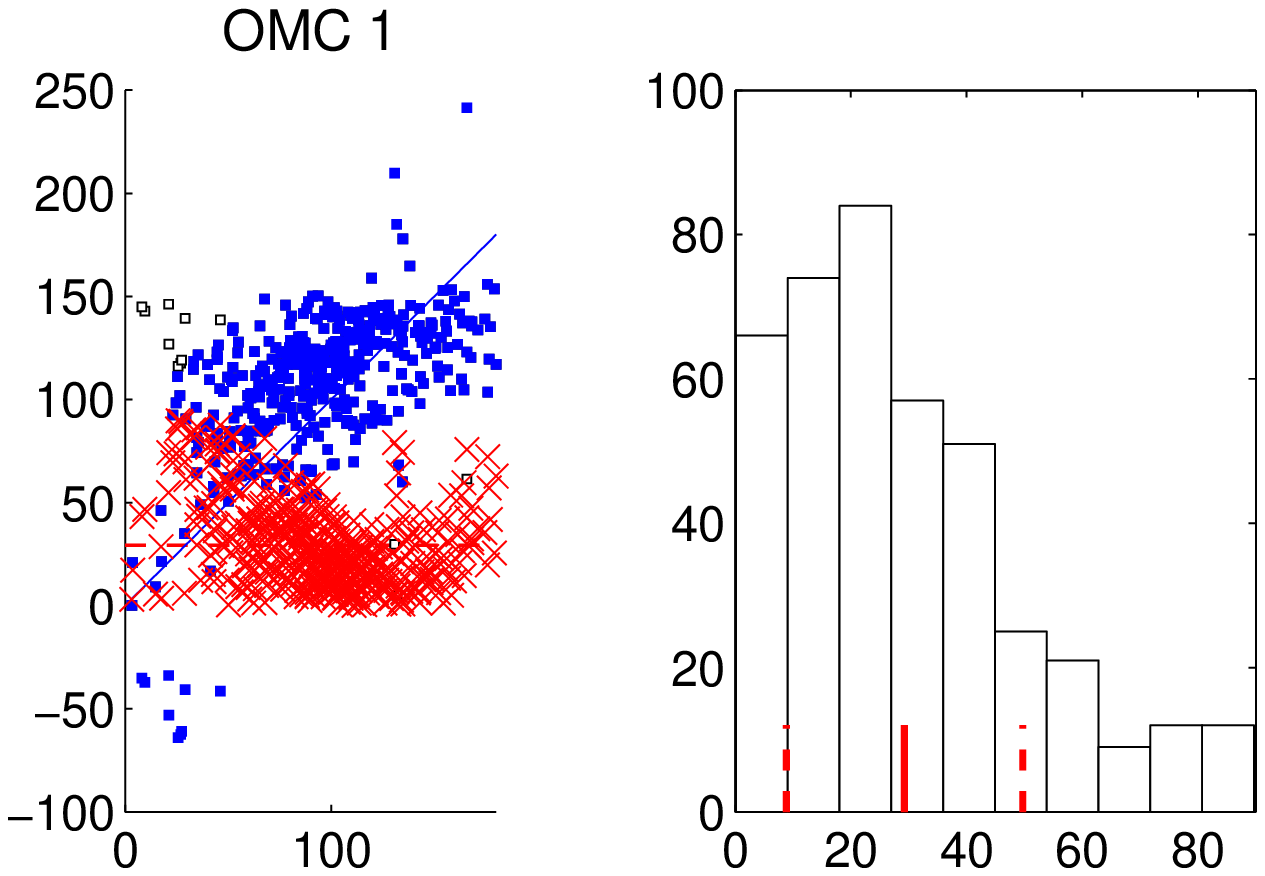}
\includegraphics[scale=0.5]{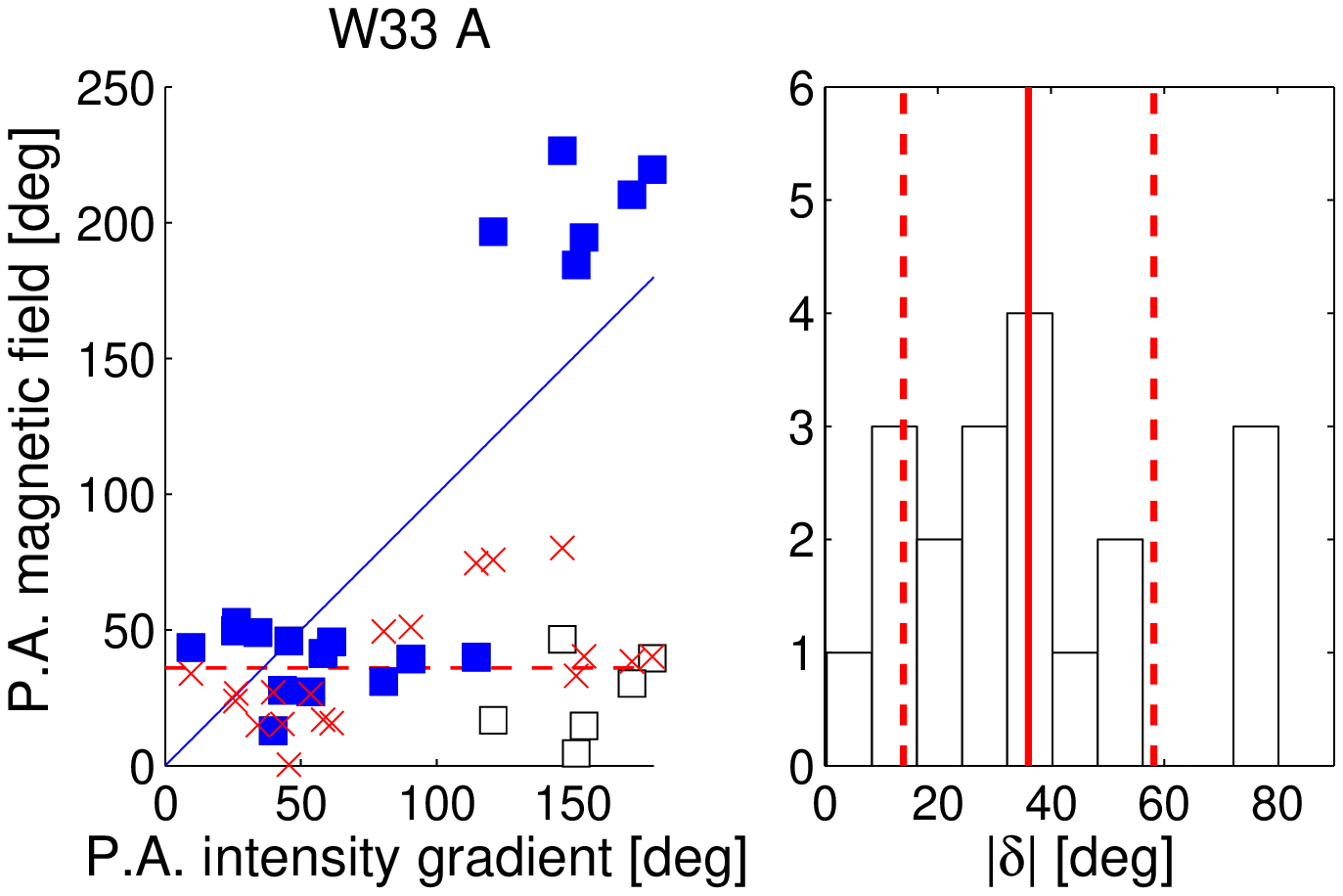}
\includegraphics[scale=0.5]{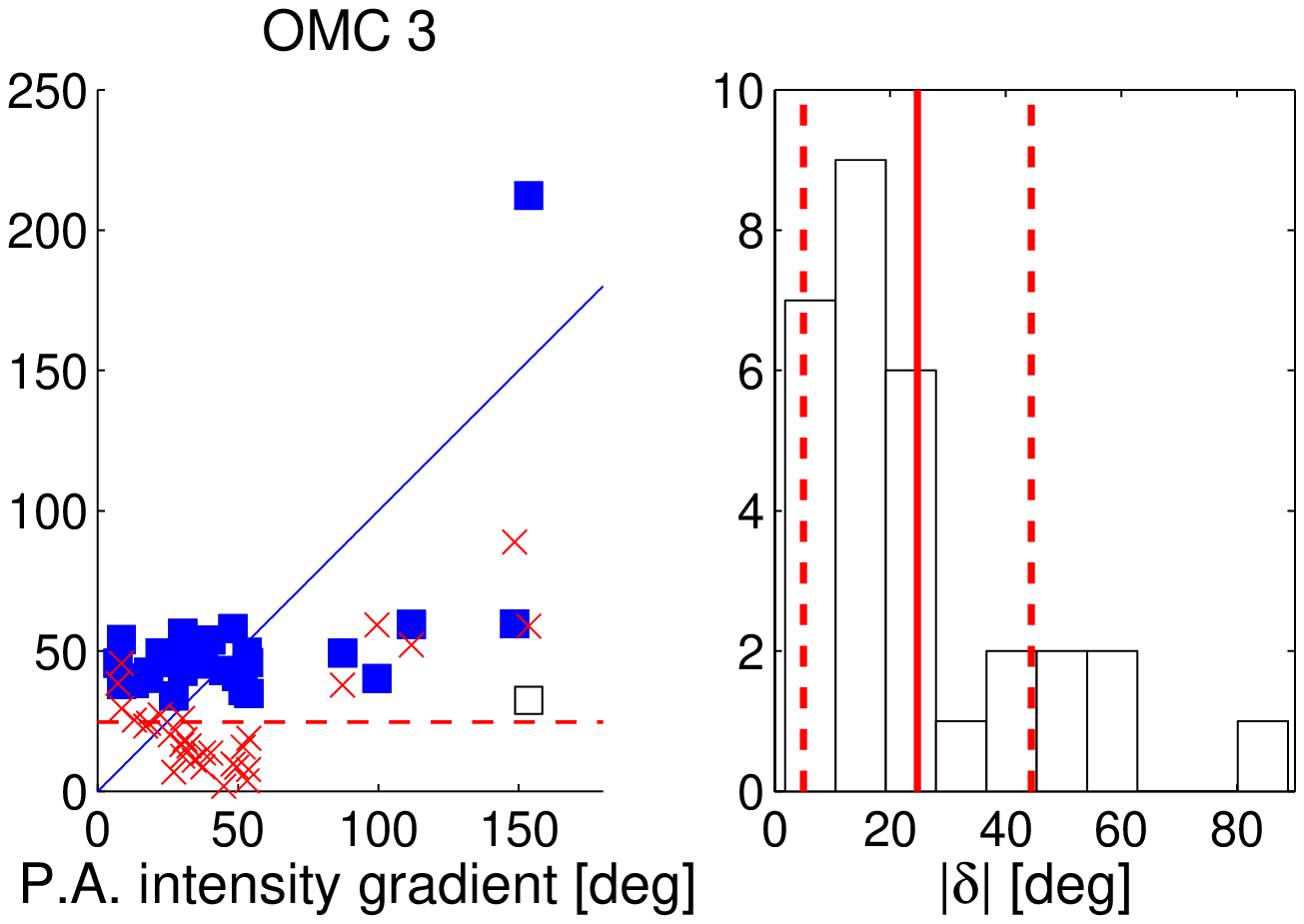}
 \caption{\label{sample_hist_2}
Same as in Figure \ref{sample_hist} for the sources in 
the Figures \ref{sample_summary_3} and \ref{sample_summary_4}.
}
\end{center}
\end{figure}

\begin{figure}
\begin{center}
\includegraphics[scale=0.5]{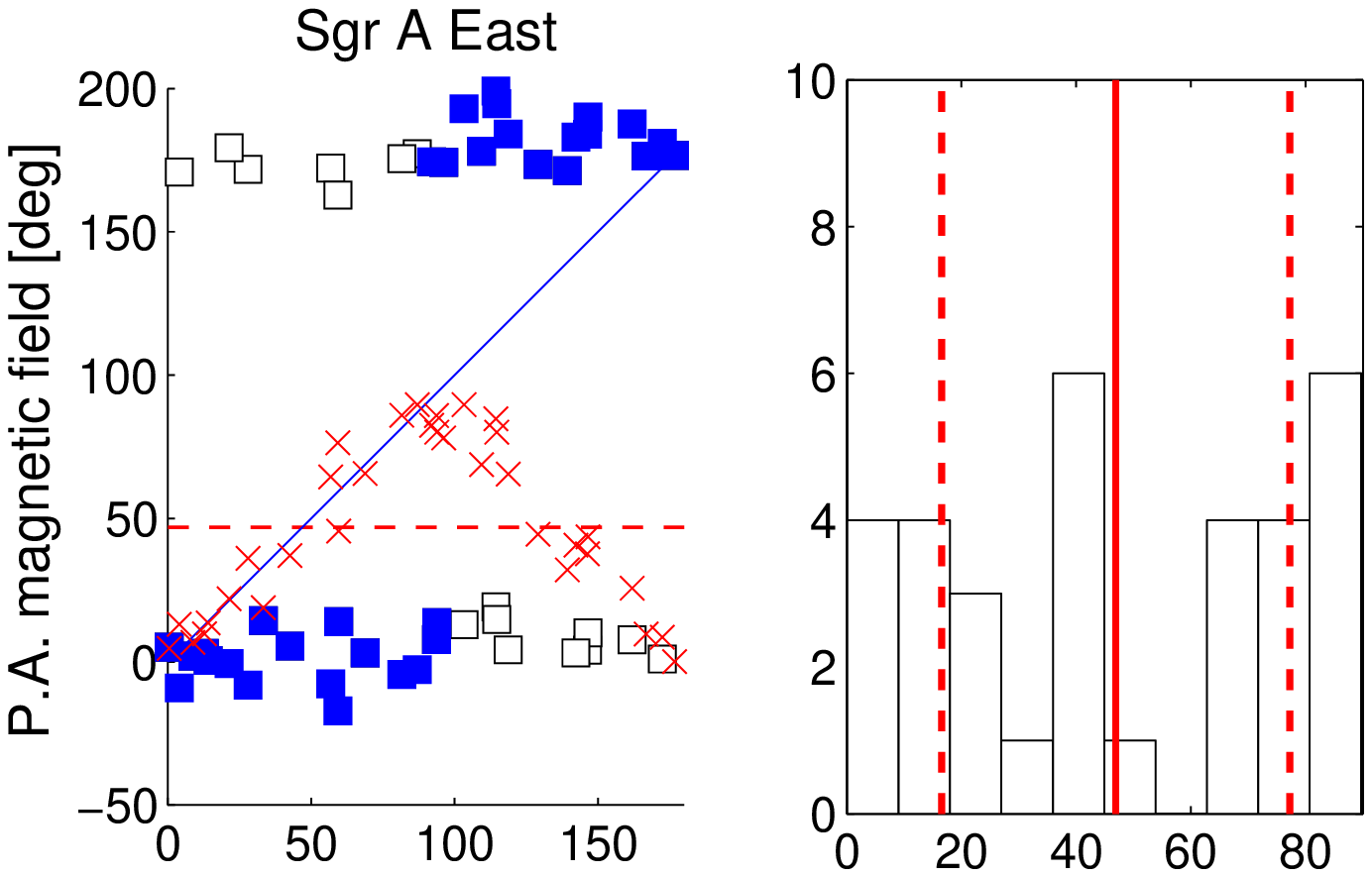}
\includegraphics[scale=0.5]{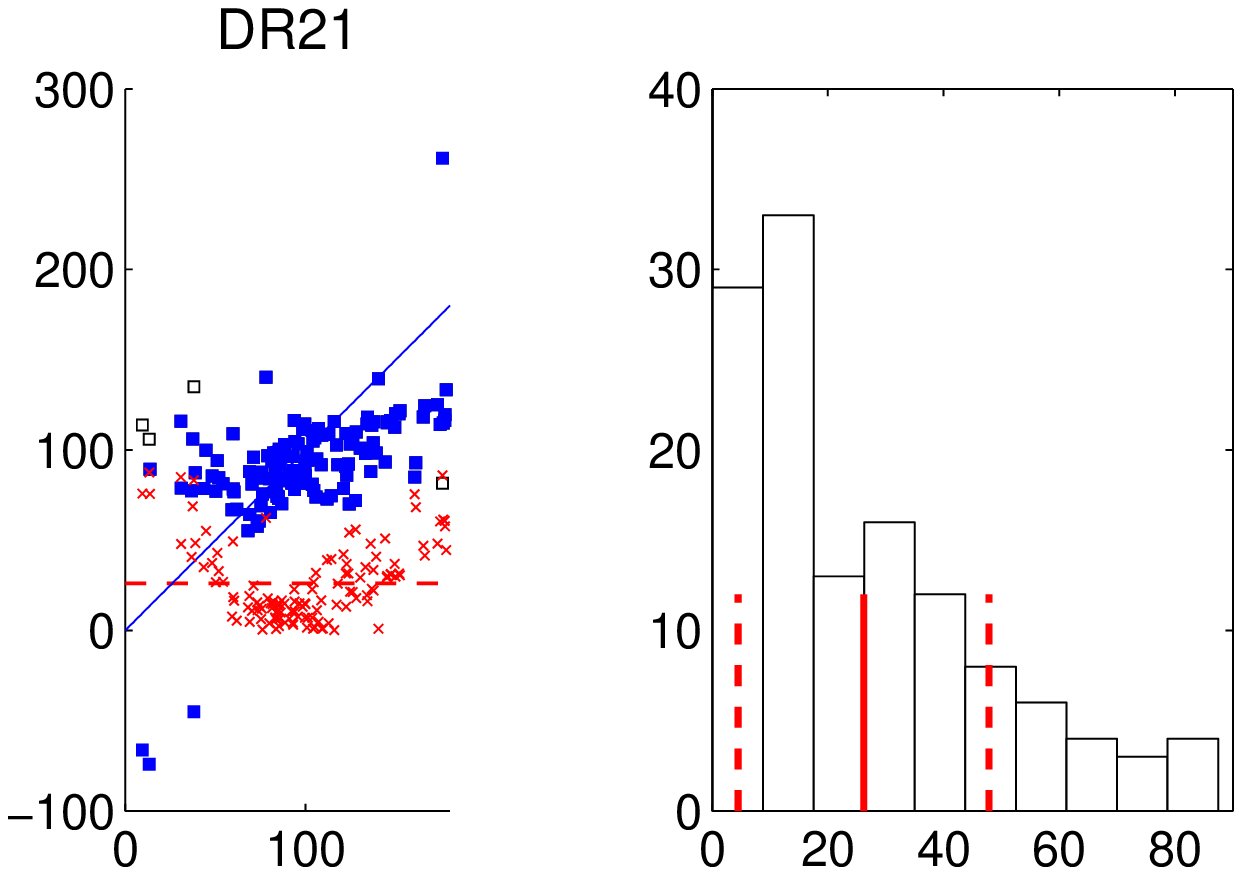}
\includegraphics[scale=0.5]{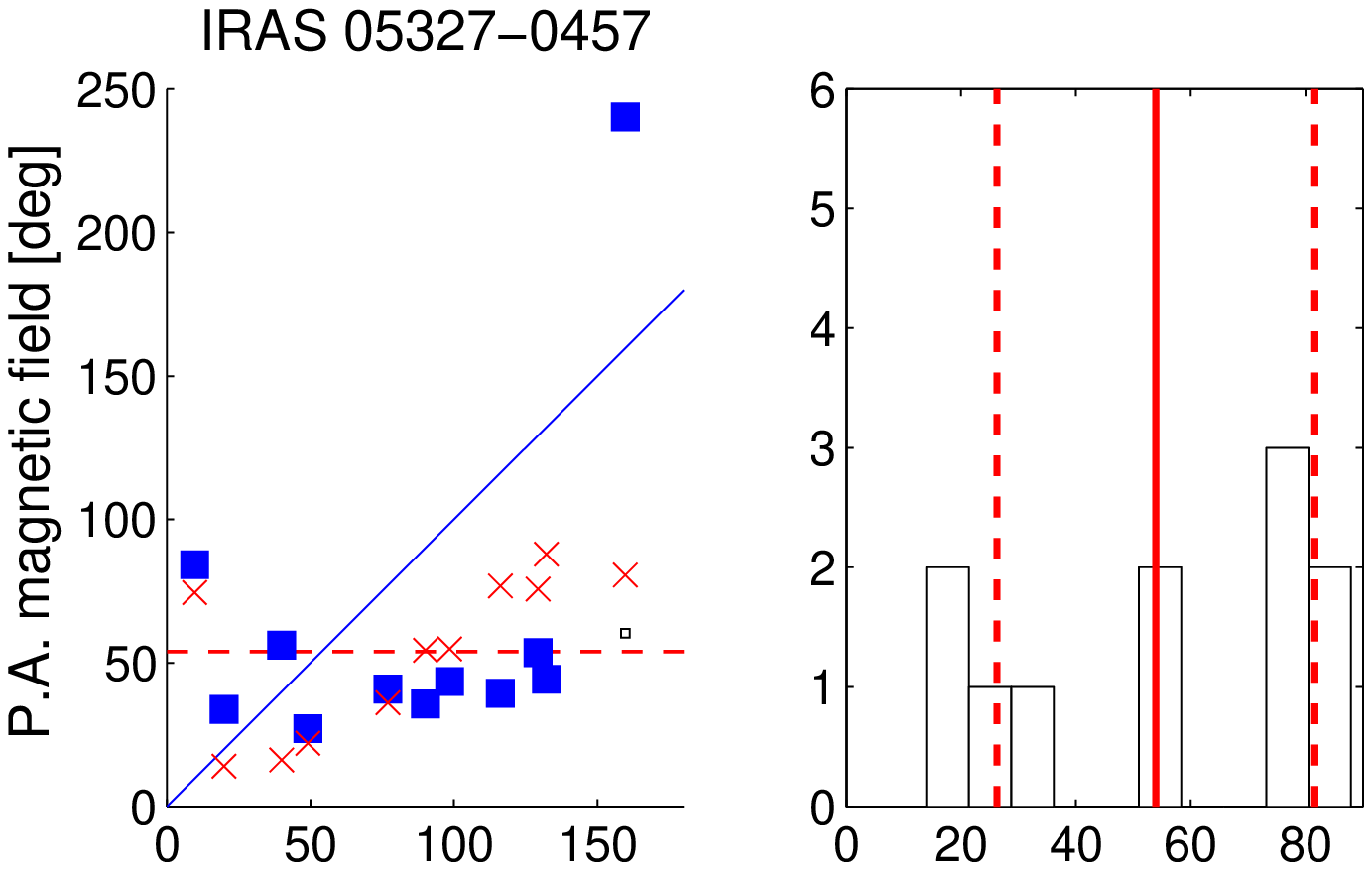}
\includegraphics[scale=0.5]{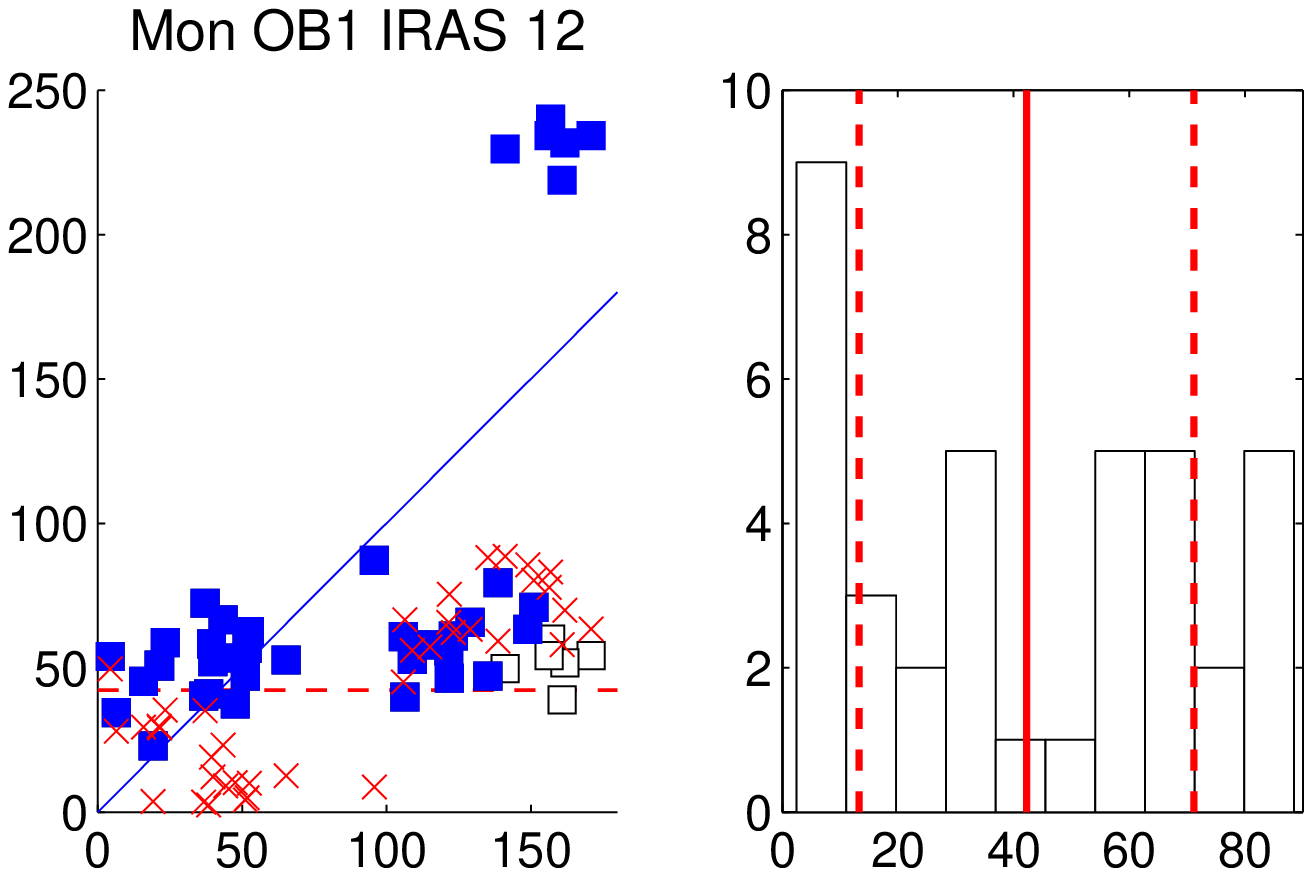}
\includegraphics[scale=0.5]{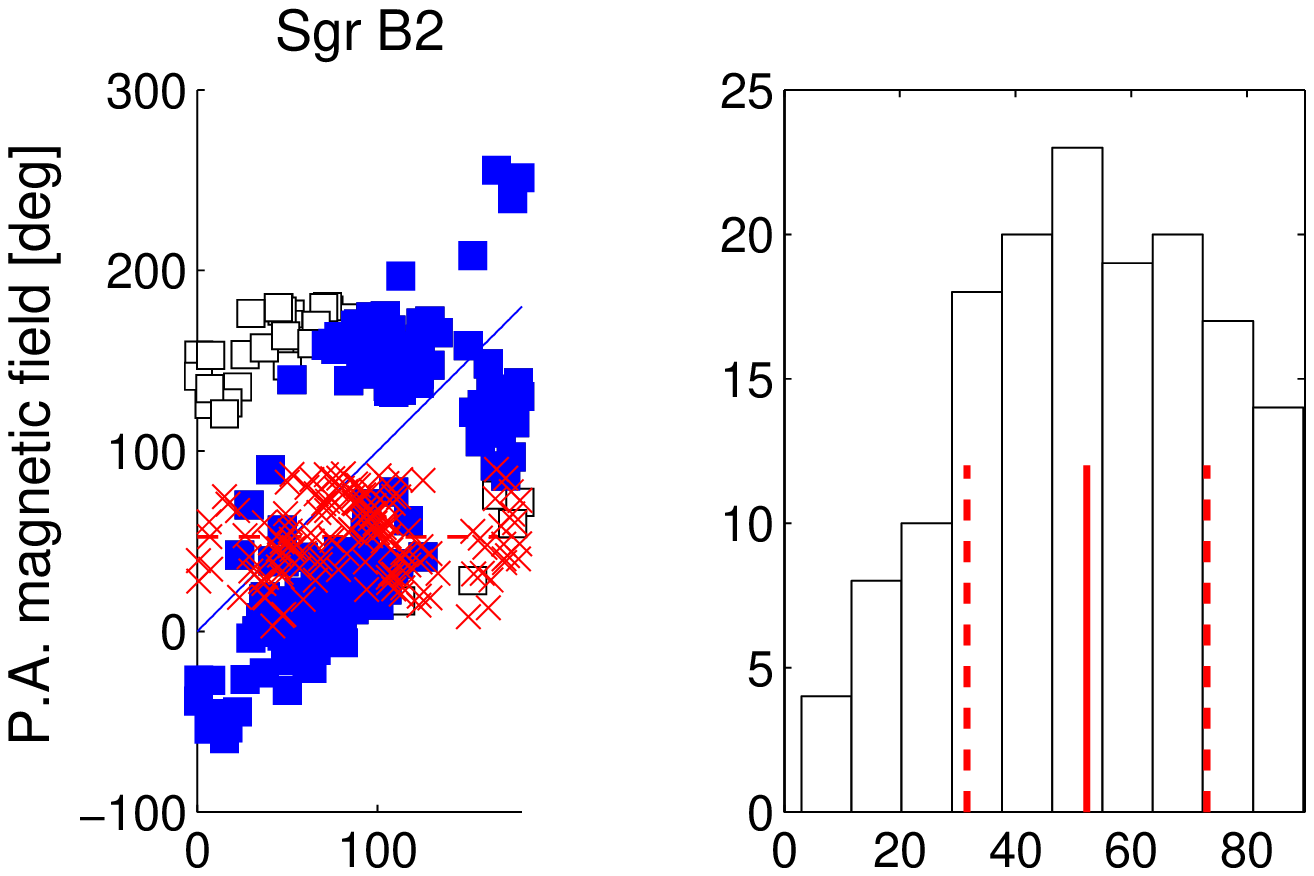}
\includegraphics[scale=0.5]{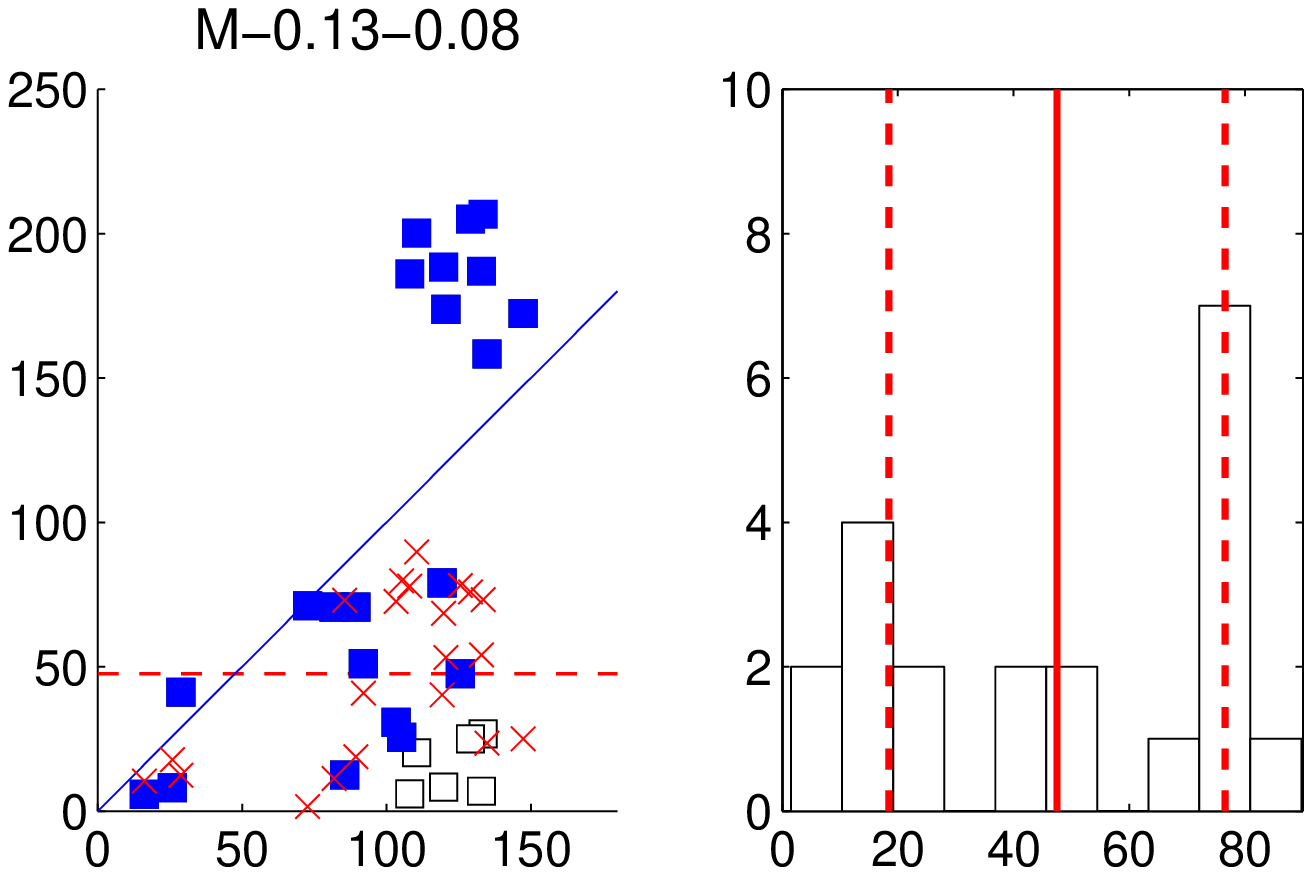}
\includegraphics[scale=0.5]{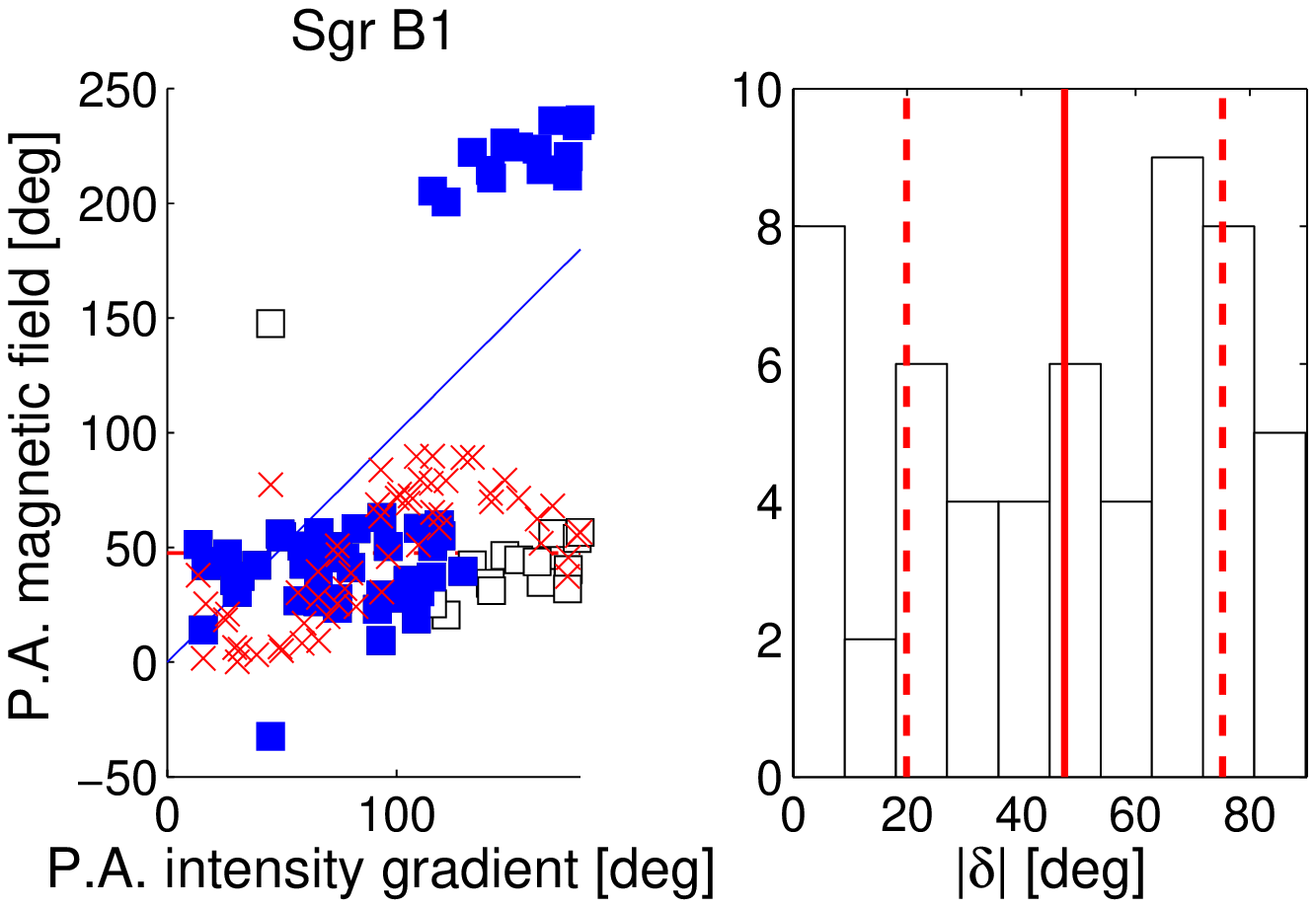}
\includegraphics[scale=0.5]{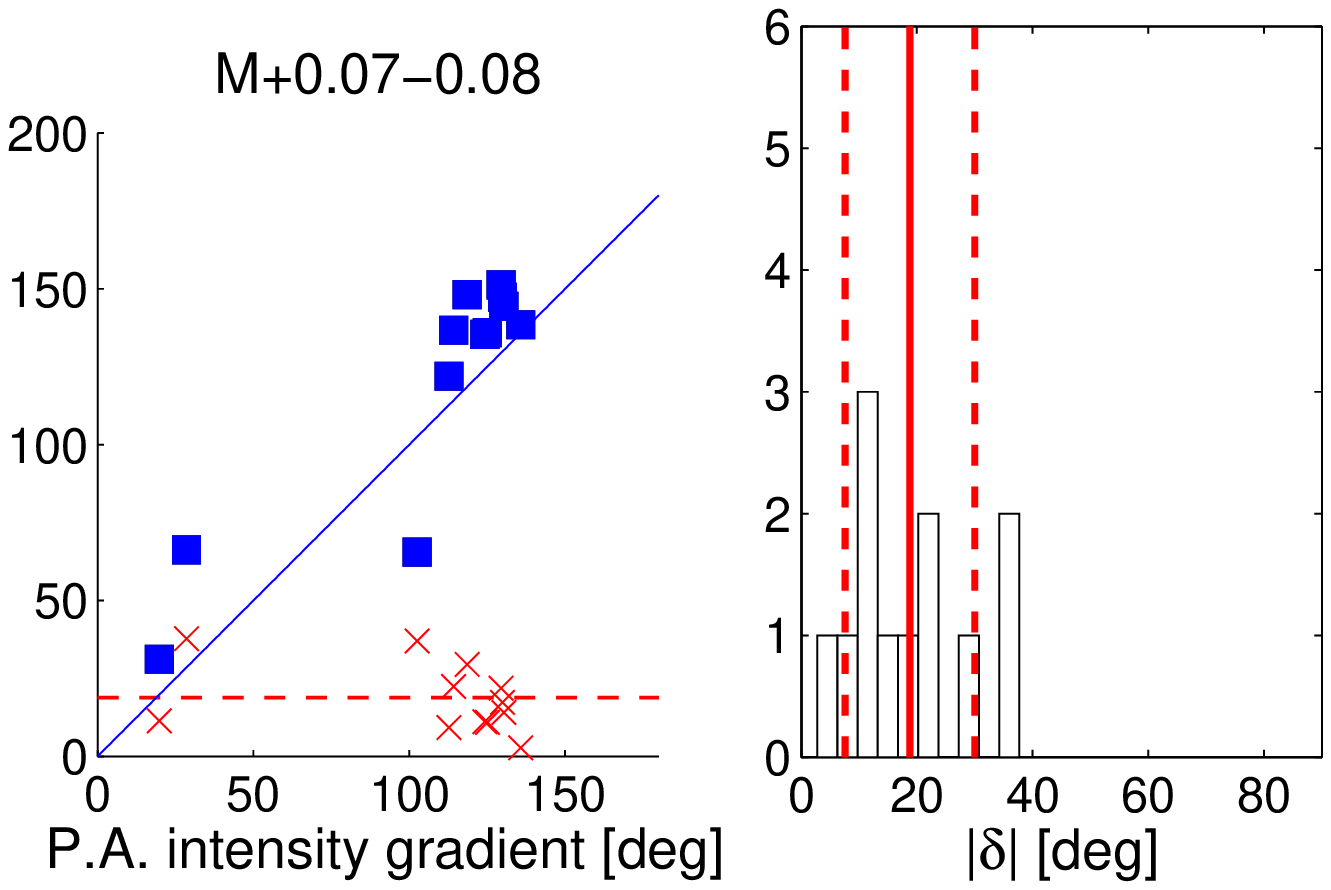}
 \caption{\label{sample_hist_3}
Same as in Figure \ref{sample_hist} for the sources in 
the Figures \ref{sample_summary_5} and \ref{sample_summary_6}.
}
\end{center}
\end{figure}

The Figures \ref{sample_summary} to \ref{sample_summary_6} display 4 panels for each source: magnetic field and intensity 
gradient segments overlaid on Stokes $I$ dust continuum, $|\delta|$-map, $\Sigma_B$-map and the 
$\Sigma_B$ versus $|\delta|$ connection. The Figures \ref{sample_hist} to \ref{sample_hist_3}
show the corresponding magnetic field versus intensity gradient correlations with the 
histograms for $|\delta|$. Basic statistical numbers are summarized in Table \ref{table_quantities}.



\clearpage

\begin{deluxetable}{ccccccc}  
 \tabletypesize{\scriptsize}
\tablewidth{0pt}
\tablecaption{Magnetic Field Quantities
                      \label{table_field_quantities}}
\tablehead{
\multicolumn{1}{c}{Quantity}
             & \multicolumn{1}{c}{Input}   & \multicolumn{1}{c}{Range} & \multicolumn{1}{c}{Projection Effect?}  
             & \multicolumn{1}{c}{Modeling?}   & \multicolumn{1}{c}{Information}
}
\startdata

$\Delta_B$   &   $\delta$     & $[0, 1]$   &   possible   &   none: $\delta$ directly measurable  &
                                              local field efficiency \\
$\Sigma_B$   &   $\alpha$, $\psi$  & $[0, \infty]$   &  none or small  & yes: angle $\psi$        &
                                              relative local field-to-gravity/pressure force\\
$B$   &   $\alpha$, $\psi$, $\rho$, mass   & $[0, \infty]$   &  none or small  & yes: angle $\psi$, $\rho$, mass        &
                                              absolute local field strength\\

\enddata
\tablecomments{Magnetic field quantities in increasing order of complexity and information as discussed 
in Section \ref{summary_parameters}.
}

\end{deluxetable}


\begin{deluxetable}{cccccccccccccccc}  \rotate
 \tabletypesize{\footnotesize}
\tablewidth{0pt}
\tablecaption{Analysis Summary
                      \label{table_quantities}}
\tablehead{
\multicolumn{5}{c}{Observation}
             & \colhead{}
             & \multicolumn{10}{c}{Analysis}\\
  \cline{1-5} \cline{7-15}
\colhead{Source / Region} & \colhead{$d$} & \colhead{$\lambda$}  & \colhead{$\theta$}  & \colhead{$\ell$} &
\colhead{}  & \colhead{$\mathcal{C}$} &
\colhead{$\delta_{max}$} &  \colhead{$\delta_{min}$} &
\colhead{$<|\delta|>$} &  \colhead{$std (|\delta|)$} &
\colhead{$\Sigma_{B,max}$} &  \colhead{$\Sigma_{B,min}$} &
\colhead{$<\Sigma_B>$} &  \colhead{$std (\Sigma_B)$} &
\colhead{Phase} \\
 \colhead{} & \colhead{(kpc)} & \colhead{($\mu$m)} &  \colhead{($\arcsec$)} & \colhead{(mpc)}  & \colhead{} &
\colhead{} & \colhead{(deg)} & \colhead{(deg)}
& \colhead{(deg)} & \colhead{(deg)}  &
 \colhead{} &  \colhead{} &  \colhead{} &  \colhead{} &  \colhead{} \\
 \colhead{} &  \colhead{(1)} &  \colhead{(2)} & \colhead{(3)} & \colhead{(4)} &  \colhead{} &  \colhead{(5)} &  \colhead{(6)} &
 \colhead{(7)} &  \colhead{(8)} &  \colhead{(9)} &
\colhead{(10)} & \colhead{(11)} & \colhead{(12)} & \colhead{(13)} & \colhead{(14)} 
}
\startdata

CO$+0.02$/ M$-0.02$$^a$  &
7.9   &
350   &
20    &
770   &
&
0.71 &
90   &
-90  &
48   &
24   &
16.8 &
0.01 &
1.48 &
3.32 &
I
\\

Mon R2  & 
0.95 &
350   &
20    &
92    &
&
0.78  &
67   &
-55  &
23  &
18   &
1.56 &
0.09 &
0.57 &
0.35 &
IIA

\\

M$+0.25+0.01$$^a$ &
7.9   &
350   &
20    &
770   &
&
0.61  &
88  &
-89 &
57  &
24   &
14.5 &
0.02 &
2.91 &
3.56 &
IIB
\\

NGC 2068 LBS10 &
0.4 &
350 &
20 &
39 &
&
0.62 &
77 &
-84 &
51 &
24 &
5.02 &
0.002 &
1.38 &
1.42 &
IIB

\\

NGC 2024 &
0.4 &
350  &
20  &
39 &
 &
0.55 &
85  &
-85 &
37 &
24 &
7.14 &
0.02 &
1.03 &
1.17 &
IIA

\\

NGC 2264 &
0.8 &
350 &
20 &
78 &
 &
0.94 &
42 &
-67 &
36 &
18 &
1.66 &
0.04 &
0.63 &
0.51 &
IIB

\\

NGC 6334A &
1.7 &
350 &
20 &
170 &
 &
0.84 &
86 &
-86 &
28 &
26 &
2.44 &
0.02 &
0.48 &
0.50 &
IIA
\\

G34.3$+0.2$$^a$ &
3.7 &
350 &
20 &
360 &
 &
0.87 &
89 &
-85 &
35 &
24 &
2.38 &
0.01 &
1.30 &
0.30 &
IIB 
 
\\

$\rho$ Oph &
0.139 &
350 &
20 &
14 &
 &
0.66 &
88 &
-78 &
27 &
24 &
5.90 &
0.09 &
0.93 &
1.00 &
IIA

\\

W49 A&
11.4 &
350 &
20 &
1110 &
&
0.55 &
77 &
-65 &
27 &
22 &
1.01 &
0.02 &
0.31 &
0.30 &
IIA

\\

GGD12&
1.7 &
350 &
20 &
170 &
&
0.98 &
56 &
-18 &
24 &
16 &
1.20 &
0.03 &
0.39 &
0.32 &
IIA 

\\

NGC 6334I&
1.7 &
350 &
20 &
170 &
&
0.63 &
88 &
-84 &
38 &
25 &
5.76 &
0.01 &
0.96 &
0.91 &
IIA
\\

M$+0.34+0.06$&
8 &
350 &
20 &
780 &
&
0.81 &
71 &
-83 &
58 &
26 &
6.19 &
0.55 &
1.96 &
1.81 &
IIB
\\

W33 C&
2.4 &
350 &
20 &
230 &
&
0.80 &
89 &
-75 &
37 &
26 &
6.52 &
0.05 &
0.80 &
1.48 &
IIA / IIB
\\

W33 A&
2.4 &
350 &
20 &
230 &
&
0.87 &
80 &
-75 &
36 &
22 &
2.57 &
0.07 &
0.66 &
0.74 &
IIA
\\

M 17$^a$ &
1.6 &
350 &
20 &
160 &
&
0.44 &
88 &
-90 &
34 &
26 &
4.46 &
0.02 &
0.78 &
1.05 &
IIA
\\

W3 &
1.95 &
350 &
20 &
190 &
&
0.91 &
78 &
-81 &
24 &
23 &
5.83 &
0.04 &
0.70 &
0.96 &
IIA
\\

OMC 1$^a$&
0.414 &
350 &
20 &
40 &
&
0.59 &
86 &
-90 &
29 &
20 &
18.0 &
0.004&
0.87 &
2.50 &
IIA
\\

OMC 3$^a$&
0.414 &
350 &
20&
40 &
&
0.59 &
59 &
-89 &
25 &
20 &
1.61 &
0.02 &
0.88 &
0.65 &
IIA
\\

Sgr A$^{\ast}$ East$^a$ &
8 &
350 &
20 &
780 &
 &
0.83 &
90 &
-90 &
47 &
30 &
8.00 &
0.13 &
2.20 &
3.00  &
I \\

IRAS 05327&
0.414 &
350 &
20 &
40 &
&
0.43 &
81 &
-88 &
54 &
28 &
12.0 &
0.39 &
2.13 &
3.39 &
IIB
\\

Sgr B2$^a$&
8 &
350 &
20 &
780 &
&
0.70 &
90 &
-87 &
52 &
21 &
15.0 &
0.002 &
1.45 &
1.80 &
IIB
\\

Sgr B1$^a$&
8 &
350 &
20 &
780 &
&
0.76 &
90 &
-90 &
48 &
28 &
9.30 &
0.03 &
1.43 &
1.75 &
I / IIA
\\

DR21 &
3 &
350 &
20 &
290 &
&
0.58 &
86 &
-87 &
26 &
22 &
4.38 &
0.00&
0.75 &
0.59 &
IIA
\\

Mon OB1 IRAS 12$^a$&
0.80 &
350 &
20 &
78 &
&
0.65 &
88 &
-88 &
42 &
29 &
6.13 &
0.005 &
1.10 &
2.39 &
IIB
\\

M$-0.13-0.08$$^a$&
8 &
350 &
20 &
780 &
&
0.71 &
90 &
-80 &
47 &
29 &
3.70 &
0.02 &
1.13 &
1.60 &
IIB 
\\

M$+0.07-0.08$&
8 &
350 &
20 &
780 &
&
0.89 &
37 &
-37 &
19 &
11 &
0.54 &
0.002 &
0.24 &
0.20 &
IIA
\\

W51 e2     & 
7   &
870       &
0.7        &
24        &
           &
0.95       & 
34         &
-48        & 
18      & 
12    &
1.28 &
0.002 &
0.22 &
0.30 &
III
        
 \\

W51 North   &
7       &
870       &
0.7        &
24        &
           &
0.89       & 
82         &
-81        & 
34       & 
23   &
4.59 &
0.07 &
0.86 &
0.78 &
III
       
     \\

Orion BN/KL$^a$   &
0.48   &
870       &
2.8        &
6.4      &
           &
0.70       & 
89         &
-90        & 
44      & 
25    &
5.28 &
0.02 &
1.18 &
1.30 &
III

 \\

g5.89$^a$   &
2          &
870       &
2.4        &
23      &
           &
0.72       & 
76         &
-84        & 
35      & 
26  &
3.03 &
0.07 &
0.79 &
1.20 &
 -         
  \\     

\enddata
\tablecomments{Statistical quantities are based on 
the Figures \ref{figure_delta_scales}, \ref{figure_delta_scales_cont} and \ref{figure_delta_cores}, and Figures
\ref{sample_summary} to \ref{sample_hist_3}. For sources with an upper index {\it 'a'}, large outliers  
are removed in order to calculate $\Sigma_{B,max}$, $\langle \Sigma_B \rangle$ and 
$std(\Sigma_B)$.
}
\tablenotetext{(1)}{Source distance. Values are from \citet{genzel81} for W51 and Orion BN/KL, \citet{acord98}
                    for g5.89, \citet{racine70} for Mon R2 and from \citet{reid09} for CO$+0.02-0.02$ / M$-0.02-0.07$
                    and M$+0.25+0.01$.
For the sources in Appendix B: NGC 2068 LBS10 and NGC 2024 \citep{anthony82},
                               NGC 2264 \citep{park02},
                               NGC 6334A and 6334I \citep{russeil12},
			       G34.3$+0.2$ \citep{kuchar94},
			       $\rho$ Oph \citep{mamajek08},
			       W49 A \citep{gwinn92},
			       GGD12 \citep{rodriguez80},
			       sources around the galactic center, i.e., M$+0.34+0.06$, Sgr A$^{\ast}$ East,
                                 Sgr B2, Sgr B1, M$-0.13-0.08$, M$+0.07-0.08$ \citep{genzel00},
			       W33 C and W33 A \citep{immer13},			      
			       M17 \citep{povich07},
			       W3 \citep{xu06},
			       OMC 1, OMC 3 and IRAS 05327 \citep{menten07},			      			      
			       DR21 \citep{campbell82},
			       Mon OB1 IRAS 12 \citep{walker56}.
			     }
\tablenotetext{(2)}{Observing wavelength.}
\tablenotetext{(3)}{Beam resolution. For elliptical beams (synthesized beams of the SMA) the geometrical mean is adopted.
                     For Hertz/CSO, the nominal beam size of $\sim$ $20\arcsec$ is listed.}  
\tablenotetext{(4)}{Physical size scales at the source distances for the resolutions $\theta$.}
\tablenotetext{(5)}{Correlation coefficient in the definition of Pearson's linear correlation coefficient
                    between magnetic field $P.A.$s and the intensity gradient $P.A.$s.}
\tablenotetext{(6)}{Maximum difference between magnetic field and intensity gradient orientation.}
\tablenotetext{(7)}{Minimum difference between magnetic field and intensity gradient orientation.}
\tablenotetext{(8)}{Mean absolute difference between magnetic field and intensity gradient orientations.}
\tablenotetext{(9)}{Standard deviation of absolute differences.}
\tablenotetext{(10)}{Maximum magnetic field significance.}
\tablenotetext{(11)}{Minimum magnetic field significance.}
\tablenotetext{(12)}{Mean magnetic field significance.}
\tablenotetext{(13)}{Standard deviation of magnetic field significance.}
\tablenotetext{(14)}{Assigned phase (I, IIA, IIB, III) according to the schematic scenario in Figure \ref{figure_schematic_delta}.
                     No phase is assigned to g5.89 because this source is probably in a later more evolved stage with expanding 
                     HII regions \citep{tang09a}.}
\end{deluxetable}

\begin{deluxetable}{ccccc}  
\tabletypesize{\footnotesize}
\tablewidth{0pt}
\tablecaption{Evolutionary Sequence and Magnetic Field Features
                      \label{table_evolution}}
\tablehead{
\multicolumn{1}{c}{Scale}
             & \multicolumn{1}{c}{Structure}   & \multicolumn{1}{c}{$|\delta|$-patterns} & \multicolumn{1}{c}{$B$}
& \multicolumn{1}{c}{Examples}  \\
}
\startdata

large (I) & 
irregular & 
irregular & 
variable: minor, balanced, dominant &
CO$+0.02-0.02$ / M$-0.02-0.07$ \\

medium (IIA, IIB) & 
elongated  & 
systematic & 
systematically varying with position &
Mon R2, M$+0.25+0.01$ \\

small (III) & 
symmetrized & 
systematic & 
systematic, increasingly radial &
W51 e2, W51 North \\

\enddata
\tablecomments{Numbers in parentheses refer to phases in the schematic evolutionary scenario
in Figure \ref{figure_schematic_delta}.
}

\end{deluxetable}

\end{document}